\documentclass[11pt,a4paper]{article}
\usepackage[utf8]{inputenc}
\usepackage{amsmath}
\usepackage{amssymb}
\usepackage[table,dvipsnames]{xcolor}
\usepackage{graphicx}
\usepackage[margin=1in]{geometry}
\usepackage{txfonts}       
\usepackage{pdfpages}

\usepackage{tablefootnote}
\usepackage{hyperref}
\hypersetup{
    colorlinks = true,
    urlcolor   = blue,
    citecolor  = black,
}
\usepackage{ragged2e}
\newcolumntype{L}[1]{>{\RaggedRight\arraybackslash}p{#1}}

\usepackage[sort&compress]{natbib}
\setcitestyle{citesep={,}}

\usepackage[font=footnotesize,labelfont=bf]{caption}

\usepackage{authblk}

\newcommand{\dt}[1]{\frac{\partial #1}{\partial t}}

\newcommand{\dxx}[1]{\frac{\partial^2 #1}{\partial x^2}}
\newcommand{\dx}[1]{\frac{\partial #1}{\partial x}}

\newcommand{\dy}[1]{\frac{\partial #1}{\partial x}}

\newcommand{\bm}[1]{\boldsymbol{#1}}
\newcommand{\Peclet}{{P\'{e}clet }}
\newcommand{\Pe}{\mathrm{Pe}}
\newcommand{\Pes}{\mathrm{Pe}^{\star}}
\newcommand{\vwall}{ {v_\mathrm{wall}} }
\newcommand{\De}{ {D_\mathrm{eff}} }
\newcommand{\Ve}{ {V_\mathrm{eff}} }
\newcommand{\Vint}{\mathcal{V}_{\rm int}}
\newcommand{\tVint}{\tilde{\mathcal{V}}_{\rm int}}
\newcommand{\Uint}{\mathcal{U}_{\rm int}}
\newcommand{\dd}{\text{d}}
\newcommand{\ddx}{\text{d}x}
\newcommand{\rt}{\rho_\mathrm{tr}}
\newcommand{\pt}{p_\mathrm{tr}}
\newcommand{\rhob}{\overline{\rho}}
\newcommand{\tdiff}{\tau_{\rm diff}}
\newcommand{\twall}{\tau_{\rm wall}}
\newcommand{\Drho}{\overline{D}_0}
\newcommand{\Pebar}{\overline{\mathrm{Pe}}}
\newcommand{\Dstar}{D^\star}

\title{Interactions enhance dispersion in fluctuating channels via emergent flows}

\author[1]{Yating Wang}
\author[2,3]{David S. Dean}
\author[1,4]{Sophie Marbach \footnote{These authors contributed equally to the supervision of the work.}}
\author[1,5,6]{Ruben Zakine$^*$}

\affil[1]{Courant Institute of Mathematical Sciences, New York University, New York, New York, USA }
\affil[2]{Univ. Bordeaux, CNRS, LOMA, UMR 5798, F-33400, Talence, France.}
\affil[3]{Team MONC, INRIA Bordeaux Sud Ouest, CNRS UMR 5251, Bordeaux INP, Univ. Bordeaux, F-33400, Talence, France.}
\affil[4]{CNRS, Sorbonne Universit\'{e}, Physicochimie des Electrolytes et Nanosyst\`{e}mes Interfaciaux, F-75005 Paris, France.}
\affil[5]{Chair of Econophysics and Complex Systems, École Polytechnique, 91128 Palaiseau Cedex, France}
\affil[6]{LadHyX UMR CNRS 7646, École Polytechnique, 91128 Palaiseau Cedex, France}

\begin{document}
\maketitle

\begin{abstract}

Understanding particle motion in narrow channels is essential to guide progress in numerous applications, from filtration to vascular transport. 
Thermal or active fluctuations of channel walls for fluid-filled channels can slow down or increase the dispersion of tracer particles. Entropic trapping in the wall bulges slows dispersion, and hydrodynamic flows induced by wall fluctuations enhance dispersion. Previous studies primarily concentrated on the case of a single Brownian tracer either embedded in an incompressible fluid or in the ideal case where the presence of fluid is ignored.
Here we address the question of what happens when there is a large ensemble of interacting Brownian tracers -- a common situation in applications. Introducing repulsive interactions between the tracer particles, while ignoring the presence of a background fluid, leads to an effective flow field. This flow field enhances tracer dispersion, a phenomenon strongly reminiscent of that seen in incompressible background fluid. We characterise the dispersion by the long-time diffusion coefficient of a single tracer, numerically and analytically with a mean-field density functional analysis. We find a surprising effect where an increased particle density enhances the diffusion coefficient, challenging the notion that crowding effects tend to reduce diffusion. Here, inter-particle interactions push particles closer to the fluctuating channel walls. Then, interactions between the fluctuating wall and the now-nearby particles drive particle mixing. Our mechanism is sufficiently general that we expect it to apply to various systems. In addition, the perturbation theory we derive quantifies dispersion in generic advection-diffusion systems with arbitrary spatiotemporal drift. 
\end{abstract}

\section*{Introduction}

Fluid transport in confined channels, and generally in porous structures, is relevant for a broad range of biological and industrial applications: from nutrient transport in vascular networks and microorganisms in soils~\citep{tomkins2021update,bhattacharjee2019bacterial}; to improved filtration of liquids or gases, including modern desalination techniques and oil recovery~\citep{werber2016materials,marbach2019osmosis}. However, narrow channels present a number of challenging features to model, to name but a few: the predominance of surface effects, the importance of spatio-temporal fluctuations as well as specific electrostatic response~\citep{kavokine2021fluids}. Significant progress in the last decade has improved our understanding of transport features in such porous environments, and we briefly review below the effects of (i) spatial and (ii) temporal fluctuations of the confining environment as well as (iii) crowding effects due to interactions. 

First, purely spatial corrugations of the confining channel reduce the long-time diffusion coefficient of isolated particles. In fact, random crossings of channel constrictions are rare events that impede overall transport: the constrictions  form effective entropic barriers~\cite{zwanzig1992diffusion}. This effect is well captured by the so-called Fick-Jacobs formalism~\citep{jacobs_diffusion_1967,reguera_kinetic_2001, kalinay_corrections_2006,rubi2019entropic} which reduced the problem to an effectively one dimensional one.  The Fick-Jacobs formalism has recently been extended to arbitrary channel geometries~\citep{chavez2018effects, dagdug2016description}. Recent work suggests that the approach is also adapted to study fluid flow through biological membranes~\citep{arango2020entropic}. Such entropic contributions induce significant corrections to transport in  microfluidic~\citep{yang2017hydrodynamic} or biological channels~\citep{rubi2017entropy}. 

Secondly, and of importance in several applications, confining channels are often not static but fluctuate in time, either due to thermal agitation or to an external forcing. Molecular dynamics simulations found, early on, enhanced diffusion of gas in microporous materials if the thermal vibrations of the material are accounted for~\citep{leroy2004self,haldoupis2012quantifying}, and more recently, enhanced water diffusion in solid state pores via phonon-fluid coupling~\citep{ma2015water,ma2016fast,cao2019water,noh2021phonon}. Recent theoretical work suggests that enhanced diffusion is universally due to longitudinally fluctuating fluid flows, driven by fluctuations of the channel walls, that convect the tracer particles~\citep{marbach2018transport}. The mechanism is thus reminiscent of Taylor-Aris dispersion, where the long-time diffusion coefficient is enhanced by a cross-sectionally inhomogeneous fluid flow profile (in the initially studied case of Poiseuille flow)~\citep{taylor1953,aris1956dispersion}. When the characteristic timescale of wall fluctuations is smaller (resp. larger) than the timescale to diffuse across typical constrictions, diffusion is enhanced (resp. decreased), a criterion which was verified numerically~\citep{chakrabarti2020} and in experiments in fluctuating porous matrices~\citep{sarfati2021enhanced}. Such enhancement of dispersion properties are relevant in a number of biological contexts, such as in blood vessels~\citep{masri2021reduced}, slime mold vasculature~\citep{marbach2019active}, the gut~\citep{codutti2022changing}, near molecular motors~\citep{evans2021cooperative} and are of general relevance in plants~\citep{tomkins2021update}.

Third, in addition to the fluctuating confinement, tracer particles are often not isolated and interact with other particles or molecules that are also diffusing in the medium. The current paradigm is that such {\em crowded} environments, in an open domain, tend to slow down diffusion at equilibrium~\citep{lekkerkerker1984calculation,lowen1993long,dean_self-diffusion_2004}. 
However, subtle effects may arise if the crowded media is driven out of equilibrium. For example, the diffusion coefficient of a tracer driven by an external force may be enhanced at low density and high forcing~\citep{benichou2013fluctuations,demery2014generalized, benichou_tracer_2018, illien_nonequilibrium_2018}. Typically, a trade-off is observed between, on the one hand, the increased diffusion induced by faster exploration of space thanks to the driving force, and on the other hand, the decreased diffusion due to spatial constrictions induced by the confining media~\citep{illien_nonequilibrium_2018}. Such effects may be exploited for active microrheology within spatially corrugated channels~\citep{puertas2018active,malgaretti2022active}.

Given its obvious importance, especially in biological systems, the coupling between the effects of fluctuating channels and inter-particle interactions has received surprisingly little attention. Since temporal channel fluctuations  increase transport coefficients and since inter-particle interactions, or crowding effects,  generally decreases diffusion, it is natural to ask if one can predict which effect dominates? Even though recent numerical work showed that diffusion enhancement could be obtained with increased particle density in microporous matrices~\citep{obliger2019methane}, and sometimes even exhibiting a maximum~\citep{pireddu2019scaling} at a certain density, it was also noticed that the effect can strongly depend on the precise details  of the system  understudy~\citep{obliger2023development}. These results thus call for a general theoretical investigation.

\begin{figure}
    \centering
    \includegraphics[width = \textwidth]{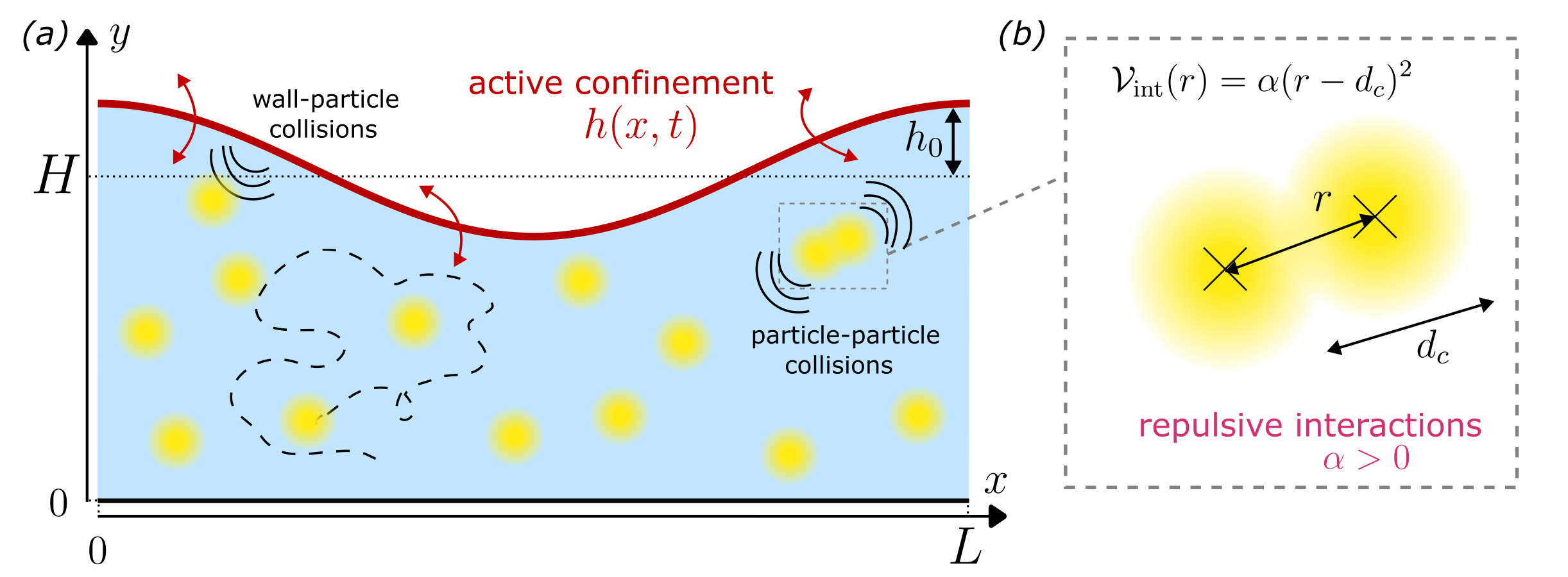}
    \caption{Set up to study the transport of tracers in an interacting system with fluctuating boundaries. (a) Tracer particles (yellow) perform a random walk within this {\em wiggling} environment, here represented by a top, moving wall (red arrows). (b) We consider pairwise, soft, repulsive interactions between particles, and we vary the interaction strength $\alpha$ and the particle number  density $\rho_0$ to investigate more or less crowded environments. 
   }
    \label{fig:sketch}
\end{figure}

In this article, we investigate the motion of diffusing particles with repulsive  interactions in a confined \textit{and} fluctuating channel (see Fig.~\ref{fig:sketch}a) which is essentially a spatially periodic profile moving with constant velocity $\vwall$ . We perform numerical simulations to quantify the effective long-time self-diffusion coefficient $D_{\rm eff}$ and the effective long-time drift $V_{\rm eff}$ of particles. We explain their behaviour for  a broad range of interaction strengths between particles and fluctuating channel speeds $\vwall = \omega_0/k_0$, where $k_0$ and $\omega_0$ are the wave number and frequency of the wall fluctuations. Using perturbation theory, we obtain simple analytical predictions for $\Ve$ and $\De$ that are in excellent agreement with simulations. 

The paper is organised as follows. 
To start (Sec.~\ref{sec:ideal}), we consider a simple ideal gas in the fluctuating channel. We find that the long-time diffusion coefficient $D_{\rm eff}$ exhibits a maximum with respect to the wall phase velocity $\vwall$, corresponding to maximal enhancement of diffusion. This result should be contrasted with that of the diffusion of tracer particles in \textit{incompressible} fluids that exhibits a monotonic increase with the wall phase velocity $\vwall$~\citep{marbach2018transport}. The non-monotonic behaviour seen in the case studied here  originates from the interplay between diffusion and advection due to the wall that increases long time diffusion only  when the diffusive timescale and the advection timescale are comparable. 
In (Sec.~\ref{sec:interactions}) we then consider soft-core interactions between particles (see Fig.~\ref{fig:sketch}b). We find that $D_{\rm eff}$ can be further enhanced with increasing repulsive interactions. Here, repulsive interactions play a role in generating a more uniform distribution of particles in the channel, even in the vicinity of rapidly fluctuating bulges. Eventually, increased wall-particle collisions, caused by now-nearby particles, enhance dispersion. This behaviour is reminiscent of the mechanism of enhanced tracer diffusion in a fluctuating channel filled with an incompressible fluid~\citep{marbach2018transport}, and indeed we show, that remarkably, analytically and numerically, transport coefficients converge to a universal incompressible fluid regime in the limit of strong repulsive interactions. 
Finally, we discuss how these mechanisms and techniques may be further used to investigate more complex situations of transport in fluctuating confined environments.

\section{Transport of ideal (isolated) tracers in a fluctuating channel}
\label{sec:ideal}

\subsection{Simulation results}
We start by exploring the motion of tracers in a fluctuating channel, where the environment is not crowded, \textit{i.e.} where tracer particles are sufficiently far away from each other that we can consider that they do not interact. The tracer particles perform a random walk  with diffusion coefficient $D_0$. 
We refer to this case as the ``ideal gas'' regime. 
Throughout this study, we will assume that the impact of the fluctuating channel boundaries on the velocity field of the supporting fluid is negligible. This is the case of particles embedded in a highly compressible fluid or gas, for which the mean free path is much smaller than any relevant lengthscale in the system, or by considering that the boundaries are not made of \textit{hard} walls but rather of potential barriers, such as (fluctuating) electrostatic potentials for charged tracers. The impact of fluctuating boundaries on the fluid's velocity field has already been characterised in the Stokes flow regime~\citep{marbach2018transport}.

We simulate the motion of these non-interacting particles in a fluctuating channel, described through a fluctuating wall at height $h(x,t) = H + h_0 \cos \left( k_0 x - \omega_0 t \right)$. The shape of the fluctuating interface is chosen to be sinusoidal as a generic interface profile  can be decomposed in terms of plane waves. The presence of the walls is incorportated via a soft boundary potential acting  on the particles. We track the motion of particles over long-times and evaluate their effective long-time self-diffusion coefficient $D_{\rm eff}$ and mean drift $V_{\rm eff}$ along the main channel axis $x$ (see Appendix~\ref{app:simulation} for details). 

\begin{figure}
    \centering
    \includegraphics[width = \linewidth]{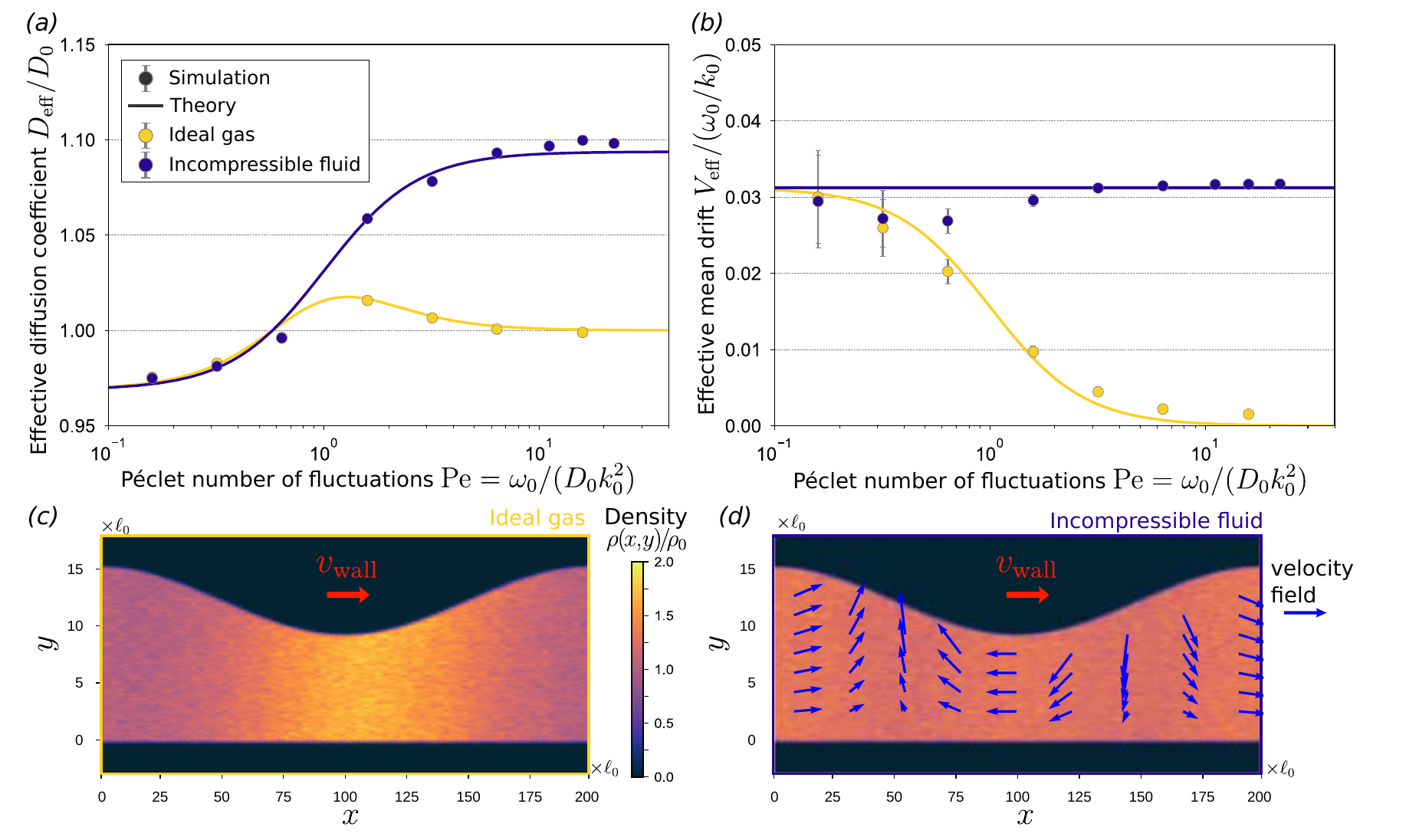}
    \caption{Transport properties of non-interacting tracers in a periodic driven channel; comparison between a compressible (ideal gas) and an incompressible supporting fluid. (a) Long-time longitudinal diffusion coefficient and (b) drift of particles with $\mathrm{Pe} = \omega_0/D_0 k_0^2 =v_\mathrm{wall}L/(2\pi D_0)$. In (a) and (b) error bars correspond to one standard over 10 independent runs, are smaller than point size, except for (b) small $\Pe$, since $V_{\rm eff} \simeq \vwall \sim 0.01 \ell_0/\tau_0$ is small compared to the noise level. Ideal gas theory corresponds to ~\eqref{eq:Deperiodic} and~\eqref{eq:Veperiodic} and incompressible fluid to~\eqref{eq:DeFlow} and \eqref{eq:VeFlow}. (c) Stationary density profile of particles for $\vwall=0.1 \ell_0/\tau_0$ (or $\mathrm{Pe} \simeq 3$) for the ideal gas and (b) 
    for an incompressible fluid for $\vwall=0.5 \ell_0/\tau_0$ ($\mathrm{Pe} \simeq 16$). 
    Blue arrows: velocity field in the channel, represented in the referential of the moving wall, with a length proportional to their magnitude (arbitrary scale).  
    The color scale for the density profiles is shared in (c) and (d) and yellow regions (resp. purple) indicate regions of high (resp. low) densities. Other numerical parameters are $L=200\ell_0$, $H=12 \ell_0$, $h_0=3\ell_0$, and $D_0=1 \ell_0^2/\tau_0$. 
}
    \label{fig:idealVsFlow}
\end{figure}

Previous work on incompressible fluids has established that a relevant parameter to analyse the system is given by the \Peclet number characterising the fluctuations \begin{equation}
    \Pe = \frac{\omega_0}{D_0 k_0^2} = \frac{\tdiff}{\twall},
\end{equation} which compares the timescale to diffuse across the length of a channel corrugation $\tdiff = 1/D_0 k_0^2$ to the period of the channel fluctuations $\twall = 1/\omega_0$~\citep{marbach2018transport}. In simulations we therefore fix the typical channel corrugation length $L = 2 \pi/ k_0$ and vary the wall phase velocity $\vwall = \omega_0/k_0$. All parameters are expressed in terms of a time unit $\tau_0$ and a length unit $\ell_0$, that are arbitrary. We present the results (yellow dots) in Fig.~\ref{fig:idealVsFlow}a and b for $\De$ and $\Ve$ with increasing $\Pe$. 

Interestingly, the long-time diffusion coefficient $\De$ exhibits a non-monotonic dependence on the \Peclet number $\Pe$ (yellow dots in Fig.~\ref{fig:idealVsFlow}a). 
At low \Peclet numbers, $\Pe \ll 1$, particles move much faster than the wall, $\twall \gg \tdiff$ and hence the wall appears {\em frozen}. Particles therefore spend time exploring the bulges before escaping the constrictions, and their diffusion coefficient is consequently decreased $D_{\rm eff} \leq D_0$. This effect is well captured by the Fick-Jacobs approximation, where transport is reduced due to constrictions acting as effective entropy barriers~\citep{jacobs1935diffusion,zwanzig1992diffusion,jacobs_diffusion_1967,burada2007biased,reguera_kinetic_2001,kalinay_corrections_2006,mangeat2017dispersion,marbach2018transport,rubi2019entropic}. At intermediate \Peclet numbers, $\Pe \gtrsim 1$,  $\twall \simeq \tdiff$ and the moving boundaries enhance particle motion: particles collide with the moving corrugated walls, which increases their diffusion coefficient overall $D_{\rm eff} \geq D_0$. At high \Peclet numbers, $\Pe \gtrsim 10$, $\twall \ll \tdiff$, the wall moves so fast that particles no longer have time to enter the bulges, and therefore behave as though they were not seeing any corrugation, the effective diffusion coefficient is unchanged and equals the bare diffusion $D_{\rm eff} = D_0$. Said otherwise, here the from the point of view of the particles, the channel is effectively flat with a height given by its minimum height.

To further understand  the behaviour at high \Peclet numbers, we plot the average density distribution $\rho(x,y)$ of particles within the channel (the average being over noise), in the reference frame  where the  channel profile is stationary, for a large value of $\vwall$ (Fig.~\ref{fig:idealVsFlow}c). We observe that particles accumulate at the constriction. Indeed, in this reference frame, the channel constriction acts as a bottleneck for transport. This can be seen as an {\em inverse} Bernoulli effect: in the channel's frame of reference, the particle flux $\rho(x) h(x) v_{\rm wall}$ is necessarily constant, where $\rho(x)$ is the cross-sectional averaged density of particles. Hence the density (and not the velocity, as in the Bernoulli effect) is largest where the channel is constricted, $\rho(x) \propto 1/h(x)$.   Eventually, since the particles only explore low vertical coordinates $ y \lesssim H - h_0$, they no longer collide with the moving wall and hence their diffusion coefficient is unchanged. 

The effective particle drift $\Ve$ monotonically decreases with the \Peclet number (yellow dots in Fig.~\ref{fig:idealVsFlow}b, noting that the vertical axis is the normalised drift relative to the wall phase velocity, $\Ve/\vwall$). At small \Peclet numbers, the density of particles is uniform in the channel and therefore the  particles within the bulge are {\em carried along} in the same direction as the wall phase velocity. At higher \Peclet numbers the fraction of particles within the bulge decreases, as seen in Fig.~\ref{fig:idealVsFlow}c. Since all particles accumulate within the bottleneck, they are no longer pushed by the moving wall and hence $\Ve/\vwall \rightarrow 0$.

\subsection{Analytic theory to account for transport in fluctuating channels}
\label{sec:methodo}

To account for this broad behaviour we build a general analytic theory that reproduces these effects. 
With the goal of making a pedagogical introduction to our perturbation method, which closely follows  that which was only briefly described in \cite{marbach2018transport}, we devote the following subsection to a detailed explanation. 
\subsubsection{Constitutive equations}

Brownian tracer particles evolve in a fluctuating environment, confined in the $y$ direction between $y = 0$ and $y = h(x,t)$, and infinitely long in the $x$ direction. Compared to the simulation, we here make no assumption on the form of $h(x,t)$, which may describe thermal motion of the wall (determined \textit{e.g.} through a Hamiltonian characterising the flexibility of the interface) or active motion (driven by an external source of energy, as in our simulations where fluctuations are imposed). The probability density function $\rt(x,y,t)$ to find a tracer particle at position $x,y$ at time $t$ obeys the Fokker-Planck equation
\begin{equation}
\begin{split}
    \dt{\rt(x,y,t)}&+ \bm \nabla \cdot \bm j (x,y,t) = 0  \\
\text{with} \,\,    \bm j(x,y,t) &\equiv - D_0  \bm \nabla \rt(x,y,t) + \bm u (x,y,t) \rt(x,y,t), 
    \label{eq:FPE1}
\end{split}
\end{equation}
and where we have assumed that the tracer's diffusion coefficient $D_0$ is uniform in space and that the tracer is advected by the field $\bm u (x,y,t)$ which can have both potential and non-potential components. In the case of ideal non-interacting particles, the underlying supporting fluid does not move due to the moving interface and there are no interactions, $\bm u (x,y,t) \equiv 0$. For now we keep $\bm u(x,y,t)$ arbitrary as this will be useful in the incompressible and interacting regimes.

\subsubsection{Boundary conditions}
\label{sec:bcs}

We impose no flux boundary conditions at both surfaces. This means that the projection of the current on the direction normal to the lower surface boundary is zero, and that similarly, the projection of the current on the direction normal to the upper boundary in the frame moving with speed $\bm U_\mathrm{wall}$, where the surface is static, is zero.
The boundary conditions thus read
\begin{eqnarray}
    j_y(x,0,t) &=& 0 \label{eq:bcbot} \\
     \bm{j} (x,h(x,t),t)  \cdot \bm n   &=& - \rho (x,h(x,t),t) \, \bm U_{\rm wall} \cdot \bm n \label{eq:bctop}.
\end{eqnarray}
Here $\bm{n}$ is the outward normal to the interface, $ \bm U_{\rm wall} = (0, \partial_t h)$ is the velocity of the wall, considering that the wall atoms move vertically about their average position, similarly to phonon modes that propagate on material structures~\citep{chaikin1995principles} or peristaltic motion of vasculature~\citep{alim2013random,marbach2019active}. We derive these boundary conditions from first principles in Supplementary Material Sec.~1.

\subsubsection{Simplified longitudinal equation for the probability distribution}
To make progress on the long-time behaviour of \eqref{eq:FPE1}, we place ourselves within the lubrication approximation. This means we consider the typical corrugation length $L$ to be much bigger than the average channel height, $\langle h(x,t) \rangle = H$, $H \ll L$, itself much bigger than the amplitude of the channel fluctuations, i.e. $\sqrt{\langle (h - H)^2 \rangle}\propto h_0  \ll H$. Within the lubrication approximation, the outward normal to the interface in \eqref{eq:bctop} simplifies to $\bm n \simeq (\partial_x h,1)$. 
In this framework, the particle density relaxes much faster on the vertical direction $y$ than on the longitudinal direction $x$. At low \Peclet number, this should notably yield a vertically uniform particle density. 
Fig.~\ref{fig:idealVsFlow}c shows that the density profile is indeed independent of $y$ in our simulations. Also, in Appendix~\ref{sec:lubrication} we show that even though the lubrication approximation should fail for \textit{e.g.} larger channel heights $H$, the results derived below remain surprisingly robust. 

We therefore look for an approximate evolution equation on the probability distribution integrated vertically, or \textit{marginal distribution}, $\pt(x,t) = \int_0^{h(x,t)} \rt(x,y,t) \, \dd y$. Taking the time derivative of $\pt(x,t)$ and using \eqref{eq:FPE1} yields 
\begin{align}
    \dt{\pt(x,t)} = &\dt{h(x,t)}\rt(x,h(x,t),t)- \int_0^{h(x,t)}  \left( \dy{j_y(x,y,t)} + \dx{j_x(x,y,t)} \right) \dd y  \nonumber \\
    = &  \dt{h(x,t)}\rt(x,h(x,t),t) - j_y(x,h(x,t),t) + j_y(x,0,t) \nonumber \\
    &+ \dx{h(x,t)} j_x(x,h(x,t),t) - \dx{} \left(\int_0^{h(x,t)}  j_x(x,y,t) \dd y\right) 
    \label{eq:marginal_firstEq}
\end{align}
where we used the simple and useful relation
\begin{equation}
    \dx{} \left( \int_0^{h(x,t)} f(x,y,t) \dd y\right) = \dx{h(x,t)} f(x,h(x,t)) + \int_0^{h(x,t)} \dx{f(x,y,t)} \dd y \nonumber
    \label{eq:trick}
\end{equation}
for any function $f$.
The above Eq.~\eqref{eq:marginal_firstEq} can be remarkably simplified, by using the no flux boundary conditions Eq.~\eqref{eq:bctop} and Eq.~\eqref{eq:bcbot} which lead to all the surface terms vanishing and we find
\begin{equation}
    \dt{\pt(x,t)} = - \dx{} \left(\int_0^{h(x,t)}  j_x(x,y,t) \dd y\right).
\label{eq:marginal_firstEq2}
\end{equation}
We look for a closed equation on $\pt(x,t)$. Writing $j_x(x,y,t) = - D_0 \dx{\rt(x,y,t)} + u_x \rt(x,y,t)$ explicitly and using the relation Eq.~\eqref{eq:trick} again we find
\begin{equation}
    \dt{\pt(x,t)} = - \dx{} \left(- D_0 \frac{\partial\pt(x,t)}{\partial x} + D_0 \dx{h(x,t)} \rt(x,h(x,t),t) + \int_0^{h(x,t)}u_x\rt(x,y,t) \dd y \right) \nonumber
\end{equation}
where the last terms still do not depend explicitly on the marginal $\pt$. We will therefore make the common first order approximation that, since the probability distribution profile is nearly uniform vertically, we may assume $\rt \sim \frac{1}{h} \pt$. This step can be made more rigorous  using a center manifold expansion, see \cite{mercer1990centre,marbach2019active}. We then obtain the following Fokker-Planck equation on the marginal distribution
\begin{equation}
    \dt{\pt} =  - \dx{} \left( -D_0 \dx{\pt} + D_0 \pt\dx{\ln h} + \overline{u_x} \pt \right),
    \label{eq:Marginal_brownianConfined}
\end{equation}
where $\overline{u_x} = \frac{1}{h} \int_0^{h(x,t)} u_x \dd y$ is the vertically-averaged longitudinal drift.
Eq.~\eqref{eq:Marginal_brownianConfined} clearly simplifies the initial problem and it is sufficient to study the long-time behaviour of $\pt$ to obtain the long-time diffusion coefficient $\De$ and drift $\Ve$. When $\overline{u_x} \equiv 0$, Eq.~\eqref{eq:Marginal_brownianConfined} corresponds exactly with the Fick-Jacobs equation~\citep{jacobs1935diffusion,zwanzig1992diffusion,jacobs_diffusion_1967,reguera_kinetic_2001} which describes the motion of (non-interacting) particles in  spatially varying but time independent channels $y \equiv h(x)$. Interestingly, the Fick-Jacobs equation is therefore also valid for fluctuating channels \textit{in time}, $h(x,t)$, regardless of the functional shape of $h(x,t)$, the only assumption being the lubrication approximation. We note that, \eqref{eq:Marginal_brownianConfined} is also consistent with the case of a background incompressible fluid, when $\overline{u_x}$ is the cross-sectionally averaged fluid velocity (Eq.~(1) of \cite{marbach2018transport}). 


\subsubsection{Perturbation theory to obtain the long-time transport behaviour}

Our goal is now to obtain the long-time transport behaviour of particles from the simplified, longitudinal evolution equation \eqref{eq:Marginal_brownianConfined}. We seek an approximate long-time equation for the marginal distribution $p_{\rm eff}(x,t) = \pt(x, t \rightarrow\infty)$ of the form
\begin{equation}
    \dt{p_{\rm eff}(x,t)} = D_{\rm eff} \dxx{p_{\rm eff}(x,t)} - V_{\rm eff} \dx{p_{\rm eff}(x,t)} + \delta(x)\delta(t)
    \label{eq:FPE_goal}
\end{equation}
such that we can naturally read off the long-time diffusion $D_{\rm eff}$ and drift $V_{\rm eff}$, and where we assumed, without loss of generality, that the particle was initially at the center of the domain. 

To keep the calculations general, we rewrite \eqref{eq:Marginal_brownianConfined} as a Fokker-Planck equation 
\begin{equation}
    \dt{\pt(x,t)} =  - \dx{} \left( - D_0 \dx{\pt(x,t)} +  v(x,t) \pt(x,t) \right) + \delta(x)\delta(t),
    \label{eq:FPEgeneral}
\end{equation}
where $v(x,t)$ is a general advection coefficient (for the ideal gas case, $v = D_0 \partial_x h / h$). 
Here, we consider that $v(x,t) = O(\varepsilon)$ is a fluctuating perturbation. Its average over realizations of the noise vanishes
$\langle v(x,t)\rangle = 0$ and its fluctuations are described in Fourier space by a spectrum $S(k,\omega)$ as
\begin{equation}
     \langle \tilde{v}(k,\omega) \tilde{v}(k',\omega') \rangle  = S(k,\omega) (2\pi)^2 \delta(k + k')\delta(\omega + \omega') 
     \label{eq:sv}
\end{equation}
where $k$ and $\omega$ are the wave number and frequency respectively, and the Fourier transform $\tilde{v}(k,\omega)$ of $v(x,t)$ is defined in \eqref{eq:FFT}.
Performing a perturbation development to solve \eqref{eq:FPEgeneral} on the small parameter $\varepsilon$ we find (see Appendix~\ref{app:perturbation})
\begin{align}
D_{\rm eff} & = D_0 \left(1- \varepsilon^2 \int \frac{\dd k \dd \omega}{(2\pi)^2} \frac{k^2( D_0^2 k^4 - 3 \omega^2)}{( D_0^2 k^4+ \omega^2 )^2} S(k,\omega)\right),
    \label{eq:De1} \\
V_{\rm eff} &= \varepsilon^2 \int \frac{\dd k \dd \omega}{(2\pi^2)} \frac{\omega}{k} \frac{k^2}{D_0^2k^4 + \omega^2} S(k,\omega).
    \label{eq:Ve1}
\end{align}
Equations \eqref{eq:De1} and \eqref{eq:Ve1} are one of the main results of this work. They predict the long-time transport properties of particles within a fluctuating channel with arbitrary fluctuating local drift $v(x,t)$. The results can be applied \textit{regardless} of the nature of the fluctuations, be they thermal or non-equilibrium, and \textit{regardless} of the strength of the interactions in-between particles and between particles and the supporting fluid. In essence, we generalise the results of \citet{marbach2018transport} which were only valid for an incompressible supporting fluid. 

\subsection{Applications of the theory to the periodic channel}
\label{sec:analytics}

\subsubsection{Ideal gas}
To use the above analytic framework to describe our simulations, we now take the case of the periodic traveling wave $h(x,t) = H + h_0 \cos \left( 2\pi (x - \vwall t ) /L \right) = H + h_0 \cos( k_0 x - \omega_0 t)$ on the surface. The Fourier transform of $h - H$ is
$\tilde{h}(k,\omega) = (h_0/2) \left( \delta(k+k_0) \delta(\omega - \omega_0) + \delta(k-k_0)\delta(\omega + \omega_0)  \right)$.
The local drift $v$ in the ideal gas case is at lowest order in $\varepsilon = h_0/H$,
\begin{equation}
    v(x,t) = D_0 \frac{1}{H} \dx{h} \,\, \,\,\text{ and in Fourier space }\,\,\,\, \tilde{v}(k,\omega) = i D_0 k \frac{\tilde{h(k,\omega)}}{H}.
    \label{eq:uideal}
\end{equation} 
This yields a spectrum for the local drift as
\begin{equation}
    S(k,\omega) = \frac{D_0^2 k^2}{(2\pi)^2} \frac{h_0^2}{4 H^2} \left( \delta(k+k_0) \delta(\omega - \omega_0) + \delta(k-k_0)\delta(\omega + \omega_0)  \right).
\end{equation}
Plugging the expression for the spectrum in \eqref{eq:De1} and \eqref{eq:Ve1} yields the long-time transport coefficients in the ideal gas case
\begin{align}
    \frac{D^{\rm \textit{ideal gas}}_{\rm eff}}{D_0} &= 1  + \frac{h_0^2}{2 H^2}\frac{3\Pe^2-1}{(\Pe^2 + 1)^2},
    \label{eq:Deperiodic} \\
    \frac{V^{\rm \textit{ideal gas}}_{\rm eff}}{v_{\rm wall}} &=  \frac{h_0^2}{2 H^2} \frac{1}{1 + \Pe^2}.
    \label{eq:Veperiodic}
\end{align}
where we recall the expression of the \Peclet number
    $\Pe = \omega_0/D_0 k_0^2$.

We present plots of \eqref{eq:Deperiodic} and \eqref{eq:Veperiodic} as lines in Fig.~\ref{fig:idealVsFlow}a and b, and find excellent agreement with simulations. This agreement is robust over a wide range of physical parameters -- see Fig.~\ref{fig:variation_channelHeight}. Analytically, we recover for $\Pe \rightarrow 0$ (corresponding to fixed channels, $\vwall = 0$) the well-known entropic trapping result where $D_{\rm eff} = D_0 \left( 1- h_0^2/2H^2\right)$~\citep{zwanzig1992diffusion, jacobs_diffusion_1967, reguera_kinetic_2001, marbach2018transport, rubi2019entropic}. The analytic computation recovers the non monotonic behaviour of the diffusion coefficient for intermediate $\Pe$, and confirms that $D_{\rm eff} = D_0$ at high $\Pe\to \infty$. Finally, the amplitude of the correction for both the diffusion and the drift scales as $h_0^2/H^2$, which can be interpreted naturally and confirms the collision mechanism in bulges: indeed, the strength of the fluctuations scales as $h_0/H$ but only a fraction $h_0/H$ of particles lie within the bulge and takes part in the enhancement of the diffusion coefficient or in the mean drift.


\subsubsection{Comparison with transport in \textit{incompressible} fluid}

We now relate our results for the ideal gas to transport within incompressible fluids~\citep{marbach2018transport}. In that case, when the channel walls fluctuate, they induce, because the supporting fluid is incompressible, fluctuations in the fluid's velocity field. Particles are thus advected by fluid flow. The channel walls here are perfectly slipping walls, which corresponds to the smooth boundary conditions considered for the gas particles, that have no specific lateral friction with the walls. The flow field $(u_x,u_y)$ can be derived in the low Reynolds number limit, that is the relevant limit to consider since we envision applications in micro to nanofluidics. We find
\begin{align}
     & u_x(x,y,t) = U_0(x,t)      \label{eq:ff1}
 \\
     & u_y(x,y,t) =  \frac{y}{h} \left( \dt{h(x,t)} +  U_0 \dx{h(x,t)} \right) 
     \label{eq:ff2}
\end{align}
where $U_0(x,t)$ is the average fluid flow. We calculate $U_0(x,t)$ assuming  \textit{peristaltic} flow, \textit{i.e.} that the flow is purely driven by channel fluctuations and that there is no mean pressure-driven flow~\citep{marbach2019active, chakrabarti2020}. The average pressure force on the fluid has to vanish, and we find $U_0(x,t) \simeq v_{\rm wall}  (h_0/H) \cos \left( 2\pi (x - v_{\rm wall} t )/L \right) $ at lowest order in $h_0/H$ (see Supplementary Material Sec.~2). 
The flow field is presented in Fig.~\ref{fig:idealVsFlow}d as blue arrows. As the channel moves towards the right hand side, fluid mass in the right hand side bulge has to swell out, consistently with the flow lines. Similarly the bulge on the left hand side opens up, allowing fluid flow to come in. 

The advection term now has two contributions, one coming from the spatial inhomogeneities, and one coming from advection by fluid flow
\begin{equation}
    v(x,t) = D_0 \frac{1}{H} \dx{h(x,t)} + U_0(x,t) \label{eq:uflow}
\end{equation}
or $\tilde{v}(k,\omega) = \left( i D_0 k + \omega/k \right) \tilde{h}(k,\omega)/H$ in Fourier space. The spectrum of the fluctuating drift is thus
\begin{equation}
    S(k,\omega) = \frac{\left( D_0^2 k^2 + \omega^2/k^2 \right)}{(2\pi)^2} \frac{h_0^2}{4 H^2} \left( \delta(k+k_0) \delta(\omega - \omega_0) + \delta(k-k_0)\delta(\omega + \omega_0)  \right).
\end{equation}
We then obtain
\begin{align}
    &    \frac{D_{\rm eff}^{\rm \textit{inc. fluid}}}{D_0} = 1 + \frac{h_0^2}{2H^2} \frac{3 \Pe^2 -1}{\Pe^2 + 1}, \label{eq:DeFlow}\\
     & \frac{V_{\rm eff}^{\rm \textit{inc. fluid}}}{\vwall} = \frac{h_0^2}{2H^2}.
    \label{eq:VeFlow}
\end{align}

We perform numerical simulations where particles are advected by the flow field defined by \eqref{eq:ff1} and \eqref{eq:ff2}. We present the numerical results as blue dots and the analytical results as blue lines in Figs.~\ref{fig:idealVsFlow}-a and b.
We find perfect agreement between simulation and theory, confirming the analytical approach of \cite{marbach2018transport}. 

In contrast with the ideal gas, when particles are surrounded by an incompressible fluid, $\De$ increases monotonically until it reaches a plateau at large $\Pe$, and $\Ve$ stays constant regardless of $\Pe$. In fact, at any value of $\Pe$ and especially at high $\Pe$, the density distribution of particles within the channel is \textit{uniform}, as can be seen in Fig.~\ref{fig:idealVsFlow}d. Therefore, even at high $\Pe$, the population of particles lying within the bulges is pushed by the moving boundaries and increases both $\De$ and $\Ve$. 

In this case it is natural to ask why  the density distribution remains homogeneous along the channel? Looking closely at the velocity field (Fig.~\ref{fig:idealVsFlow}d) shows that particles are carried away from the bottleneck by the flow field, and into the bulges. The flow field therefore facilitates re-circulation of accumulated particles. 
This naturally raises the question of how the supporting fluid's compressibility changes the transport properties of particles within fluctuating channels.

\section{Interactions increase diffusion and drift in fluctuating channels}
\label{sec:interactions}

 To investigate the impact of the compressibility of the supporting fluid, we introduce interactions between the particles  (see Fig.~\ref{fig:sketch}b), and tune the interaction strength to vary the effective compressibility of the system.

\subsection{Pairwise interactions and compressibility}

We simulate the dynamics of interacting particles within a simple periodic fluctuating channel $h(x,t) = H + h_0\cos(k_0 x - \omega_0t)$, as in Sec.~\ref{sec:ideal}. We use a pairwise interaction potential between particles, 
\begin{equation}
    \Vint(r) =\begin{cases} \alpha (r - d_c)^2  &\text{ if $r < d_c$,}\\
    0 &\text{if $r \geq d_c$}
    \end{cases}
    \label{eq:VintSoft}
\end{equation}
where $r$ is the interparticle distance, $d_c$ is a critical distance characterising the radius of the interaction (see Fig.~\ref{fig:sketch} b) and $\alpha$ is a spring constant that characterises the strength of the interaction. Note that $\alpha>0$ corresponds to repulsive interactions while $\alpha<0$ corresponds to attractive interactions.
We use a soft-core potential (instead of a hard-core potential as in \textit{e.g.} \cite{benichou2013fluctuations,suarez2015transport}) as it facilitates numerical integration over long timescales, which is necessary to obtain reliable statistics on the diffusion coefficient. Soft-core potentials are commonly used and also simplify analytic treatments~\citep{pamies_hertzian_2009,demery2014generalized, antonov_driven_2021}.
We will show later that the numerical results are well reproduced by our theory whose predictions are robust to strong changes of the interaction potential. The mix of interacting particles and the surrounding (compressible) fluid forms a fluid of interacting particles, and we study its properties below. 

The choice of interactions is well adapted to tune the compressibility $\chi^{\star}_T$ of the fluid of interacting particles, and hence probe different compressibility regimes, in between the {\em ideal gas} and the {\em incompressible} fluid cases. Indeed, $\chi^{\star}_T$  is related to the structure factor $S_f(k)$ of the gas at zero wavelength~\citep{hansen2013theory}:
\begin{equation}
    \chi^{\star}_{\scriptstyle{T}}(\rho_0,\alpha) = \chi_T^\mathrm{id}\lim_{k \rightarrow 0} S_f(k),
\end{equation}
where $\chi_T^\mathrm{id} = 1/(\rho_0 k_BT)$ is the compressibility of the {\em ideal gas}, which is infinite when $\rho_0 \rightarrow 0$. 
In the so-called and broadly used random phase approximation, one can relate the structure factor $S_f(k)$ to the interaction potential, as 
\begin{equation}
    S_f(k) = \frac{1}{1 - \rho_0 \left( - \frac{1}{k_BT} \tilde{\mathcal{V}}_{\rm int}(k)\right)}.
\end{equation}
We calculate
\begin{equation}
     \lim_{k\rightarrow 0} \tilde{\mathcal{V}}_{\rm int}(k) = \iint \mathcal{V}_{\rm int}(x,y) \dd x \dd y = \frac{\pi}{6} \alpha d_c^4.
\end{equation}
We therefore obtain the compressibility of the fluid of interacting particles as
\begin{equation}
    \chi^{\star}_T(\rho_0,\alpha) =\frac{ \chi_T^\mathrm{id}}{1 + \frac{\pi}{6} \frac{\rho_0 \alpha d_c^4}{k_B T}} \equiv  \frac{ \chi_T^\mathrm{id}}{1 + \frac{E_0}{k_B T}},
\end{equation}
where we introduced $E_0 = (\pi/6) \alpha d_c^4\rho_0$ that can be understood as the energy contribution of interactions contained in a typical volume $d_c^d$, with $d$ the dimension of space (here $d=2$). For $\alpha<0$, one has $E_0<0$ and the fluid of particles ends up with a higher compressibility than the one of the ideal gas. In what follows we will mainly focus on repulsive interactions ($\alpha>0$), but we keep in mind that our analytic computation does not assume the sign of $\alpha$.
For $\alpha>0$, the compressibility $\chi^{\star}_T(\rho_0,\alpha)$ decreases with the gas density $\rho_0$ and with increasing interaction strength. Therefore, any of the parameters $\rho_0$ or $\alpha$ are good candidates to probe the intermediate regime between the ideal gas and the incompressible fluid for which $\chi_T^\star\to 0$. In the next section we explore varying values of the density $\rho_0$, and we will find similar results when varying $\alpha$ (reported in Appendix \ref{sec:addDataCompressible}). 

Finally, it is known that the long-time self-diffusion coefficient of interacting particles, in the bulk, \textit{i.e.} in an open domain, $\Drho(\rho_0)$ is decreased in a crowded environment in equilibrium, compared to the infinitely dilute case~\citep{demery2014generalized} (although it may be non-monotonic at high densities as the energy landscape is {\em flattened} when the local density is large for soft interactions, see Fig.~\ref{fig:D0_a}). We therefore first perform simulations within fixed flat walls $(h_0 = 0, \vwall = 0)$, where the system is in equilibrium at temperature $T$. The confinement plays no particular role in that case because our channels are wide enough compared to the typical size of a particle ($d_c \ll H$). We measure the diffusion coefficients $\Drho(\rho_0)$ for various particle densities in the channel $\rho_0$ and interaction strengths $\alpha$, and our results agree well with existing theories~\citep{demery2014generalized} (see Appendix~\ref{sec:appD0}). We can thus now measure changes in the self-diffusion coefficient when there are fluctuations relative to this bulk value $\Drho(\rho_0)$. 

\subsection{Increasing interactions enhances diffusion and drift}

\begin{figure}
    \centering
    \includegraphics[width = \textwidth]{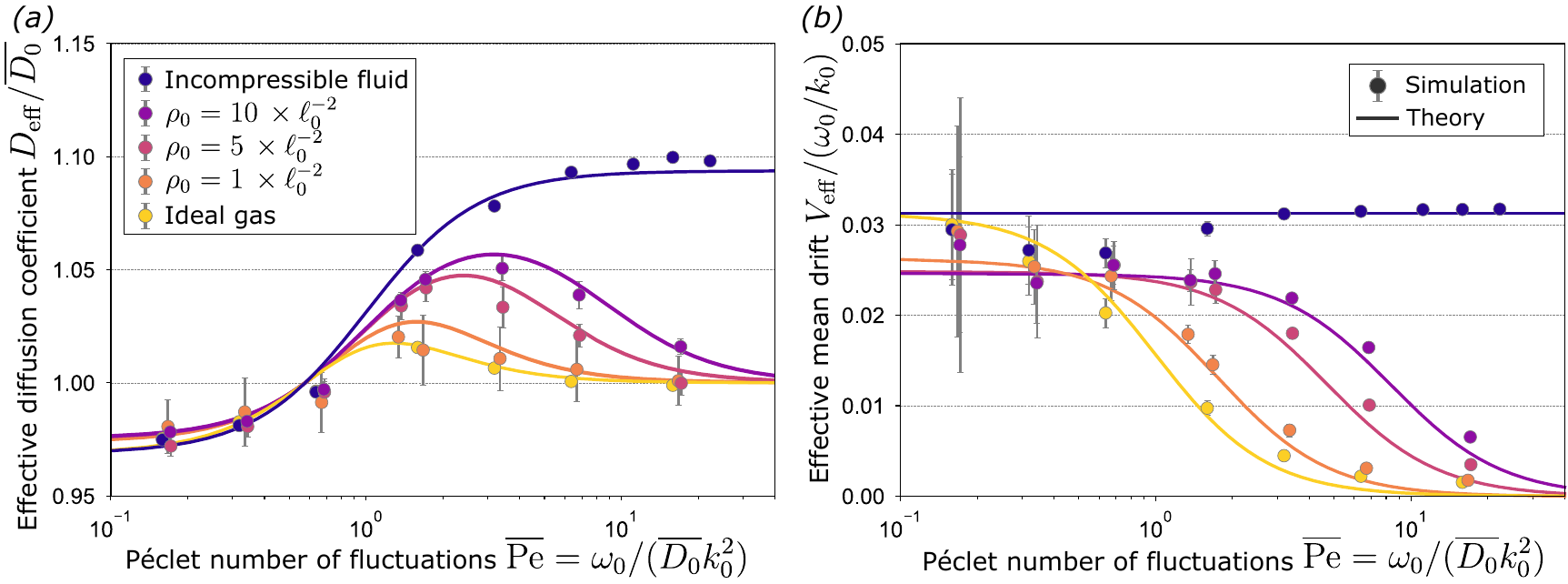}
    \caption{Transport properties of a fluid of interacting particles. Effective diffusion (a) and mean drift (b) for a compressible fluid of soft-core interacting Brownian particles with varying density $\rho_0$. Numerical parameters used here are similar to Fig.~\ref{fig:idealVsFlow} with additionally $L=200 d_c$, $d_c = 2^{1/6}\ell_0 \simeq 1.12\ell_0$, $\alpha =1 k_B T/\ell_0^2$. Error bars correspond to one standard deviation over 10 independent runs. Error bars for lower density values are larger due to a smaller number of tracked particles (see Table.~\ref{tab:tab1}). Theory curves for the ideal gas and incompressible fluid are the same as for Fig.~\ref{fig:idealVsFlow} a and b. Theory curves for the fluid of interacting particles correspond to \eqref{eq:DeUnified} and \eqref{eq:VeUnified}. }
    \label{fig:DeVsRho}
\end{figure}

We perform simulations of interacting particles in fluctuating channels ($h_0 = H/4$, varying $\vwall > 0$).  We present our numerical results for the long-time self diffusion coefficient $\De/\Drho(\rho_0)$ and mean drift $\Ve/\vwall$ with increasing particle density $\rho_0$ in Fig.~\ref{fig:DeVsRho}a and~\ref{fig:DeVsRho}b. We find striking variations with increasing density. At low $\Pe$, systems of interacting particles differ very little from systems with no interactions: we still observe entropic slow-down. At intermediate $\Pe$ interestingly, the enhancement of the diffusion coefficient increases with particle density $\rho_0$. The turnaround point at a critical $\Pe$ increases also with increasing particle density. Similarly, mean particle drift is significant at larger $\Pe$ numbers, the more so with increasing particle density. Increasing particle density thus does appear to bridge the gap between the ideal gas case and the incompressible fluid. 

\subsection{Increasing interactions impacts the density profile}

\begin{figure}
    \centering
    \includegraphics[width = \textwidth]{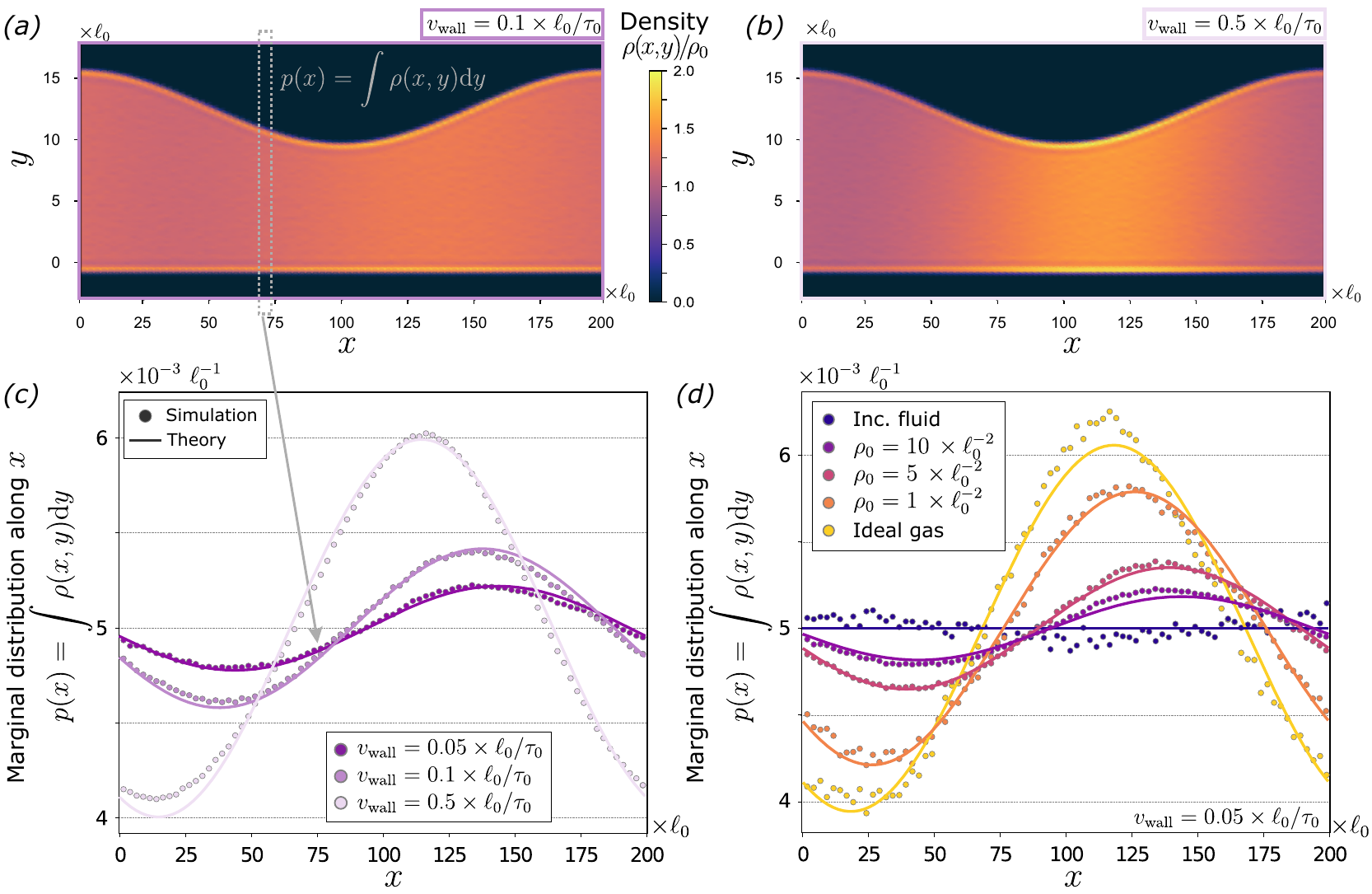}
    \caption{Particle distribution within the channel for varying particle densities. (a,b): Particle distribution in a fluid of interacting particles at high density $\rho_0=10 \ell_0^{-2}$, (a) for intermediate $\vwall=0.1 \ell_0/\tau_0$ and (b) high $\vwall=0.5 \ell_0/\tau_0$ wall velocity. The color scale for the density profiles is shared between (a) and (b) and yellow regions (resp. purple) indicate regions of high (resp. low) densities. (c,d): Marginal density $p(x)=\int\rho(x,y) \dd y$ of particles in the channel, calculated by integrating the 2D density profile in (a) over vertical slabs as with the dashed gray box. (c)  $p(x)$ for different values of $\vwall$, and (d) for different values of particle density $\rho_0$. Numerical parameters used here are similar to Fig.~\ref{fig:DeVsRho}. Theory curves correspond to \eqref{eq:rho1approx}. }
    \label{fig:RhoVsV}
\end{figure}

To understand the behaviour of $\De$ and $\Ve$, we need to understand first how particles rearrange due to inter-particle forces. Typically, a tracer particle will be forced down density gradients because of repulsive pairwise interactions. 
We therefore measure and analyse the particle density distributions in the channel for various average densities (see Fig.~\ref{fig:RhoVsV}a and \ref{fig:RhoVsV}b). 
First, we find that the particle distribution is uniform in the vertical direction, as expected within the lubrication approximation: particles diffuse sufficiently fast on the vertical axis compared to all other relevant timescales. 
Secondly, compared to the ideal gas case  at the same wall speed (Fig.~\ref{fig:idealVsFlow}c \textit{vs} Fig.~\ref{fig:RhoVsV}a) the particle distribution is also quite homogeneous on the horizontal direction, much like in the incompressible case (Fig.~\ref{fig:idealVsFlow}d). However, at higher wall velocities (Fig.~\ref{fig:RhoVsV}b) particles accumulate again at the bottleneck region. Such accumulation in narrow channels is also observed in simulations of driven, interacting, tracers in corrugated channels~\citep{suarez2015transport,suarez2016current}, which share a similar geometry in the reference frame where the wall is static. 

To further quantify the particle distributions, we study the marginal distribution profiles $p (x)=\int_{-\infty}^{\infty}\rho(x,y) \dd y$ along the channel, obtained in the simulations. As expected, $p(x)$ presents a peak, which indicates particles accumulating near the bottleneck. We find again that the profile $p(x)$ is more peaked with increasing wall velocity (Fig.~\ref{fig:RhoVsV}c). It flattens out with increasing particle density $\rho_0$ (see Fig.~\ref{fig:RhoVsV}d), indicating that density profiles indeed converge to the incompressible fluid case.  

\section{Analytic theory with pairwise interactions}

\subsection{Model for particle density profiles along the channel}

To understand how the average density profile depends on interaction parameters, we expand our analytical theory. 
Our starting point is the equation governing the evolution of the density of particles $\rho(x,y,t)$, which can be written formally as \citep{dean1996langevin,kawasaki1998microscopic}
\begin{equation}
    \dt{\rho} = D_0 \bm \nabla^2 \rho + \frac{D_0}{k_B T} \bm \nabla \cdot ( \rho \bm \nabla ( \Vint * \rho ) ) 
    + \bm\nabla \cdot ( \sqrt{2 D_0 \rho}\bm \xi ),\label{dk}
\end{equation}
where $\bm \xi$ is a 2-component Gaussian white noise field, such that $\langle \xi_\mu(x,y,t) \rangle = 0$ and $\langle \xi_\mu(x,y,t) \xi_\nu(x',y',t') \rangle = \delta_{\mu \nu} \delta(x-x')\delta(y-y')\delta(t-t')$. Here the convolution product is $(\Vint * \rho)(x,y) = \iint \dd x' \dd y' \Vint(\sqrt{x'^2 + y'^2}) \rho(x- x', y-y')$, where $\Vint$ is any pairwise interaction potential. The term $D_0$ represents the microscopic diffusion constant of the individual Brownian particles. In an infinite domain and in the absence of interactions the long time diffusion constant of a tracer will be identical to $D_0$. The effect of interactions and geometry in the problem at hand here both play a role in modifying the late time diffusion constant with respect to the microscopic one. 

To make progress we first decompose the density into an average and fluctuating component
\begin{equation}
\rho(x,y,t) = \rhob(x,y,t)+ \psi(x,y,t),
\end{equation}
where $\psi(x,y,t)$ is the noise field such that $\langle \psi(x,y,t) \rangle = 0$. 
The noise perturbation $\psi(x,y,t)$ can be inferred, as in \cite{demery2014generalized}, by assuming a large, flat, fixed environment and in the case where the background concentration $\rhob(x,y,t) = \rho_0$ is uniform. The short distance fluctuations in the density field can then be absorbed into a renormalization of the diffusion constant which now depends on the average density. We will denote by $\Drho(\rho_0)$ the diffusion coefficient renormalised by the interactions. Here, we will assume that the density fluctuations are short enough in time and space, such that $\rhob(x,y,t)$ may be treated as a constant and such that the same renormalization applies here quickly enough to yield a \textit{local} diffusion constant depending on the \textit{local} average density $\Drho(\rhob)$. This assumption is basically the same as that used in the macroscopic fluctuation theory by \cite{mft}, where it is argued that a coarse grained hydrodynamic description of a system 
simply leads to an equation of the type Eq. (\ref{dk}) but with a local diffusion constant and mobility (describing the coupling the external forces) which depends on the local density field. 
Here,  we assume that we  can write a coarse grained equation for the evolution of the average density $\rhob$ as
\begin{equation}
    \dt{\rhob} = - \bm \nabla \cdot\left( - \Drho(\rhob) \bm \nabla \rhob - \frac{\Drho(\rhob)}{k_B T} \rhob \bm \nabla ( \Vint * \rhob ) \right). 
\end{equation}
As the field $\rhob(x,y,t)$ is smooth and the interaction is short range compared to all the other length scales in the system, we can make the local approximation
\begin{equation}
    \Vint(x,y) = \frac{E_0}{\rho_0}\delta(x)\delta(y)
\end{equation}
such that the Fokker-Planck equation for the particle density simplifies to
\begin{equation}
    \dt{\rhob} = - \bm \nabla\cdot \left( - \Drho(\rhob) \bm \nabla \rhob - \Drho(\rhob)\frac{E_0}{k_B T} \frac{\rhob}{\rho_0} \bm \nabla \rhob \right)
\end{equation}
which is nonlinear in $\rhob$.

Our goal is now to obtain an expression for the average density $\rhob(x,y,t)$. We seek a solution beyond trivial mean field (where $\bar{\rho}(x,y,t) = \rho_0$), which makes our approach similar in essence to other derivations of particles on  lattices~\citep{illien_nonequilibrium_2018,rizkallah2022microscopic}. Notice that for the height fluctuation profile considered here, the density field is stationary in the frame of reference where the wall is static, \textit{i.e.} making the change of variables $x' = x - v_{\rm wall} t$ 
and dropping the prime signs yields 
\begin{equation}
   0 = \vwall \dx{\rhob(x,y)} - \bm \nabla \cdot\left( - \Drho(\rhob) \bm \nabla \rhob(x,y) - \Drho(\rhob)\frac{E_0}{k_B T} \frac{\rhob}{\rho_0} \bm \nabla \rhob \right). 
\end{equation}
Integrating vertically, and using exactly the same arguments as in Sec.~\ref{sec:bcs} for the boundary conditions shows that 
\begin{equation}
    J = \int_0^{h(x,t)} \dd y \left(  - \vwall \rhob(x,y) - \Drho(\rhob) \dx {\rhob(x,y)}  - \Drho(\rhob)\frac{E_0}{k_B T} \frac{\rhob (x,y)}{\rho_0} \dx {\rhob(x,y)}  \right) ,
    \label{eq:rho1dint}
\end{equation}
where $J$, the vertically integrated longitudinal flux, is a constant to be determined.

To solve \eqref{eq:rho1dint}, we first assume that we can make the Fick-Jacobs approximation $\rhob(x,y) \simeq \rhob(x)$, where the density profile mostly depends only on the longitudinal coordinate, which is correct to first order in $H/L$ and can be demonstrated using lubrication theory. Second, we assume that the average density profile has only small variations around its mean value $\rho_0$, and the variations are of order $\varepsilon = h_0/H$. We therefore seek a perturbative solution as $\rhob(x,y) \simeq \rho_0 + \varepsilon \rho_1(x) + \cdots $, where $\rho_0 = N/(L H)$ is the mean particle density in the simulation. We can then expand \eqref{eq:rho1dint} in powers of $\varepsilon$. At the lowest order we find the value of the vertically integrated drift  $J = - \vwall \rho_0 H$. At the next order we obtain a closed set of equations for the perturbation $\rho_1$
\begin{equation}
    \begin{cases} & - H \vwall \rho_0 \varepsilon \cos(k_0 x) =  \displaystyle H\left[- \Drho(\rho_0)\left( 1 + \frac{E_0}{k_B T}\right)\dx{\rho_1(x)} -  v_{\rm wall}  \rho_1(x)  \right]   \\
    & \int_0^L \rho_1 (x) H \ddx = 0 \\
    &\rho_1 (x + L) = \rho_1 (x)
    \end{cases}
    \label{eq:forRho1}
    \end{equation}
    with relevant periodic boundary conditions and vanishing integral of $\rho_1$ since the average density should be conserved. We abbreviate $\Drho = \Drho(\rho_0)$ in the following.

Interestingly, we find in \eqref{eq:forRho1} that the density profile relaxes with an effective diffusion coefficient, 
\begin{equation}
    \Dstar \equiv \Drho \left( 1 + \frac{E_0}{k_B T}\right)= \Drho \frac{\chi_T^\mathrm{id}}{\chi_T^{\star}},
    \label{eq:D0star}
\end{equation}
that is the \textit{collective diffusion coefficient}, as it characterises how the fluid of interacting particles relaxes and not how a single particle diffuses. Eq.~\eqref{eq:D0star} is also expected for hard spheres~\citep{lahtinen2001diffusion,hess1983generalized}.
If interactions are repulsive (i.e. $ E_0>0$), the fluid becomes more incompressible with $\chi_T^\star<\chi_T^\mathrm{id}$, and density inhomogeneities relax faster than they would in the ideal gas, since $\Dstar>\Drho$. Conversely, attractive interactions (i.e. $E_0<0$) increase compressibility and stabilise density inhomogeneities.
This property is absolutely essential to understand the long-time transport properties in a fluctuating medium.

Solving \eqref{eq:forRho1} we find 
\begin{equation}
\overline{\rho}(x) = \rho_0 + \varepsilon\rho_1(x) = \rho_0\left( 1  - \frac{h_0}{H} \frac{\Pes}{1 + (\Pes)^2}\left[ \sin(k_0x) + \Pes \cos(k_0x) \right] \right).
\label{eq:rho1approx}
\end{equation}
and a new and natural P\'{e}clet number characterising the fluctuations emerges, 
\begin{equation}
    \Pes = \frac{\vwall}{\Dstar k_0}  = \frac{\omega_0}{\Dstar k_0^2},
\end{equation} and takes into account the enhanced collective diffusion $\Dstar$ due to repulsive interactions. 
Notice that the solution can be further rewritten as 
\begin{equation}
\overline{\rho}(x) = \rho_0 + \varepsilon\rho_1(x) = \rho_0\left( 1  + \frac{h_0}{H} \frac{\Pes}{\sqrt{1 + (\Pes)^2}} \cos(k_0x + \varphi) \right),
\label{eq:rho1approxphase}
\end{equation}
where $ \varphi = \pi - \arctan \left( 1/\Pes \right)$. This shows that the perturbation in the density of particles  due to the fluctuating interface propagates with a phase $\varphi$ that is characterised by how fast particles relax as a group.

We plot the resulting longitudinal probability density profiles $p(x) =  \left[\rho_0 + \varepsilon \rho_1(x)\right]/L\rho_0$, where $\rho_1(x)$ is given by \eqref{eq:rho1approx}, in Fig.~\ref{fig:RhoVsV}c and d, and find excellent agreement with the numerical results. More specifically, particles accumulate at the constriction at high forcings $\vwall$ (Fig.~\ref{fig:RhoVsV}c, light purple) or at low particle densities (Fig.~\ref{fig:RhoVsV}d, yellow and orange). 
If collective effects are weak, the timescale for density fluctuations to diffuse across a bulge $1/\Dstar k_0^2$ is larger than the time it takes a bulge to move, $1/\vwall k_0 = 1/\omega_0$, or equivalently $\Pes \gg 1$. Clearly, from Eq.~\eqref{eq:rho1approxphase} the phase $\varphi \simeq \pi$ and the interface squeezes particles exactly out-of-phase, \textit{i.e.} in the constriction.  
This can also be understood in the conservation of mass Eq.~\eqref{eq:forRho1}, where at large $\vwall$ we simply have $ (H - h(x)) \vwall \rho_0 = \varepsilon\rho_1(x) \vwall$ such that $\rho_1(x) \sim 1/h(x)$. Similarly as for the isolated particles in Sec.~\ref{sec:ideal}, this is a ``traffic jam'' effect: in the referential where the wall is fixed, particles trying to move with the fast velocity $-\vwall$ accumulate where the road is narrow, \textit{i.e.} at the constriction. 

However, there is a trade-off between accumulation due to wall speed, and increased local repulsion of particles in the accumulation region. We observe that at higher particle densities $\rho_0$, the accumulation effect is tampered (Fig.~\ref{fig:RhoVsV}d, orange to purple). When particle repulsion increases (either due to increased particle density $\rho_0$ or interaction strength $\alpha>0$), the local repulsion dominates, resisting compression of the gas in the narrow constriction. This is coherent with the fact that the compressibility $\chi_T^{\star}/\chi_T^\mathrm{id}$ decreases with particle density $\rho_0$ (or with the strength of the repulsive interactions $\alpha$). As a result, collective effects are strong, $\Pes \ll 1$, density fluctuations relax faster than the interface's motion, and the density profile flattens out. 

We now turn to understand how this local distribution of particles affects the long-time transport properties of a tracer.

\subsection{Effective diffusion and drift with interacting particles}

\subsubsection{Equation for the diffusion of the tracer}

We now consider the motion of a tracer in this gas of soft-core interacting Brownian particles and we infer how the confined fluctuating environment modifies the long-time tracer motion. 
Based on  similar reasoning to that given above, the probability distribution $\rt(x,y,t)$ of finding the tracer particle at position $x,y$ at time $t$ satisfies
\begin{equation}
    \dt{\rt(x,y,t)} =  - \bm \nabla \cdot \left(- \Drho(\rhob)\nabla \rt  -  \Drho(\rhob) \frac{E_0}{k_B T} \frac{\rt}{\rho_0}  \bm \nabla \rhob \right).
    \label{eq:rtmotion}
\end{equation}
Again looking at the longitudinal transport within the Fick-Jacobs framework, the equation on the marginal distribution $\pt(x,t) = \int_0^{h(x,t)} \rt(x,y,t) \dd y$ is given by \eqref{eq:Marginal_brownianConfined},
\begin{equation}
    \dt{\pt(x,t)} =  - \dx{} \left(- \Drho(\rhob) \dx{\pt} + \Drho(\rhob)\pt \dx{\ln h(x,t)} -  \Drho(\rhob) \frac{E_0}{k_B T} \frac{\pt}{\rho_0}  \dx{\rhob} \right).
    \label{eq:pmotion}
\end{equation}

To understand how the tracer's motion is modified by interactions with other particles and by the fluctuating boundary, we seek to expand \eqref{eq:pmotion} in powers of $\varepsilon=h_0/H$, and obtain the particle's long-time diffusion coefficient and drift, at order $\varepsilon^2$ as was done in Sec.~\ref{sec:analytics}. Since the average particle density depends on space, the local diffusion coefficient of the tracer also depends on space as $\Drho(\rhob(x,t)) = \Drho(\rho_0) + \varepsilon \Drho'(\rho_0) \frac{\rho_1(x,t)}{\rho_0} + O(\varepsilon^2)$. Here, since the prescribed wall fluctuations $h(x,t)$ are periodic in space, it is  possible to obtain \textit{implicit} expressions for the long-time effective diffusion coefficient $D_{\rm eff}$ and drift $V_{\rm eff}$ using the approach reported in~\cite{reimann2001giant} and in~\cite{guerin2015Kubo}, and resolve their \textit{explicit} expressions after cumbersome expansions in the small parameter $\varepsilon$ (see Supplementary Material Sec.~3). In the cases explored here, the local change in diffusion coefficient is small, $\Drho'(\rho_0)/\Drho(\rho_0) \rho_0 \ll 1$ and we find that its impact on the tracer dynamics can be neglected.
We can therefore use the \textit{explicit} perturbation theory results in Sec.~\ref{sec:analytics}, assuming $\Drho(\rhob(x,t)) \simeq \Drho(\rho_0) \equiv \Drho$. As expected, the perturbation theory and the periodic framework approach of \cite{reimann2001giant} provide exactly the same results, to leading order in $\varepsilon$.
For the sake of completeness we will nonetheless derive, in a future work, a general perturbation theory with explicit results for Fokker-Planck equations with \textit{diffusion} and \textit{drift} with arbitrary dependence on space and time in \textit{any dimension}. 

\subsubsection{Local drift of a tracer in a bath of interacting particles}
\label{sec:localdrift}

\begin{figure}
    \centering
    \includegraphics[width = \textwidth]{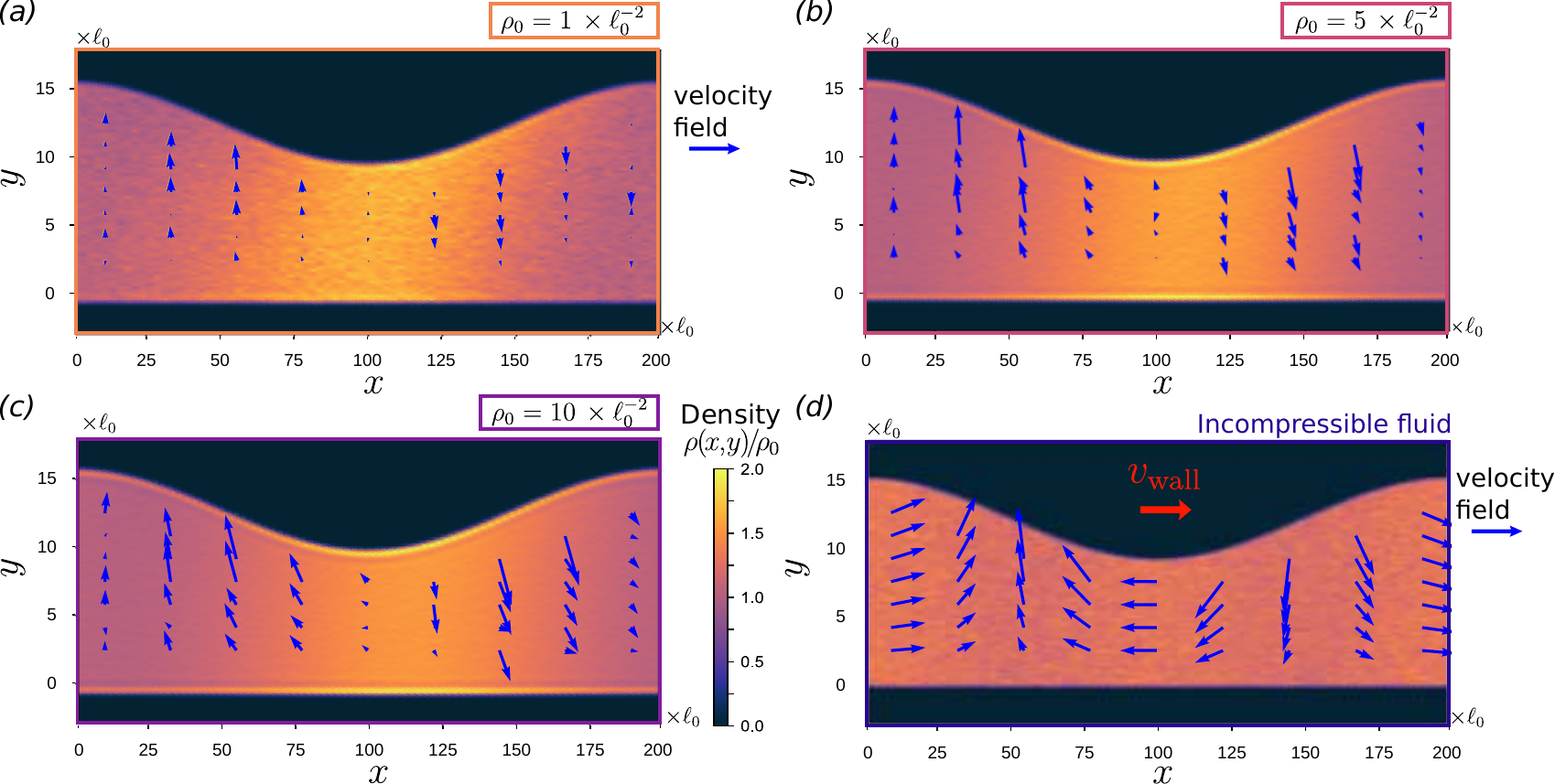}
    \caption{Local drift and particle density profiles within the channel for varying mean particle densities. (a-c) Simulation results for increasing mean density for simulations with interacting particles and (d) for the incompressible fluid case. Note that (d) here corresponds to Fig.~\ref{fig:DeVsRho}d and is shown as a reminder. 
    The color scale for the density profiles is shared between all plots. Yellow regions (resp. purple) indicate regions of high (resp. low) densities. The blue arrows correspond to the local velocity of particles (averaged over time) and the length of the arrows is proportional to the magnitude of the velocity. The scale of the arrows is arbitrary but is the same across all plots, hence, small arrows in (a) indeed indicate very weak velocity fields compared to (c) or (d). The interaction strength for (a-c) is $\alpha = 1 k_B T/\ell_0^2$ and other numerical parameters are similar to Fig.~\ref{fig:DeVsRho}.}
    \label{fig:velocityField}
\end{figure}

The equation of motion \eqref{eq:pmotion} can be simplified to 
\begin{equation}
    \dt{\pt(x,t)} =  - \dx{} \left(- \Drho \dx{\pt(x,t)} + v(x,t)\pt(x,t) \right).
    \label{eq:pmotion2}
\end{equation}
where the local longitudinal drift is 
\begin{equation}
    v(x,t) =   \Drho \frac{1}{H} \dx{h(x,t)} -  \Drho \frac{E_0}{k_B T} \frac{h_0}{H} \frac{1}{\rho_0} \dx{\rho_1(x,t)}
    \label{eq:uForInteracting}
\end{equation}
at lowest order in $\varepsilon = h_0/H$. We can verify that terms in $O(\varepsilon^2)$ in the local drift vanish when averaged over one period and hence do not contribute to the renormalization of the long-time transport properties. 
The last term of \eqref{eq:uForInteracting} quantifies the effect of interactions on \textit{local} tracer motion, and was not present in the ideal gas case. It indicates that particles drift away from accumulation regions because of repulsive interactions. Finally, the magnitude of the effect is proportional to the interaction strength, $E_0/k_B T = (\chi_T^\mathrm{id} - \chi_T^{\star})/\chi_T^{\star}$, or, to the relative compressibility of the fluid. 

We compute in simulations the mean local particle velocity in the interacting gas of particles, see Fig.~\ref{fig:velocityField}. With increasing incompressibility (increasing $\rho_0$ in  Fig.~\ref{fig:velocityField} a-c) a local drift emerges that carries particles away from the accumulation region located at the channel constriction. As expected with increasing interactions, the velocity field approaches that of the incompressible fluid (reported for comparison in Fig.~\ref{fig:velocityField} d, already shown in Fig.~\ref{fig:DeVsRho} d). 

These results are confirmed by the analytical prediction, using the expression for $\rho_1$ in \eqref{eq:rho1approx}. Recalling that the \emph{collective} \Peclet number is $\Pes = \omega_0^2 / \Dstar k_0$, and denoting $\Pebar=\omega_0^2 / \Drho k_0$ the \Peclet number of a tracer in medium of density $\rho_0$, we obtain in Fourier space
\begin{equation}
    \tilde{v}(k,\omega) =   \frac{\tilde{h}}{H} \frac{i \Drho k  (1 + \Pes \Pebar) +   (\Pes/\Pebar - 1)\omega/k }{{1 + (\Pes)^2}}.
    \label{eq:uInt}
\end{equation}
Eq.~\eqref{eq:uInt} perfectly interpolates between the ideal gas and the incompressible fluid. When interactions vanish, $\Dstar=\Drho=D_0$ and $\Pes=\Pebar=\Pe$, which yields the ideal gas expression for the local velocity \eqref{eq:uideal}, where $\tilde{v}(k,\omega) = i  D_0 k \tilde{h}/H$. Reciprocally, the incompressible fluid limit is $\Dstar\to\infty$, which implies $\Pes \rightarrow 0$, hence $\tilde{v}(k,\omega) = (i \Drho k - \omega/k)\tilde{h}/H $, as already found in \eqref{eq:uflow} but with $\Drho$ instead of $D_0$ since the interactions have renormalised the bare diffusion coefficient of the tracer.

\subsubsection{Effective long-time diffusion and mean drift}

We use the perturbation results in \eqref{eq:De1} and \eqref{eq:Ve1} to obtain the long-time diffusion coefficient and drift. From \eqref{eq:uInt}, the spectrum of the local velocity fluctuations is
\begin{equation}
    S(k,\omega) = \frac{\Drho^2 k^2}{(2\pi)^2}\frac{1 + \Pebar^2}{1 + (\Pes)^2} \frac{h_0^2}{4 H^2} \left( \delta(k+k_0) \delta(\omega - \omega_0) + \delta(k-k_0)\delta(\omega + \omega_0)  \right).
\end{equation}
and the long-time transport properties are
\begin{align}
&    \frac{D_{\rm eff}^{\rm \textit{int. gas}}}{ \Drho} = 1 - \frac{h_0^2}{2H^2} \frac{1 - 3 \Pebar^2}{1 + \Pebar^2}  \frac{1}{1 + (\Pes)^2},  \label{eq:DeUnified} \\
 & \frac{V_{\rm eff}^{\rm \textit{int. gas}}}{\vwall} = \frac{h_0^2}{2H^2}  \frac{1}{ 1 + (\Pes)^2}.   
 \label{eq:VeUnified}
\end{align} 

Equations~\eqref{eq:DeUnified} and \eqref{eq:VeUnified} form the main analytic result of the paper and give the transport properties of tracers in a fluid with arbitrary compressibility. We recall that the modified \Peclet number is connected to the compressibility $\Pes =  \Pebar \chi_T^{\star}/\chi_T^{\rm id}$, where $\Pebar$ is the \Peclet number characterising the fluctuations. For weak interactions, where $\rho_0, \alpha \rightarrow 0$, the compressibility is that of the ideal gas $\chi_T^{\star} \rightarrow \chi_T^{\rm id}$, and $\Pes \rightarrow \Pe$. In this limit, we recover the ideal gas results \eqref{eq:Deperiodic} and \eqref{eq:Veperiodic}. Reciprocally, for strong interactions, $\rho_0, \alpha \rightarrow \infty$, $\chi_T^{\star} \rightarrow 0$ hence $\Pes \rightarrow 0$ and we recover the incompressible fluid results of \eqref{eq:DeFlow} and \eqref{eq:VeFlow}. Note that in the limit $\rho_0\to\infty$, $\Drho(\rho_0)$ is still finite, see Appendix~\ref{sec:appD0}. Equations~\eqref{eq:DeUnified} and \eqref{eq:VeUnified} therefore interpolate between the ideal gas and the incompressible fluid cases. 

Despite the mean field like assumptions made, our analytic theory summarised in \eqref{eq:DeUnified} and \eqref{eq:VeUnified} perfectly captures the simulation results, see solid lines in Fig.~\ref{fig:DeVsRho} and Fig.~\ref{fig:DeVsA}, and confirms transport mechanisms in this complex environment. With increasing particle density, particle collisions push particles away from the accumulation region, further into the bulges. Particle-wall collisions then push and disperse particles. In this complex environment, and in contrast with the paradigm where crowded environments slow down diffusion~\citep{lekkerkerker1984calculation,lowen1993long,dean_self-diffusion_2004}, here particle collisions or interactions favour mixing.

\section{Discussion and Conclusion}

In this work we have explored the impact of crowding on the long-time transport properties of particles in fluctuating channels. Our simulation results show a broad range of behaviours that are well captured by our explicit analytic theory based on a perturbation expansion in the wall fluctuation amplitude $h_0/H$. The results are best described in terms of  a \Peclet number characterising the fluctuations, $\Pe= \omega_0/D_0 k_0^2 = 2\pi \vwall/D_0 L$. At low $\Pe \ll 1$, corresponding to fixed interfaces, all fluids behave similarly: particles are slowed down by constrictions, the effective diffusion is decreased. At intermediate $\Pe \simeq 1$, particle-wall collisions stir particles and the effective diffusion is increased. This effect persists until the wall moves so fast that particles no longer have the time to reach the moving bulges and accumulate in the constrictions. In this final regime, $\Pe \gg 1$, the effective diffusion is unchanged. The accumulation regime arises for higher and higher $\Pe$ numbers for increasing particle-particle interactions, \textit{i.e.} for increasing incompressibility which resists accumulation.

\subsection{Collisions enhance diffusion}

One of the main findings  of our work is that here, both numerically and analytically, we have demonstrated that increasing repulsive interactions or \textit{collisions} between particles can enhance the  late time diffusion coefficient and the mean drift characterising the dispersion of a tracer particle in fluctuating channels. This is in contrast with the intuition that collisions in equilibrium reduce the diffusion coefficient~\citep{lekkerkerker1984calculation,lowen1993long}. The mechanism of diffusion enhancement is in fact rather simple: collisions or repulsive interaction help to push particles closer to the walls. Eventually wall-particle collisions help mixing. Since this mechanism is rather straightforward, we expect it to be broadly applicable; for example beyond lubrication approximation or for hard-core repulsive interactions. Such effects could be explored within our framework or alternatively using dynamic density functional theory~\cite{marconi1999dynamic}. In more detailed physical settings, such as with hard-core interactions, other effects would likely also come into play; for example the accessible volume in the channel is smaller for larger particles~\citep{suarez2015transport,riefler2010entropic} and hydrodynamic effects become important~\citep{yang2017hydrodynamic}. 

We now put our results in a broader context. In the introduction, we recalled that diffusion of a driven, out-of-equilibrium, tracer in a bath of interacting particles is enhanced by repulsive interactions~\citep{benichou2013fluctuations,demery2014generalized, benichou_tracer_2018, illien_nonequilibrium_2018}. In a confined channel, \textit{thermal} fluctuations of the wall could possibly enhance the diffusion coefficient of particles, as we have seen in the limiting case of an incompressible fluid~\cite{marbach2018transport}. Another recent work finds that diffusion of odd-diffusing interacting particles is enhanced with increasing densities~\citep{kalz2022collisions}. While the physical setup in \cite{kalz2022collisions} is very different from ours, the mathematical similarities that lead to diffusion enhancement are striking (for example comparing their Eq.~(9) with our \eqref{eq:DeFlow}), and one might speculate that there exists a universal framework to understand these effects under the same light. 

\subsection{Open questions on fluctuating channels}

Beyond the question asked here, namely of understanding how crowding affects transport in fluctuating channels, there are many open fundamental questions. For example, boundaries are not necessarily repulsive and smooth. Surface rugosity would lead to Taylor dispersion in an incompressible fluid~\citep{marbach2019active,kalinay2020taylor}, but how would surface rugosity of the wall potential induce Taylor dispersion in the fluid of interacting particles? Attraction at the boundaries also leads to surprising speed up of diffusion in corrugated static channels~\citep{alexandre2022stickiness}. Is this speed up further enhanced by fluctuations? 
Down the scales, molecular~\citep{yoshida2018dripplons} or quantum~\citep{kavokine2022fluctuation,coquinot2023quantum} effects enhance the mobility of individual molecules; how would these effects combine with mechanical fluctuations?
With the advent of highly sensitive techniques to probe the motion of particle near surfaces in soft and increasingly complex environments~\citep{zhang2020direct,sarfati2021enhanced,vilquin2022nanoparticle}, one might expect to answer some of these questions in the light of further experimental results.

\appendix

\section{Simulation details}
\label{app:simulation}

\subsection{Simulation algorithm}
All simulations are performed using a forward Euler stochastic scheme to discretise the overdamped Langevin dynamics of the particles.
For a particle $i$ at position $\bm X_i(t) = (x_i(t),y_i(t))$, the following position at time $t + \Delta t$ is computed as
\begin{align}
    \bm X_i(t+\Delta t) = \bm X_i(t) + \Delta t \bm U_i + \Delta t \frac{D_0}{k_B T} (\bm F_i(t))+\sqrt{2D_0\Delta  t} \bm G_i(t),
\end{align}
where $\bm U_i = \bm U (x_i(t),y_i(t))$ is the background flow field (that is non zero only in the incompressible case), $\bm F_i = \bm F_{\rm wall} + \sum_{j\neq i} \bm F_{\mathrm{int}, ij}$ is the sum of the forces exerted by the channel walls $\bm F_{\rm wall}$ on the particle and by the neighboring particles $ \bm F_{\mathrm{int}, ij}$ when interactions are present and $\bm G_i(t)$ is a vector of two independent random numbers drawn from a Gaussian distribution of zero mean and variance 1. 

Particles are confined in the channel by means of a potential that exerts a force on the particles only if they reach the boundaries. More precisely, the force exerted by the wall on a particle with coordinate $(x,y)$ is given by $\bm F_{\rm wall}=-\bm \nabla \mathcal{V}_{\rm wall}(x,y)$, with
\begin{align}
    \mathcal{V}_{\rm wall}(x,y) = 
    \begin{cases}
    \lambda (y-Y_\mathrm{upper}(x,t))^4 &\text{ if $y>Y_\mathrm{upper}(x,t)$,}\\
    0 &\text{ if $Y_\mathrm{lower}(x,t)<y<Y_\mathrm{upper}(x,t)$,}\\
     \lambda (y-Y_\mathrm{lower}(x,t))^4 &\text{ if $y<Y_\mathrm{lower}(x,t)$},
    \end{cases}
\end{align}
with $\lambda>0$ a stiffness coefficient (with dimensions of energy over length to the power 4), and where the boundary equations are, unless stated otherwise, given by
\begin{align}
    Y_\mathrm{upper}(x,t) &= H + h_0 \cos\left( \frac{2\pi}{L}(x-\vwall t)\right)\\
    Y_\mathrm{lower}(x,t) &=0
\end{align}
where $H$ and $h_0$ represent the average height and the variation amplitude of the channel height, respectively.
Using a soft confining potential to model the wall is convenient for simulation purposes, as it avoids dealing with reflecting Brownian walks, which carries some challenges~\citep{scala2007event}. It also allows one to keep a rather large integration time step. Note that our simulations have been tested with a time step twice as small and yielded no significant difference. Such boundary models have been extensively used in theoretical active matter systems~\citep{solon_pressure_2015, zakine_surfaceTension2020, ben_dor_disordered_2022}, and as our theory only relies on the presence of a boundary, the results are largely unaffected by a change of potential, as long as the boundary layer of particles at the wall and subjected to the potential repulsion is small compared to all other relevant length scales in the system. 

In Section~\ref{sec:interactions} we have performed simulations with pairwise interacting particles. The force on particle $i$ exerted by particle $j$ is $ \bm F_{\mathrm{int}, ij} = - \bm \nabla \Vint(r_{ij}) $ where $r_{ij}$ is the distance between $i$ and $j$. We chose simple repulsive interactions described by the soft potential \eqref{eq:VintSoft}, that we recall here $\Vint(r) = \alpha (r - d_c)^2 \Theta(r < d_c)$,
where $\alpha$ is the interaction strength in units of a spring constant and $d_c$ is the typical particle diameter. For this quadratic interaction $\Vint$, the expression of its vertically integrated version is given by
\begin{align}
    \Uint(x) &=  \int_{-\infty}^{+\infty}  \Vint(x,y) \dd y\\
    &= 2 \alpha \bigg(\frac{1}{3} \left(d_c^2+2 x^2\right) \sqrt{d_c^2-x^2} +d_c x^2 \log \left(\frac{x}{\sqrt{d_c^2-x^2}+d_c}\right)\bigg) \Theta(d_c > |x|).
\end{align}
This expression is used to compute the density profiles $\rho(x)$, but we always use the radial potential $\Vint$ in the Monte Carlo simulations.
\begin{figure}
    \centering
    \includegraphics[width=0.35\textwidth]{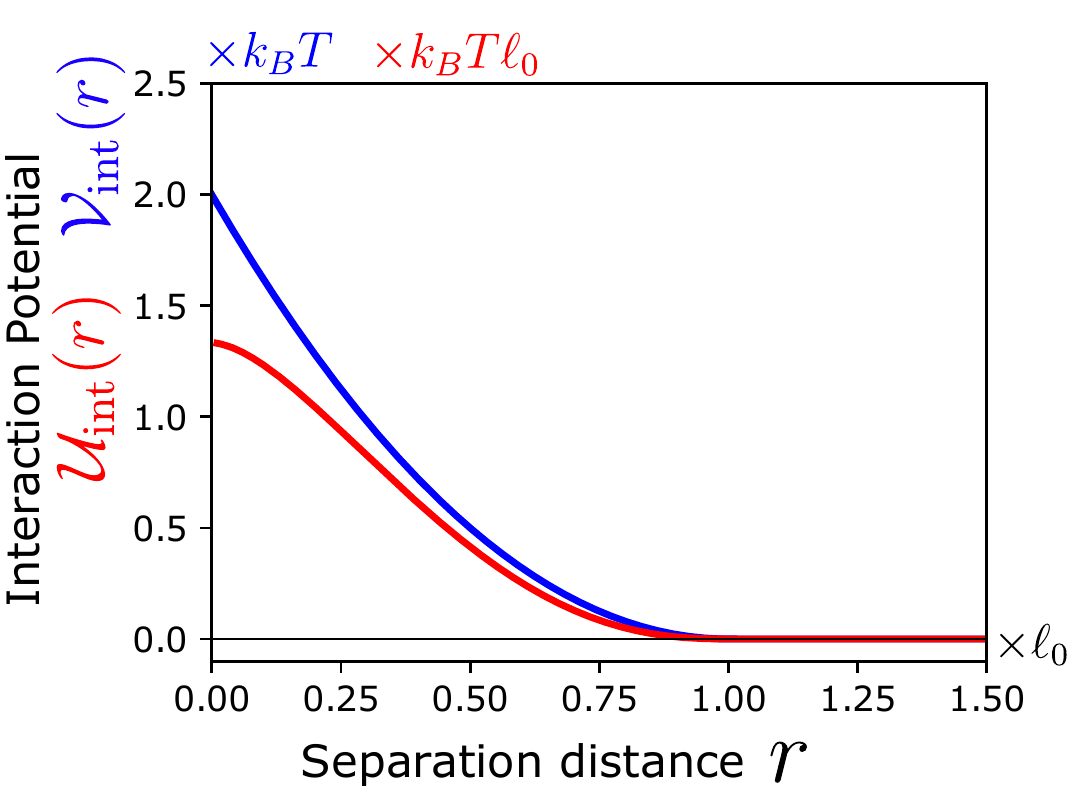}
    \caption{Radial potential $\Vint$ (blue) and its one dimensional smoothed out version $\Uint$ (red). Parameters: $d_c=1 \ell_0$, $\alpha=2 k_B T / \ell_0^2$.}
    \label{fig:V_and_U}
\end{figure}

The system size along the $x$ direction is $L$ and we take periodic boundary conditions in the $x$ direction. We keep track of the full dynamics of the $N$ particles with their absolute position $\bm X_\mathrm{true}$ to compute their mean-square displacement and mean drift, but interactions are computed using the folded positions in the periodic domain $[0,L)\times (-\infty, \infty)$. 
The mean density of particles is $\rho_0 = N/(L H)$. To access higher particle densities we therefore change the total number of particles $N$ in the simulation (see Table~\ref{tab:tab1}).

We initialise systems from a uniform distribution of particles in the bottom part of the channel ($ 0< y < H- h_0$). We first let the system equilibrate for a time of $t_{\rm eq}\sim L_x^2/D_0\sim 4\times 10^4$ within a fixed undulated channel ($\vwall = 0$). Then the actual simulation starts with a positive value of $\vwall$.

\subsection{Simulation parameters}

The domain characteristics are $L=200\ell_0$, $H=12\ell_0$, $h_0=3\ell_0$ for all the simulations.
To simulate non-interacting particles, we take a wall stiffness $\lambda=300 k_B T/\ell_0^4$ and the integration time step is $\Delta t=4\times 10^{-3} \tau_0$. For a typical simulation, the total number of iterations is $N_{\rm iter}=5\times 10^6$, and the number of particles is $N=10^5$. 
To simulate interacting particles,
we take $\lambda=10 k_B T/\ell_0^4$ and the integration time step is $\Delta t=4\times 10^{-3} \tau_0$. Typical values of the number of particles $N$ and total number of iterations $N_\mathrm{iter}$ are shown in Table~\ref{tab:tab1}.

\begin{table}
    \centering
    \begin{tabular}{ c|c|c }
         $\rho (\times \ell_0^{-2})$ & $N$ & $N_{\rm iter}$  \\ 
         1 & 2400 & $10^7$  \\
         5 & 12000 & $2 \times 10^6$\\
         10 & 24000 & $10^6$  \\
    \end{tabular}
    \caption{Typical simulation parameters for interacting particle system}
    \label{tab:tab1}
\end{table}

\subsection{Simulation analysis}
We perform multiple independent simulations to increase statistical resolution. For each value of $\vwall$ we perform 10 independent simulations starting from different initial configurations (and different seeds). Symbols in all graphs represent the mean value of the observable, and error bars correspond to one standard deviation over these 10 independent measurements.
The mean drift is simply calculated as $V_{\rm eff} = \frac{1}{N}\sum_{i=1}^N \langle \frac{x_{\mathrm{true}, i} (t) - x_{\rm true}(0)}{t} \rangle_t$. 
The mean squared displacement of particles is calculated as $\mathrm{MSD}(t) =  \frac{1}{N}\sum_{i=1}^N \langle (x_{\mathrm{ true}, i} (t + t_0) - x_{\mathrm{true}}(t_0))^2 \rangle_{t_0}$ where the average is done over initial times $t_0$. The self-diffusion coefficient is then obtained as a least-squares linear fit of the mean-squared displacement. 
The parameters we choose allow us to neglect the finite size corrections due to periodic boundary conditions. Indeed, such corrections scale as $\sim d_c/L=0.005$~\citep{dunweg1991,dunweg1993molecular}, or as $\sim (H/L)^2=0.004$~\citep{simonnin2017diffusion}, thus negligible in the numerical measurements performed here.

\begin{figure}
    \centering
    \includegraphics[width = \textwidth]{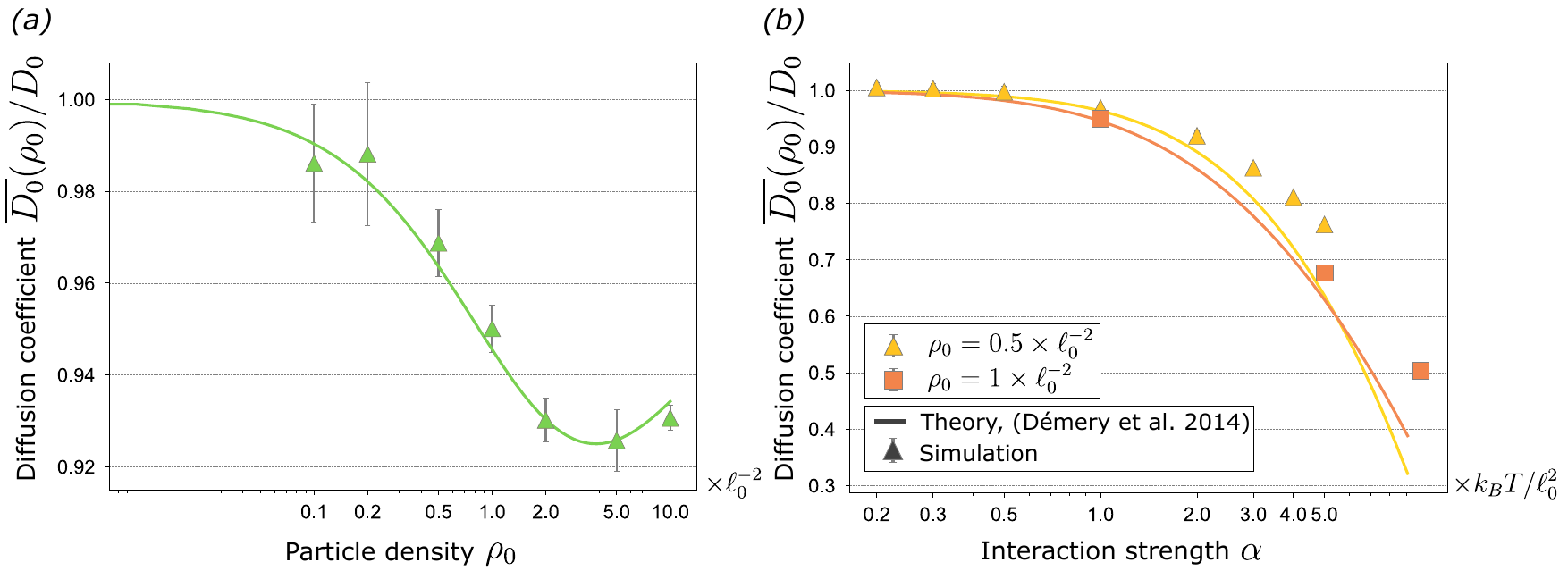}
    \caption{Effective diffusion of a tracer in a homogeneous bath of soft-core interacting particles as a function of (a) the particle density $\rho_0$ and (b) the interaction strength $\alpha$. The tracer is identical to the particles of the bath. Solid lines are obtained from the computation in \cite{demery2014generalized}, and symbols correspond to simulation results. 
    For (a), we fix $\alpha = 1 k_BT/\ell_0^2$ and channel height $H=20\ell_0$, and we simulate for $N=[1,2,5,10,20,10,20] \times 10^3$ particles in a 2-d flat channel of length  $L=[500,500,500,500,500,100,100] \ell_0$ with periodic boundary conditions. For (b) a couple values of $\rho_0$ indicated in the caption are used. Error bars correspond to one standard deviation over 10 independent runs.}
    \label{fig:D0_a}
\end{figure}

\subsection{Simulation calibration: self diffusion coefficient $ D_0(\rho_0)$ of soft-core interacting particles }
\label{sec:appD0}

To calibrate our model, we calculate the long-time self-diffusion coefficient of particles in a fluid of soft-core interacting particles.
In Fig.~\ref{fig:D0_a} we report the results of the computation of the diffusion coefficient $\Drho(\rho_0)$ for fixed flat walls according to different parameters of the interaction: the particle density $\rho_0$ and the interaction strength $\alpha$.

As a self-consistent check we also compare our simulations to the analytic mean-field  theory of \cite{demery2014generalized} and find good agreement between simulation and theory. We briefly recall the analytic formula here. 
The correction to the bare diffusion coefficient is given by
\begin{align}
    \frac{\Drho(\rho_0)}{D_0} \simeq 1-\frac{1}{2 d \rho_0 }\int \frac{\left(\frac{\rho_0 \tVint(\bm k)}{k_B T}\right)^2}{\left[1+\frac{\rho_0 \tVint(\bm k)}{k_B T}\right] \left[1+\frac{\rho_0 \tVint(\bm k)}{2 k_B T} \right]} \frac{\dd^d\bm k}{(2\pi)^d},
    \label{eq:D0bar}
\end{align}
where $\rho_0=N/V$ is the particle density, $d$ the number of spatial dimensions and $\tilde \Vint(\bm k)$ is the Fourier transform of the interaction pair potential, here
\begin{align}
    \tVint(\bm k) = \int \dd^d \bm r e^{-i \bm k\cdot \bm r} \Vint(\bm r).
\end{align}
Note that at large densities $\rho_0$ or interaction strengths, the formula Eq.~\eqref{eq:D0bar} is not valid anymore and it should be approached by~\citep{dean_self-diffusion_2004} 
\begin{equation}
    \frac{\Drho(\rho_0)}{D_0} \simeq \exp \left( -\frac{1}{2 d \rho_0 }\int \frac{\left(\frac{\rho_0 \tVint(\bm k)}{k_B T}\right)^2}{\left[1+\frac{\rho_0 \tVint(\bm k)}{k_B T}\right] \left[1+\frac{\rho_0 \tVint(\bm k)}{2 k_B T} \right]} \frac{\dd^d\bm k}{(2\pi)^d} \right).
    \label{eq:D0barexp}
\end{equation}

In our setup for $d=2$ and $k=|\bm k|$, we have
\begin{align}
   \tVint(\bm k) = 
   \frac{2 \pi  \alpha d_c^2  (-\pi   H_1(d_c k) J_0(d_ck)+\pi  H_0(d_c k) J_1(d_c k)+2  J_0(d_c k)-4 J_1(d_c k)/(d_c k))}{k^2},
\end{align}
where $J_\nu(z)$ is the special Bessel function, and $H_\nu(z)$ is the Struve function
\begin{align}
    H_\nu(z) = \left( \frac{z}{2}\right)^{\nu+1}\sum_{n=0}^\infty \frac{(-1)^n (z/2)^{2n}}{\Gamma(n+3/2)\Gamma(n+\nu+3/2)}.
\end{align}

We plot \eqref{eq:D0bar} in Fig.~\ref{fig:D0_a} and compare it to our numerical results. 
Unsurprisingly, the mean-field approximation starts to fail as the interaction strength $\alpha$ increases and prevents particles from overlapping. To present the results of the long-time diffusion coefficients $D_{\rm eff}$ in confined wiggling spaces relative to $D_0$ (Figs.~\ref{fig:DeVsRho} and \ref{fig:DeVsA}), we always use the numerically calculated diffusion coefficient $\Drho(\rho_0)$ in the flat fixed space.

In the main manuscript, we investigate limit behaviours as $\rho_0 \rightarrow \infty$ and $\alpha \rightarrow \infty$.
A convenient Gaussian interaction potential can be considered to gain analytical insights on the diffusion. With $\Vint(\bm r)=\alpha e^{-\bm r^2/(2d_c^2)}$, we have $\tVint(\bm k)=2\pi\alpha d_c^2 e^{-d_c^2\bm k^2}$ in dimension $d=2$ for instance. The integral in \eqref{eq:D0barexp} can thus be approximated when integrating only in the range $\frac{\rho_0\tVint(\bm k)}{k_B T}\gg1$. The interaction-and-density dependent cutoff is given by $\Lambda(\rho_0)\sim\frac{1}{d_c}\sqrt{\ln\left(\alpha d_c^2 \rho_0/k_B T\right)}$, and the diffusion in thus given by
\begin{equation}
    \frac{\Drho(\rho_0)}{D_0} \simeq 
    \exp \left( -\frac{(2\Lambda)^d}{d \rho_0 (2\pi)^d } \right), 
    \label{eq:D0barexp_cutoff}
\end{equation}
which yields, first, $\Drho(\rho_0)\to D_0$ for $\rho_0\to\infty$ and $\alpha$ fixed (as expected since the potential landscape becomes flat), and second, $\Drho(\rho_0)\to 0$ for $\alpha\to\infty$ and $\rho_0$ fixed (as expected for jamming).

\subsection{Simulation calibration: confinement with the soft wall potential}
\label{sec:penetration}

\begin{figure}
    \centering
    \includegraphics[width = \textwidth]{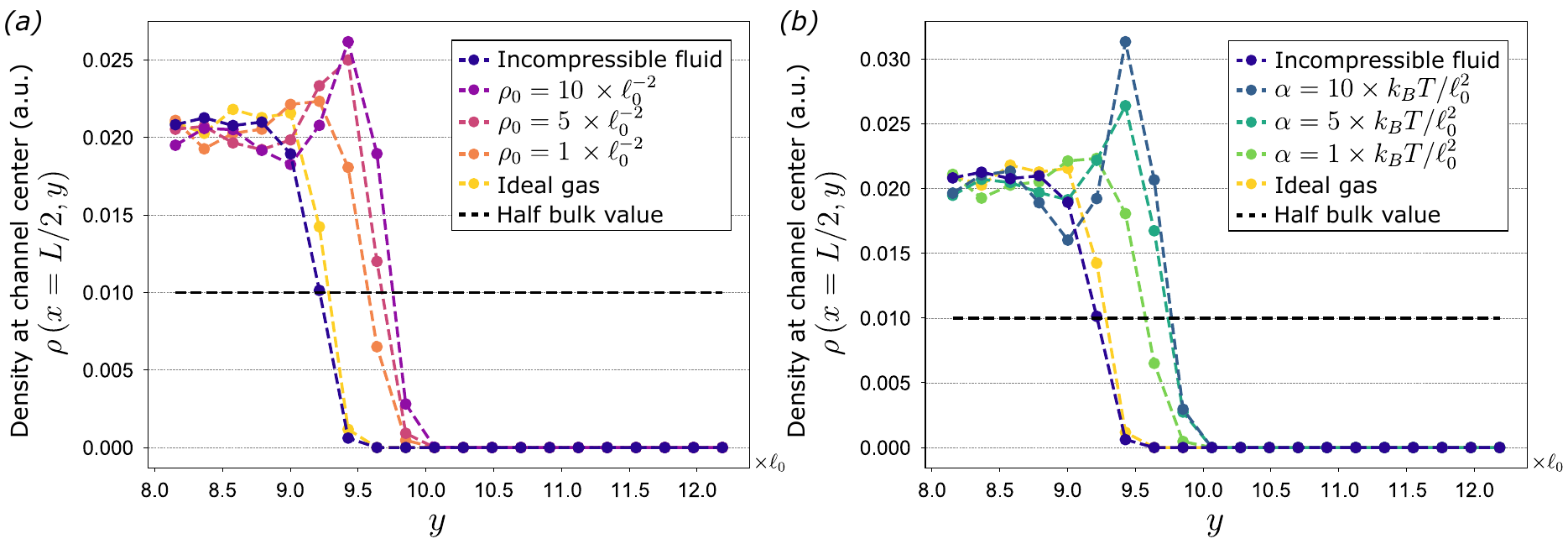}
    \caption{ Vertical density profile near the repulsive confining wall of particle systems for (a) various particle densities and (b) various interaction strengths. Numerical systems correspond to those detailed in Figs.~\ref{fig:DeVsRho} and \ref{fig:DeVsA}, and the confining wall is set in both cases in $y = h(x = L/2) = H - h_0 = 12 - 3 = 9\ell_0$. Hence particle systems extend up to $\delta h \sim \ell_0$ into the confining wall.}
    \label{fig:penetrationdepth}
\end{figure}

For each simulation type, we check the penetration depth of the particles in the confining wall. With increasing repulsive interactions (with increased $\rho_0$ or $\alpha$), and since the wall is ``soft'', particles may be squeezed into the confining soft wall. 
We then estimate the penetration depth of each fluid within the confining wall by looking at the probability density at the center of the channel where the constriction is (in the frame of reference where the wall is fixed), see Fig.~\ref{fig:penetrationdepth}. We find that indeed, with increasing interactions the penetration depth $\delta h = 0.2\ell_0 - 1 \ell_0$ increases. We record the penetration depth $\delta h$ from Fig.~\ref{fig:penetrationdepth} for each set of numerical parameters, corresponding to the depth for which the probability density is half of its bulk value (dashed black horizontal line). We then use $H = H + 2 \delta h$ (since there is an upper and a bottom wall) in all analytical formulas. 

\section{Additional data for the ideal gas}
\label{sec:lubrication}

\begin{figure}
    \centering
    \includegraphics[width = \textwidth]{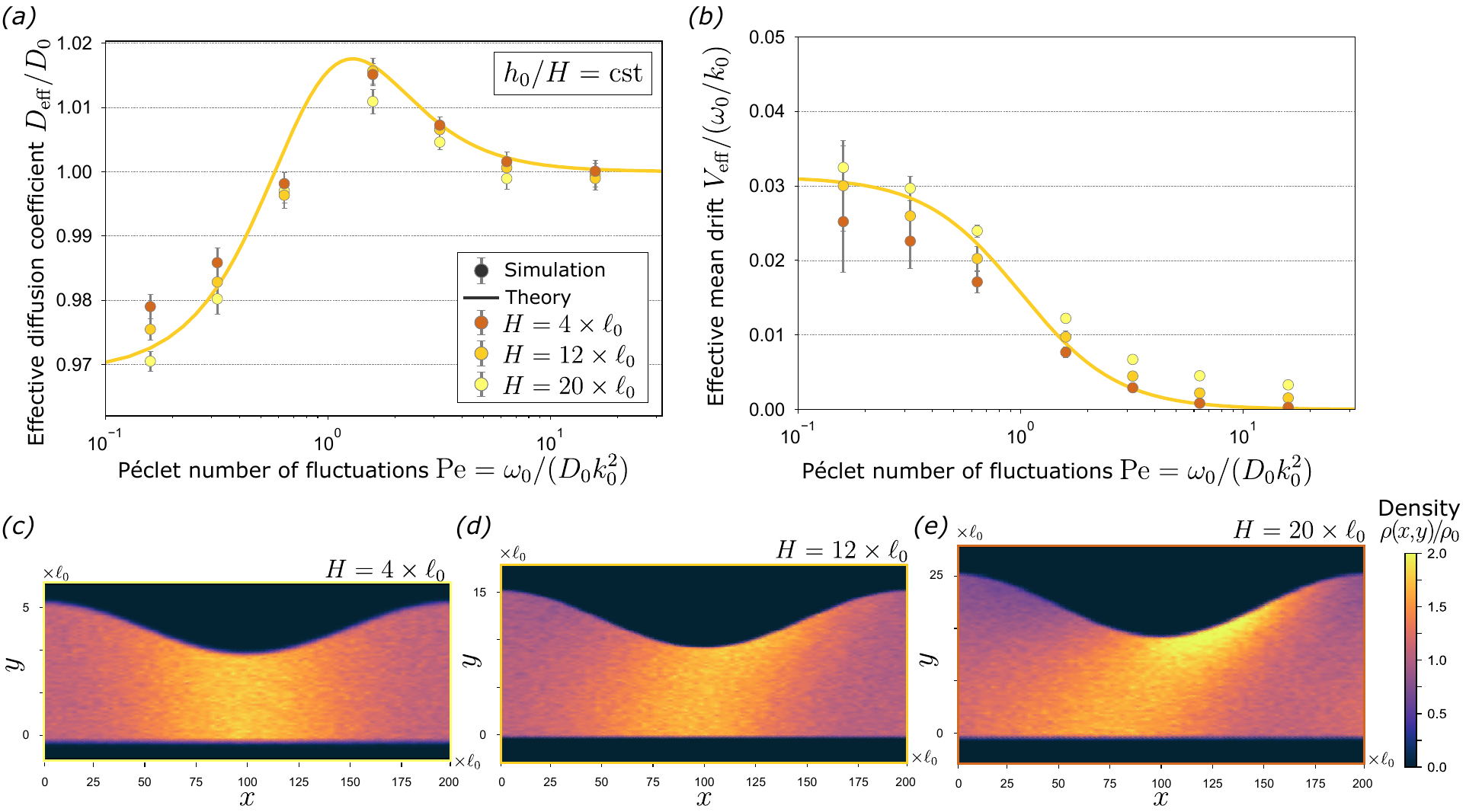}
    \caption{Top panels: Effective diffusion (a) and advection normalised by wall velocity (b) for an ideal Brownian gas as a function of the \Peclet number, for different values of the channel height $H$. Error bars correspond to one standard deviation over 10 independent runs. Theory curves correspond to \eqref{eq:Deperiodic} and \eqref{eq:Veperiodic}.
    Bottom panels: Stationary density profiles in the periodic channel for different values of $H$: (c) $H=4\ell_0$, (d) $H=12\ell_0$, (e) $H=20\ell_0$. Bottom panels share the same colorscale where yellow (resp. purple) regions indicate regions of high (resp. low) density. Bottom panels are all presented for $\vwall = 0.5 \ell_0/\tau_0$ corresponding to $\Pe \simeq 16$. Numerical parameters are the same as for Fig.~\ref{fig:idealVsFlow}, in particular $L = 200\ell_0$.}
    \label{fig:variation_channelHeight}
\end{figure}

To test the validity of the analytic derivation in Sec.~\ref{sec:ideal} for the transport of isolated particles, we explore here a broader range of simulation parameters. In particular we go beyond the lubrication approximation and investigate systems for which $H/L \simeq 0.02$ up to $H/L \simeq 0.1$. We report the measured long-time diffusion coefficients $D_{\rm eff}$ and mean drift $V_{\rm eff}$ in Fig.~\ref{fig:variation_channelHeight} along with representative plots of the density profile in the frame of reference where the channel wall is fixed. We find that as the width $H$ increases, a $y$-dependence of the stationary density emerges (see Fig.~\ref{fig:variation_channelHeight}e). In fact, diffusion across the channel width can no longer be considered fast with respect to diffusion along length $x$. This corresponds to the progressive break down of the lubrication approximation. 

Surprisingly, the analytic formulas \eqref{eq:Deperiodic} and \eqref{eq:Veperiodic} are still in remarkable agreement with simulations, up to $H/L \simeq 0.1$. Slight deviations may be observed for $H/L \simeq 0.1$ (corresponding to $H = 20 \times \ell_0$), on the mean drift, where $V_{\rm eff}/\vwall > 0$ even at large $\Pe$. This is due to accumulation of particles in the upstream bulge, as they collide and leave a wake of particles, instead of having the time to distribute vertically. As a result, some particles are still in the bulge even at large $\Pe$ and are therefore carried, which produces a net mean drift. 
Interestingly, at very small $H/L \simeq 0.02$ (corresponding to $H = 4 \times \ell_0$), we observe slight deviation from the theory this time at small $\Pe$. This is due to the fact that for such systems the penetration depth in the wall $\delta h \simeq 0.2 \ell_0$ becomes more and more comparable with the channel height $H$. The effective vertical accessible space $H$ is therefore larger and the value in \eqref{eq:Deperiodic} and \eqref{eq:Veperiodic} should be appropriately modified (see Appendix~\ref{sec:penetration}). 

\section{Perturbation theory to obtain long-time transport coefficients}
\label{app:perturbation}

Here we perform the perturbation theory to obtain the long-time transport coefficients $\De$ and $\Ve$ of the general Fokker-Planck equation \eqref{eq:FPEgeneral}.

In the following it will be easier to work in Fourier space, and we therefore define, for any function $f(x,t)$, the Fourier transform 
\begin{equation}
    \tilde{f}(k,\omega) = \int \dd x \dd t \, e^{-i(kx + \omega t)} f(x,t),
    \label{eq:FFT}
\end{equation}
where the $\int$ sign encompasses integration over space and time. Conversely, the reverse Fourier transform is given by
\begin{equation}
    f(x,t) = \int \frac{\dd k\dd \omega}{(2\pi)^2} e^{i(kx + \omega t)} \tilde{f}(k,\omega).
\end{equation} 
Performing a Fourier transform on \eqref{eq:FPEgeneral} yields
\begin{equation}
    i \omega \tilde{p}(k,\omega) = - D_0 k^2 \tilde{p}(k,\omega) + 1 - i \int \frac{\dd k' \dd\omega' }{(2\pi)^2}  k \tilde{v}(k', \omega') \tilde{p}(k-k',\omega - \omega'),
\end{equation}
which can be written, using the notation $1/(D_0 k^2 + i \omega)\equiv \tilde{p}_0(k,\omega)$, as
\begin{equation}
    \tilde{p}(k,\omega) = \tilde{p}_0(k,\omega)  - i \tilde{p}_0(k,\omega)  \int \frac{\dd k' \dd\omega' }{(2\pi)^2}  k \tilde{v}(k', \omega') \tilde{p}(k-k',\omega - \omega').
    \label{eq:perturbative0}
\end{equation}

We would like to eventually simplify \eqref{eq:perturbative0} in a form where diffusion and advection coefficients can be easily read, as suggested by the Fourier transform of \eqref{eq:FPE_goal} that yields
\begin{equation}
  \tilde{p}_{\rm eff}(k,\omega) = \frac{1}{D_{\rm eff} k^2 + i (V_{\rm eff} k + \omega) }.
  \label{eq:target}
\end{equation}
The lubrication approximation enables us to further simplify the effective equation on $\tilde{p}(k,\omega)$.
Indeed, we consider that the fluctuations at the channel boundary are a perturbation, with small relative amplitude $\varepsilon = h_0/H$. Hence the last term of \eqref{eq:perturbative0}, containing the advection $v$, can be considered as a perturbation. For example, in the case of the ideal gas
\begin{equation}
    \tilde{v}(k,\omega) \simeq i D_0 k \frac{\tilde{h}(k,\omega)}{H} = O(\varepsilon), \label{eq:uidealgas2}
\end{equation} where we denote $\tilde{h}(k,\omega)$ the Fourier transform of the non constant part of $h(x,t)$, that is $h(x,t)- H$. We therefore seek a solution to \eqref{eq:perturbative0} as an expansion in $\varepsilon$, namely, $\tilde{p}(k,\omega) = \tilde{p}_0(k,\omega) + \varepsilon \tilde{p}_1(k,\omega) + \varepsilon^2 \tilde{p}_2(k,\omega) +\cdots $. 

Additionally, since we seek the behaviour of the solution at long-times, it is natural to calculate the noise-averaged solution $\langle \tilde{p}(k,\omega) \rangle$ where $\langle . \rangle$ denotes the usual average over realizations of the noise. \footnote{Note that we can also treat the propagating wave case which is deterministic, in terms of a random field. Consider that we define $h(x,t) = H + h_0 \cos(k_0 x - \omega_0 t + \theta)$ where $\theta$ is a random phase uniformly distributed on $[0,2\pi]$. Clearly the value of $\theta$ cannot affect the result at late times as it just fixes the initial configuration of the height at time $t = 0$ when the advection diffusion process starts. This choice of $\theta$ is also equivalent for instance to choosing an arbitrary initial time $\tau_0 = \theta/\omega_0$. We thus define $\langle . \rangle$ here as a uniform average over $\theta$ on $[0,2\pi]$. Using this convention we immediately see that $\langle h \rangle = H$. } 
Here we consider additionally Gaussian fluctuations with mean 0, that is to say that the first two moments of the noise completely specify the problem. Thus $\langle \tilde{v}(k,\omega)\rangle = 0$ and 
\begin{equation}
     \langle \tilde{v}(k,\omega) \tilde{v}(k',\omega') \rangle  = S(k,\omega) (2\pi)^2 \delta(k + k')\delta(\omega + \omega') 
\end{equation}
where $S(k,\omega)$ corresponds to the spectrum of the fluctuating advection, as defined in \eqref{eq:sv}. Of course, $\langle \tilde{p}_0(k, \omega) \rangle = \tilde p_0(k, \omega)$. 
  
We now solve iteratively for $\tilde{p}(k,\omega)$. At first order in $\varepsilon$ we obtain
$$ \tilde{p}_1(k,\omega)  = - i \tilde{p}_0(k,\omega)  \int \frac{\dd\omega' \dd k'}{(2\pi)^2}  k \tilde{v}(k', \omega') \tilde{p}_0(k-k',\omega - \omega') $$ 
and $\langle \tilde{p}_1(k,\omega) \rangle = 0$. At second order in $\varepsilon$ we have
\begin{align*} 
\tilde{p}_2(k,\omega) =  &(- i)^2 \tilde{p}_0(k,\omega)  \int \frac{\dd k' \dd\omega' }{(2\pi)^2} k  \bigg[ \tilde{v}(k', \omega')  \tilde{p_0}(k-k',\omega - \omega') \\
& \,\,\, \times \int \frac{\dd k'' \dd\omega'' }{(2\pi)^2}  (k-k') \tilde{v}(k'', \omega'')  \tilde{p_0}(k-k'-k'',\omega - \omega'- \omega'')\bigg],
\end{align*} 
and averaging over noise
$$ \langle \tilde{p}_2(k,\omega) \rangle = \tilde{p}_0(k,\omega)  \left( \int \frac{\dd k' \dd\omega' }{(2\pi)^2} k (k-k')  \tilde p_0(k-k',\omega - \omega')  \frac{S(k',\omega')}{h_0^2}  \right)  \tilde{p}_0(k,\omega).  $$
We observe that we can define $\Sigma(k,\omega)$ such that 
\begin{equation}
    \Sigma(k,\omega) = \int \frac{ \dd k' \dd\omega'}{(2\pi)^2}  \frac{k (k-k') }{D_0 (k-k')^2 + i (\omega - \omega')}    \frac{S(k',\omega')}{h_0^2}
    \label{eq:sigma_general}
\end{equation}
and $\langle \tilde{p}_2(k,\omega) \rangle   =  \tilde{p}_0(k,\omega) \Sigma(k,\omega) \tilde{p}_0(k,\omega)$. 

Pursuing the derivation one can show that the full solution is a geometric series $\langle \tilde{p}\rangle = \tilde{p}_0 + \varepsilon^2 \tilde{p}_0  \Sigma \tilde{p}_0 + \varepsilon^4 \tilde{p}_0  \Sigma \tilde{p}_0   \Sigma \tilde{p}_0 +\dots $ that can be resummed to obtain 
\begin{equation}
    \langle \tilde{p}(k,\omega) \rangle =\frac{\tilde p_0(k,\omega)}{1-\varepsilon^2\Sigma(k,\omega)\tilde p_0(k,\omega)}=\frac{1}{D_0 k^2 +\omega^2-\varepsilon^2\Sigma(k,\omega)}.
    \label{eq:final_propagator}
\end{equation}
Additional steps may be found in the supplementary information of \cite{marbach2018transport}. 

\subsubsection{Long-time results}
From the target long-time expression \eqref{eq:target} and from the resummed propagator \eqref{eq:final_propagator}, one can read off the effective long-time diffusion coefficient as 
\begin{equation}
    D_{\rm eff} = D_0 - \frac{\varepsilon^2}{2} \lim_{k\rightarrow0} \partial_{kk} \Sigma(k,0),
    \label{eq:Deff_sigma}
\end{equation}
and the mean drift as
\begin{equation}
    V_{\rm eff} = i \varepsilon^2\lim_{k\rightarrow0} \partial_{k} \Sigma(k,0).
    \label{eq:Veff_sigma}
\end{equation}
Injecting $\Sigma(k,\omega)$ from \eqref{eq:sigma_general} into \eqref{eq:Deff_sigma} and \eqref{eq:Veff_sigma}
(and dropping the $'$ in the integrals for simplicity), we obtain 
\begin{equation}
    D_{\rm eff} = D_0 - D_0 \varepsilon^2 \int \frac{\dd k \dd\omega }{D_0(2\pi)^2}  \frac{(D_0 k^2 + i\omega)}{(D_0 k^2  - i\omega)^2} S(k,\omega)
    \label{eq:Deff_beforeSimple}
\end{equation}
and
\begin{equation}
    V_{\rm eff} = -i\varepsilon^2 \int \frac{\dd k \dd\omega }{(2\pi)^2} \frac{k}{D_0 k^2  - i\omega}  S(k,\omega) .
    \label{eq:Veff_beforeSimple}
\end{equation}
Now, assuming that the problem is translationally invariant in space and time (which is reasonable considering that the channel is assumed to be infinitely long, and that such an assumption still allows one to model many different situations), this implies that the height correlations satisfy $\langle h(x,t) h(x',t') \rangle = C(|x-x'|,|t-t'|)$ and hence $S(k,\omega) = S(-k,-\omega)$. In addition, as the correlation function $C$ is real, then the conjugate $S^*(k,\omega) = S(-k,-\omega)$. Using time and space invariance we obtain that also $S$ is real. Using the fact that $S$ is even with respect to both its variables, and plugging back the expression of $\varepsilon = h_0/H$, \eqref{eq:Deff_beforeSimple} simplifies to \eqref{eq:De1} and \eqref{eq:Veff_beforeSimple} to \eqref{eq:Ve1} of the main text.


\section{Additional data for a fluid of interacting particles}
\label{sec:addDataCompressible}

\begin{figure}
    \centering
    \includegraphics[width = \textwidth]{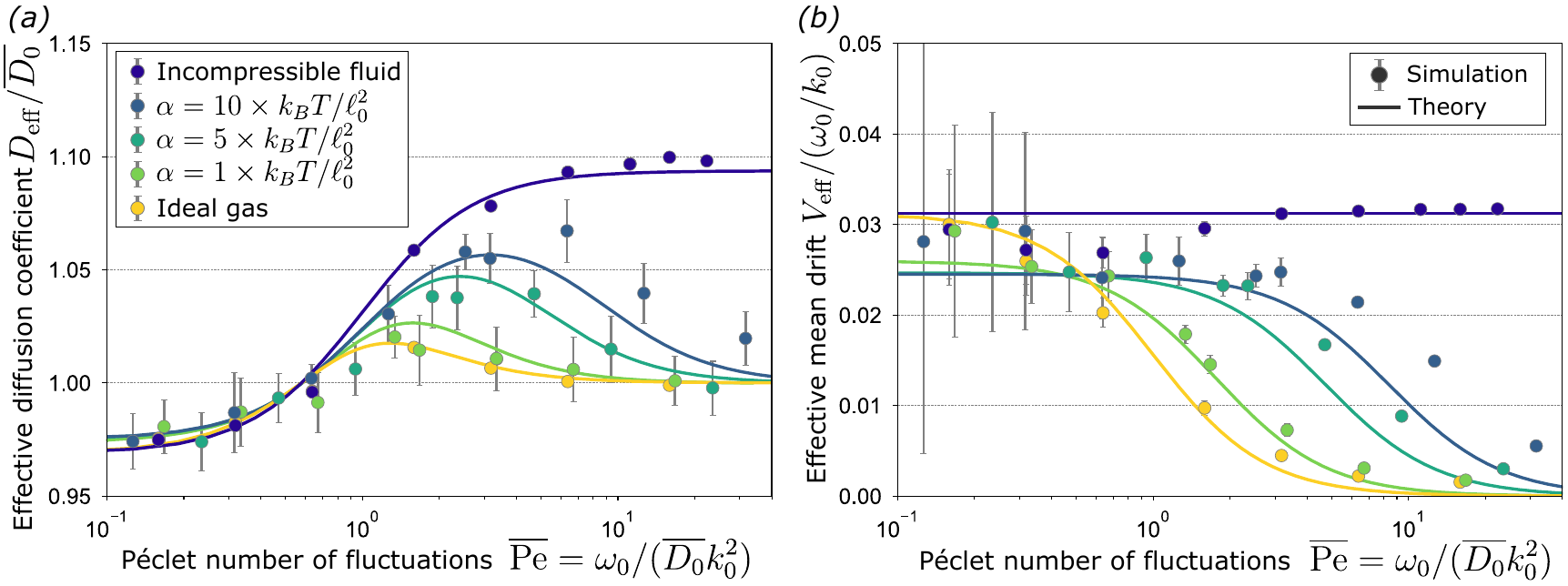}
    \caption{  Effective diffusion (a) and mean drift (b) for a compressible fluid of soft-core interacting Brownian particles with varying interaction strength $\alpha$. Numerical parameters used here are similar to that in Fig.~\ref{fig:DeVsA} and $\rho_0 =1 \ell_0^{-2}$. Error bars correspond to one standard deviation over 10 independent runs. Theory curves for the ideal gas and incompressible fluid are the same as for Fig.~\ref{fig:idealVsFlow} a and b. Theory curves for the fluid of interacting particles correspond to \eqref{eq:DeUnified} and \eqref{eq:VeUnified}.  }
    \label{fig:DeVsA}
\end{figure}

We present here additional data for interacting particles. 
We inspect various values of the interaction strength $\alpha$. We present in Fig.~\ref{fig:DeVsA} the results of the normalised effective diffusion $\De/\Drho(\rho_0)$ and the normalised mean drift $\Ve/\vwall$. We find very similar results whether probing increasing interaction strength $\alpha$ or probing increasing particle density $\rho_0$ (see Fig.~\ref{fig:DeVsRho}). In fact, increasing $\alpha$ also increases the energy scale $E_0(\alpha,\rho_0)$ contained in a volume $d_c^d$ (with $d$ the dimension of space). Hence, since the theory mainly depends on the value of $E_0$, similar effects are naturally observed. 

\section*{Acknowledgement}
The authors are grateful for fruitful discussions with Lyd\'eric Bocquet, Jean-Pierre Hansen, Pierre Levitz,  Minh-Th\^e Hoang Ngoc, Giovanni Pireddu, Benjamin Rotenberg, Brennan Sprinkle and Alice Thorneywork. S.M, R.Z., and Y.W. thank the Applied Math Summer Undergraduate Research Experience Program of the Courant Institute for putting them in contact. S.M. would like to thank the Institut d'\'{E}tudes Scientifiques de Carg\'{e}se for hosting the Transport in Narrow Channels workshop that led to inspiring discussions for this work. R.Z. would like to thank the Laboratoire MSC Paris and the Center for Data Science ENS Paris for hospitality.

\section*{Funding}
This work was in part supported by the MRSEC Program of the National Science Foundation under Award Number DMR-1420073. S.M. was also supported by the European Union’s Horizon 2020 research and innovation program under the Marie Skłodowska-Curie grant agreement 839225 MolecularControl. R.Z. was also supported by Grant
No. NSF DMR-1710163. 



\bibliographystyle{jfm}
\bibliography{main}

\begin{thebibliography}{78}
\expandafter\ifx\csname natexlab\endcsname\relax\def\natexlab#1{#1}\fi
\def\au#1{#1} \def\ed#1{#1} \def\yr#1{#1}\def\at#1{#1}\def\jt#1{\textit{#1}}
  \def\bt#1{#1}\def\bvol#1{\textbf{#1}} \def\vol#1{#1} \def\pg#1{#1}
  \def\publ#1{#1}\def\arxiv#1{#1}\def\org#1{#1}\def\st#1{\textit{#1}}

\bibitem[Alexandre {\em et~al.\/}(2022)Alexandre, Mangeat, Gu{\'e}rin \&
  Dean]{alexandre2022stickiness}
{\sc \au{Alexandre, Arthur}, \au{Mangeat, Matthieu}, \au{Gu{\'e}rin, Thomas} \&
  \au{Dean, DS}} \yr{2022}  \at{How stickiness can speed up diffusion in
  confined systems}.  \jt{Physical Review Letters}  \bvol{128}~(21),
  \pg{210601}.

\bibitem[Alim {\em et~al.\/}(2013)Alim, Amselem, Peaudecerf, Brenner \&
  Pringle]{alim2013random}
{\sc \au{Alim, Karen}, \au{Amselem, Gabriel}, \au{Peaudecerf, Fran{\c{c}}ois},
  \au{Brenner, Michael~P} \& \au{Pringle, Anne}} \yr{2013}  \at{Random network
  peristalsis in physarum polycephalum organizes fluid flows across an
  individual}.  \jt{Proceedings of the National Academy of Sciences}
  \bvol{110}~(33),  \pg{13306--13311}.

\bibitem[Antonov {\em et~al.\/}(2021)Antonov, Ryabov \&
  Maass]{antonov_driven_2021}
{\sc \au{Antonov, Alexander~P.}, \au{Ryabov, Artem} \& \au{Maass, Philipp}}
  \yr{2021}  \at{Driven transport of soft {Brownian} particles through
  pore-like structures: {Effective} size method}.  \jt{The Journal of Chemical
  Physics}  \bvol{155}~(18),  \pg{184102}.

\bibitem[Arango-Restrepo \& Rubi(2020)]{arango2020entropic}
{\sc \au{Arango-Restrepo, A} \& \au{Rubi, JM}} \yr{2020}  \at{Entropic
  transport in a crowded medium}.  \jt{The Journal of Chemical Physics}
  \bvol{153}~(3),  \pg{034108}.

\bibitem[Aris(1956)]{aris1956dispersion}
{\sc \au{Aris, Rutherford}} \yr{1956}  \at{On the dispersion of a solute in a
  fluid flowing through a tube}.  \jt{Proceedings of the Royal Society of
  London. Series A. Mathematical and Physical Sciences}  \bvol{235}~(1200),
  \pg{67--77}.

\bibitem[Ben~Dor {\em et~al.\/}(2022)Ben~Dor, Ro, Kafri, Kardar \&
  Tailleur]{ben_dor_disordered_2022}
{\sc \au{Ben~Dor, Ydan}, \au{Ro, Sunghan}, \au{Kafri, Yariv}, \au{Kardar,
  Mehran} \& \au{Tailleur, Julien}} \yr{2022}  \at{Disordered boundaries
  destroy bulk phase separation in scalar active matter}.  \jt{Physical Review
  E}  \bvol{105}~(4),  \pg{044603}.

\bibitem[B{\'e}nichou {\em et~al.\/}(2013)B{\'e}nichou, Illien, Oshanin \&
  Voituriez]{benichou2013fluctuations}
{\sc \au{B{\'e}nichou, O}, \au{Illien, P}, \au{Oshanin, G} \& \au{Voituriez,
  R}} \yr{2013}  \at{Fluctuations and correlations of a driven tracer in a
  hard-core lattice gas}.  \jt{Physical Review E}  \bvol{87}~(3),  \pg{032164}.

\bibitem[Bertini {\em et~al.\/}(2015)Bertini, De~Sole, Gabrielli, Jona-Lasinio
  \& Landim]{mft}
{\sc \au{Bertini, Lorenzo}, \au{De~Sole, Alberto}, \au{Gabrielli, Davide},
  \au{Jona-Lasinio, Giovanni} \& \au{Landim, Claudio}} \yr{2015}
  \at{Macroscopic fluctuation theory}.  \jt{Rev. Mod. Phys.}  \bvol{87},
  \pg{593--636}.

\bibitem[Bhattacharjee \& Datta(2019)]{bhattacharjee2019bacterial}
{\sc \au{Bhattacharjee, Tapomoy} \& \au{Datta, Sujit~S}} \yr{2019}
  \at{Bacterial hopping and trapping in porous media}.  \jt{Nature
  communications}  \bvol{10}~(1),  \pg{1--9}.

\bibitem[Burada {\em et~al.\/}(2007)Burada, Schmid, Reguera, Rubi \&
  H{\"a}nggi]{burada2007biased}
{\sc \au{Burada, Poornachandra~Sekhar}, \au{Schmid, Gerhard}, \au{Reguera,
  David}, \au{Rubi, JM} \& \au{H{\"a}nggi, Peter}} \yr{2007}  \at{Biased
  diffusion in confined media: Test of the fick-jacobs approximation and
  validity criteria}.  \jt{Physical Review E}  \bvol{75}~(5),  \pg{051111}.

\bibitem[Bénichou {\em et~al.\/}(2018)Bénichou, Illien, Oshanin, Sarracino \&
  Voituriez]{benichou_tracer_2018}
{\sc \au{Bénichou, O}, \au{Illien, P}, \au{Oshanin, G}, \au{Sarracino, A} \&
  \au{Voituriez, R}} \yr{2018}  \at{Tracer diffusion in crowded narrow
  channels}.  \jt{Journal of Physics: Condensed Matter}  \bvol{30}~(44),
  \pg{443001}.

\bibitem[Cao {\em et~al.\/}(2019)Cao, Wang \& Ma]{cao2019water}
{\sc \au{Cao, Wei}, \au{Wang, Jin} \& \au{Ma, Ming}} \yr{2019}  \at{Water
  diffusion in wiggling graphene membranes}.  \jt{The Journal of Physical
  Chemistry Letters}  \bvol{10}~(22),  \pg{7251--7258}.

\bibitem[Chaikin {\em et~al.\/}(1995)Chaikin, Lubensky \&
  Witten]{chaikin1995principles}
{\sc \au{Chaikin, Paul~M}, \au{Lubensky, Tom~C} \& \au{Witten, Thomas~A}}
  \yr{1995} {\em Principles of condensed matter physics\/}, ,  \vol{vol.~10}.
  \publ{Cambridge university press Cambridge}.

\bibitem[Chakrabarti \& Saintillan(2020)]{chakrabarti2020}
{\sc \au{Chakrabarti, Brato} \& \au{Saintillan, David}} \yr{2020}
  \at{Shear-induced dispersion in peristaltic flow}.  \jt{Physics of Fluids}
  \bvol{32}~(11),  \pg{113102}.

\bibitem[Ch{\'a}vez {\em et~al.\/}(2018)Ch{\'a}vez, Chac{\'o}n-Acosta \&
  Dagdug]{chavez2018effects}
{\sc \au{Ch{\'a}vez, Yoshua}, \au{Chac{\'o}n-Acosta, Guillermo} \& \au{Dagdug,
  Leonardo}} \yr{2018}  \at{Effects of curved midline and varying width on the
  description of the effective diffusivity of brownian particles}.  \jt{Journal
  of Physics: Condensed Matter}  \bvol{30}~(19),  \pg{194001}.

\bibitem[Codutti {\em et~al.\/}(2022)Codutti, Cremer \&
  Alim]{codutti2022changing}
{\sc \au{Codutti, Agnese}, \au{Cremer, Jonas} \& \au{Alim, Karen}} \yr{2022}
  \at{Changing flows balance nutrient absorption and bacterial growth along the
  gut}.  \jt{bioRxiv} .

\bibitem[Coquinot {\em et~al.\/}(2023)Coquinot, Bocquet \&
  Kavokine]{coquinot2023quantum}
{\sc \au{Coquinot, Baptiste}, \au{Bocquet, Lyd{\'e}ric} \& \au{Kavokine,
  Nikita}} \yr{2023}  \at{Quantum feedback at the solid-liquid interface:
  Flow-induced electronic current and its negative contribution to friction}.
  \jt{Physical Review X}  \bvol{13}~(1),  \pg{011019}.

\bibitem[Dagdug {\em et~al.\/}(2016)Dagdug, Garc{\'\i}a-Chung \&
  Chac{\'o}n-Acosta]{dagdug2016description}
{\sc \au{Dagdug, Leonardo}, \au{Garc{\'\i}a-Chung, Angel~A} \&
  \au{Chac{\'o}n-Acosta, Guillermo}} \yr{2016}  \at{On the description of
  brownian particles in confinement on a non-cartesian coordinates basis}.
  \jt{The Journal of Chemical Physics}  \bvol{145}~(7),  \pg{074105}.

\bibitem[Dean(1996)]{dean1996langevin}
{\sc \au{Dean, David~S}} \yr{1996}  \at{Langevin equation for the density of a
  system of interacting langevin processes}.  \jt{Journal of Physics A:
  Mathematical and General}  \bvol{29}~(24),  \pg{L613}.

\bibitem[Dean \& Lefèvre(2004)]{dean_self-diffusion_2004}
{\sc \au{Dean, D.~S.} \& \au{Lefèvre, A.}} \yr{2004}  \at{Self-diffusion in a
  system of interacting {Langevin} particles}.  \jt{Physical Review E}
  \bvol{69}~(6),  \pg{061111}.

\bibitem[D{\'e}mery {\em et~al.\/}(2014)D{\'e}mery, B{\'e}nichou \&
  Jacquin]{demery2014generalized}
{\sc \au{D{\'e}mery, Vincent}, \au{B{\'e}nichou, Olivier} \& \au{Jacquin,
  Hugo}} \yr{2014}  \at{Generalized langevin equations for a driven tracer in
  dense soft colloids: construction and applications}.  \jt{New Journal of
  Physics}  \bvol{16}~(5),  \pg{053032}.

\bibitem[D{\"u}nweg \& Kremer(1991)]{dunweg1991}
{\sc \au{D{\"u}nweg, Burkhard} \& \au{Kremer, Kurt}} \yr{1991}  \at{Microscopic
  verification of dynamic scaling in dilute polymer solutions: A
  molecular-dynamics simulation}.  \jt{Physical review letters}
  \bvol{66}~(23),  \pg{2996}.

\bibitem[D{\"u}nweg \& Kremer(1993)]{dunweg1993molecular}
{\sc \au{D{\"u}nweg, Burkhard} \& \au{Kremer, Kurt}} \yr{1993}  \at{Molecular
  dynamics simulation of a polymer chain in solution}.  \jt{The Journal of
  chemical physics}  \bvol{99}~(9),  \pg{6983--6997}.

\bibitem[Evans {\em et~al.\/}(2021)Evans, Krause \&
  Feringa]{evans2021cooperative}
{\sc \au{Evans, Jack~D}, \au{Krause, Simon} \& \au{Feringa, Ben~L}} \yr{2021}
  \at{Cooperative and synchronized rotation in motorized porous frameworks:
  impact on local and global transport properties of confined fluids}.
  \jt{Faraday Discussions}  \bvol{225},  \pg{286--300}.

\bibitem[Gu{\'e}rin \& Dean(2015)]{guerin2015Kubo}
{\sc \au{Gu{\'e}rin, T.} \& \au{Dean, D.~S.}} \yr{2015}  \at{Kubo formulas for
  dispersion in heterogeneous periodic nonequilibrium systems}.  \jt{Physical
  Review E}  \bvol{92}~(6),  \pg{062103}.

\bibitem[Haldoupis {\em et~al.\/}(2012)Haldoupis, Watanabe, Nair \&
  Sholl]{haldoupis2012quantifying}
{\sc \au{Haldoupis, Emmanuel}, \au{Watanabe, Taku}, \au{Nair, Sankar} \&
  \au{Sholl, David~S}} \yr{2012}  \at{Quantifying large effects of framework
  flexibility on diffusion in mofs: Ch4 and co2 in zif-8}.  \jt{ChemPhysChem}
  \bvol{13}~(15),  \pg{3449--3452}.

\bibitem[Hansen \& McDonald(2013)]{hansen2013theory}
{\sc \au{Hansen, Jean-Pierre} \& \au{McDonald, Ian~Ranald}} \yr{2013} {\em
  Theory of simple liquids: with applications to soft matter\/}.
  \publ{Academic press}.

\bibitem[Hess \& Klein(1983)]{hess1983generalized}
{\sc \au{Hess, W} \& \au{Klein, R}} \yr{1983}  \at{Generalized hydrodynamics of
  systems of brownian particles}.  \jt{Advances in Physics}  \bvol{32}~(2),
  \pg{173--283}.

\bibitem[Illien {\em et~al.\/}(2018)Illien, Bénichou, Oshanin, Sarracino \&
  Voituriez]{illien_nonequilibrium_2018}
{\sc \au{Illien, Pierre}, \au{Bénichou, Olivier}, \au{Oshanin, Gleb},
  \au{Sarracino, Alessandro} \& \au{Voituriez, Raphaël}} \yr{2018}
  \at{Nonequilibrium {Fluctuations} and {Enhanced} {Diffusion} of a {Driven}
  {Particle} in a {Dense} {Environment}}.  \jt{Physical Review Letters}
  \bvol{120}~(20),  \pg{200606}.

\bibitem[Jacobs(1935)]{jacobs1935diffusion}
{\sc \au{Jacobs, Merkel~Henry}} \yr{1935}  \at{Diffusion processes}.  \bt{In
  {\em Diffusion Processes\/}},  \pg{pp. 1--145}.  \publ{Springer}.

\bibitem[Jacobs(1967)]{jacobs_diffusion_1967}
{\sc \au{Jacobs, M.~H.}} \yr{1967} {\em Diffusion {Processes}\/}.
  \publ{Berlin, Heidelberg: Springer Berlin Heidelberg}.

\bibitem[Kalinay(2020)]{kalinay2020taylor}
{\sc \au{Kalinay, Pavol}} \yr{2020}  \at{Taylor dispersion in poiseuille flow
  in three-dimensional tubes of varying diameter}.  \jt{Physical Review E}
  \bvol{102}~(4),  \pg{042606}.

\bibitem[Kalinay \& Percus(2006)]{kalinay_corrections_2006}
{\sc \au{Kalinay, P.} \& \au{Percus, J.~K.}} \yr{2006}  \at{Corrections to the
  {Fick}-{Jacobs} equation}.  \jt{Physical Review E}  \bvol{74}~(4),
  \pg{041203}.

\bibitem[Kalz {\em et~al.\/}(2022)Kalz, Vuijk, Abdoli, Sommer, L{\"o}wen \&
  Sharma]{kalz2022collisions}
{\sc \au{Kalz, Erik}, \au{Vuijk, Hidde~Derk}, \au{Abdoli, Iman}, \au{Sommer,
  Jens-Uwe}, \au{L{\"o}wen, Hartmut} \& \au{Sharma, Abhinav}} \yr{2022}
  \at{Collisions enhance self-diffusion in odd-diffusive systems}.
  \jt{Physical Review Letters}  \bvol{129}~(9),  \pg{090601}.

\bibitem[Kavokine {\em et~al.\/}(2022)Kavokine, Bocquet \&
  Bocquet]{kavokine2022fluctuation}
{\sc \au{Kavokine, Nikita}, \au{Bocquet, Marie-Laure} \& \au{Bocquet,
  Lyd{\'e}ric}} \yr{2022}  \at{Fluctuation-induced quantum friction in
  nanoscale water flows}.  \jt{Nature}  \bvol{602}~(7895),  \pg{84--90}.

\bibitem[Kavokine {\em et~al.\/}(2021)Kavokine, Netz \&
  Bocquet]{kavokine2021fluids}
{\sc \au{Kavokine, Nikita}, \au{Netz, Roland~R} \& \au{Bocquet, Lyd{\'e}ric}}
  \yr{2021}  \at{Fluids at the nanoscale: From continuum to subcontinuum
  transport}.  \jt{Annual Review of Fluid Mechanics}  \bvol{53},
  \pg{377--410}.

\bibitem[Kawasaki(1998)]{kawasaki1998microscopic}
{\sc \au{Kawasaki, Kyozi}} \yr{1998}  \at{Microscopic analyses of the dynamical
  density functional equation of dense fluids}.  \jt{Journal of statistical
  physics}  \bvol{93},  \pg{527--546}.

\bibitem[Lahtinen {\em et~al.\/}(2001)Lahtinen, Hjelt, Ala-Nissila \&
  Chvoj]{lahtinen2001diffusion}
{\sc \au{Lahtinen, JM}, \au{Hjelt, T}, \au{Ala-Nissila, T} \& \au{Chvoj, Z}}
  \yr{2001}  \at{Diffusion of hard disks and rodlike molecules on surfaces}.
  \jt{Physical Review E}  \bvol{64}~(2),  \pg{021204}.

\bibitem[Lekkerkerker \& Dhont(1984)]{lekkerkerker1984calculation}
{\sc \au{Lekkerkerker, HNW} \& \au{Dhont, JKG}} \yr{1984}  \at{On the
  calculation of the self-diffusion coefficient of interacting brownian
  particles}.  \jt{The Journal of chemical physics}  \bvol{80}~(11),
  \pg{5790--5792}.

\bibitem[Leroy {\em et~al.\/}(2004)Leroy, Rousseau \& Fuchs]{leroy2004self}
{\sc \au{Leroy, F}, \au{Rousseau, B} \& \au{Fuchs, AH}} \yr{2004}
  \at{Self-diffusion of n-alkanes in silicalite using molecular dynamics
  simulation: A comparison between rigid and flexible frameworks}.
  \jt{Physical Chemistry Chemical Physics}  \bvol{6}~(4),  \pg{775--783}.

\bibitem[Lowen \& Szamel(1993)]{lowen1993long}
{\sc \au{Lowen, Hartmut} \& \au{Szamel, Grzegorz}} \yr{1993}  \at{Long-time
  self-diffusion coefficient in colloidal suspensions: theory versus
  simulation}.  \jt{Journal of Physics: Condensed Matter}  \bvol{5}~(15),
  \pg{2295}.

\bibitem[Ma {\em et~al.\/}(2015)Ma, Grey, Shen, Urbakh, Wu, Liu, Liu \&
  Zheng]{ma2015water}
{\sc \au{Ma, Ming}, \au{Grey, Fran{\c{c}}ois}, \au{Shen, Luming}, \au{Urbakh,
  Michael}, \au{Wu, Shuai}, \au{Liu, Jefferson~Zhe}, \au{Liu, Yilun} \&
  \au{Zheng, Quanshui}} \yr{2015}  \at{Water transport inside carbon nanotubes
  mediated by phonon-induced oscillating friction}.  \jt{Nature nanotechnology}
   \bvol{10}~(8),  \pg{692--695}.

\bibitem[Ma {\em et~al.\/}(2016)Ma, Tocci, Michaelides \& Aeppli]{ma2016fast}
{\sc \au{Ma, Ming}, \au{Tocci, Gabriele}, \au{Michaelides, Angelos} \&
  \au{Aeppli, Gabriel}} \yr{2016}  \at{Fast diffusion of water nanodroplets on
  graphene}.  \jt{Nature materials}  \bvol{15}~(1),  \pg{66--71}.

\bibitem[Malgaretti {\em et~al.\/}(2022)Malgaretti, Puertas \&
  Pagonabarraga]{malgaretti2022active}
{\sc \au{Malgaretti, Paolo}, \au{Puertas, Antonio~M} \& \au{Pagonabarraga,
  Ignacio}} \yr{2022}  \at{Active microrheology in corrugated channels:
  Comparison of thermal and colloidal baths}.  \jt{Journal of Colloid and
  Interface Science}  \bvol{608},  \pg{2694--2702}.

\bibitem[Mangeat {\em et~al.\/}(2017)Mangeat, Gu{\'e}rin \&
  Dean]{mangeat2017dispersion}
{\sc \au{Mangeat, Matthieu}, \au{Gu{\'e}rin, Thomas} \& \au{Dean, David~S}}
  \yr{2017}  \at{Dispersion in two dimensional channels—the fick--jacobs
  approximation revisited}.  \jt{Journal of Statistical Mechanics: Theory and
  Experiment}  \bvol{2017}~(12),  \pg{123205}.

\bibitem[Marbach \& Alim(2019)]{marbach2019active}
{\sc \au{Marbach, Sophie} \& \au{Alim, Karen}} \yr{2019}  \at{Active control of
  dispersion within a channel with flow and pulsating walls}.  \jt{Physical
  Review Fluids}  \bvol{4}~(11),  \pg{114202}.

\bibitem[Marbach \& Bocquet(2019)]{marbach2019osmosis}
{\sc \au{Marbach, Sophie} \& \au{Bocquet, Lyd{\'e}ric}} \yr{2019}  \at{Osmosis,
  from molecular insights to large-scale applications}.  \jt{Chemical Society
  Reviews}  \bvol{48}~(11),  \pg{3102--3144}.

\bibitem[Marbach {\em et~al.\/}(2018)Marbach, Dean \&
  Bocquet]{marbach2018transport}
{\sc \au{Marbach, Sophie}, \au{Dean, David~S} \& \au{Bocquet, Lyd{\'e}ric}}
  \yr{2018}  \at{Transport and dispersion across wiggling nanopores}.
  \jt{Nature Physics}  \bvol{14}~(11),  \pg{1108--1113}.

\bibitem[Marconi \& Tarazona(1999)]{marconi1999dynamic}
{\sc \au{Marconi, Umberto Marini~Bettolo} \& \au{Tarazona, Pedro}} \yr{1999}
  \at{Dynamic density functional theory of fluids}.  \jt{The Journal of
  chemical physics}  \bvol{110}~(16),  \pg{8032--8044}.

\bibitem[Masri {\em et~al.\/}(2021)Masri, Puelz \& Riviere]{masri2021reduced}
{\sc \au{Masri, Rami}, \au{Puelz, Charles} \& \au{Riviere, Beatrice}} \yr{2021}
   \at{A reduced model for solute transport in compliant blood vessels with
  arbitrary axial velocity profile}.  \jt{International Journal of Heat and
  Mass Transfer}  \bvol{176},  \pg{121379}.

\bibitem[Mercer \& Roberts(1990)]{mercer1990centre}
{\sc \au{Mercer, GN} \& \au{Roberts, AJ}} \yr{1990}  \at{A centre manifold
  description of contaminant dispersion in channels with varying flow
  properties}.  \jt{SIAM Journal on Applied Mathematics}  \bvol{50}~(6),
  \pg{1547--1565}.

\bibitem[Noh \& Aluru(2021)]{noh2021phonon}
{\sc \au{Noh, Yechan} \& \au{Aluru, NR}} \yr{2021}  \at{Phonon-fluid coupling
  enhanced water desalination in flexible two-dimensional porous membranes}.
  \jt{Nano letters}  \bvol{22}~(1),  \pg{419--425}.

\bibitem[Obliger {\em et~al.\/}(2023)Obliger, Bousige, Coasne \&
  Leyssale]{obliger2023development}
{\sc \au{Obliger, Ama\"el}, \au{Bousige, Colin}, \au{Coasne, Benoit} \&
  \au{Leyssale, Jean-Marc}} \yr{2023}  \at{Development of atomistic kerogen
  models and their applications for gas adsorption and diffusion: A
  mini-review}.  \jt{Energy \& Fuels} .

\bibitem[Obliger {\em et~al.\/}(2019)Obliger, Valdenaire, Ulm, Pellenq \&
  Leyssale]{obliger2019methane}
{\sc \au{Obliger, Ama{\"e}l}, \au{Valdenaire, Pierre-Louis}, \au{Ulm,
  Franz-Josef}, \au{Pellenq, Roland J-M} \& \au{Leyssale, Jean-Marc}} \yr{2019}
   \at{Methane diffusion in a flexible kerogen matrix}.  \jt{The Journal of
  Physical Chemistry B}  \bvol{123}~(26),  \pg{5635--5640}.

\bibitem[Pireddu {\em et~al.\/}(2019)Pireddu, Pazzona, Demontis \&
  Za{\l}uska-Kotur]{pireddu2019scaling}
{\sc \au{Pireddu, Giovanni}, \au{Pazzona, Federico~G}, \au{Demontis,
  Pierfranco} \& \au{Za{\l}uska-Kotur, Magdalena~A}} \yr{2019}  \at{Scaling-up
  simulations of diffusion in microporous materials}.  \jt{Journal of Chemical
  Theory and Computation}  \bvol{15}~(12),  \pg{6931--6943}.

\bibitem[Puertas {\em et~al.\/}(2018)Puertas, Malgaretti \&
  Pagonabarraga]{puertas2018active}
{\sc \au{Puertas, Antonio~M}, \au{Malgaretti, Paolo} \& \au{Pagonabarraga,
  Ignacio}} \yr{2018}  \at{Active microrheology in corrugated channels}.
  \jt{The Journal of chemical physics}  \bvol{149}~(17),  \pg{174908}.

\bibitem[Pàmies {\em et~al.\/}(2009)Pàmies, Cacciuto \&
  Frenkel]{pamies_hertzian_2009}
{\sc \au{Pàmies, Josep~C.}, \au{Cacciuto, Angelo} \& \au{Frenkel, Daan}}
  \yr{2009}  \at{Phase diagram of {Hertzian} spheres}.  \jt{The Journal of
  Chemical Physics}  \bvol{131}~(4),  \pg{044514}.

\bibitem[Reguera \& Rubí(2001)]{reguera_kinetic_2001}
{\sc \au{Reguera, D.} \& \au{Rubí, J.~M.}} \yr{2001}  \at{Kinetic equations
  for diffusion in the presence of entropic barriers}.  \jt{Physical Review E}
  \bvol{64}~(6),  \pg{061106}.

\bibitem[Reimann {\em et~al.\/}(2001)Reimann, Van~den Broeck, Linke,
  H{\"a}nggi, Rubi \& P{\'e}rez-Madrid]{reimann2001giant}
{\sc \au{Reimann, Peter}, \au{Van~den Broeck, Christian}, \au{Linke, H},
  \au{H{\"a}nggi, Peter}, \au{Rubi, JM} \& \au{P{\'e}rez-Madrid, Agust{\'\i}n}}
  \yr{2001}  \at{Giant acceleration of free diffusion by use of tilted periodic
  potentials}.  \jt{Physical review letters}  \bvol{87}~(1),  \pg{010602}.

\bibitem[Riefler {\em et~al.\/}(2010)Riefler, Schmid, Burada \&
  H{\"a}nggi]{riefler2010entropic}
{\sc \au{Riefler, Wolfgang}, \au{Schmid, Gerhard}, \au{Burada,
  Poornachandra~Sekhar} \& \au{H{\"a}nggi, Peter}} \yr{2010}  \at{Entropic
  transport of finite size particles}.  \jt{Journal of Physics: Condensed
  Matter}  \bvol{22}~(45),  \pg{454109}.

\bibitem[Rizkallah {\em et~al.\/}(2022)Rizkallah, Sarracino, B{\'e}nichou \&
  Illien]{rizkallah2022microscopic}
{\sc \au{Rizkallah, Pierre}, \au{Sarracino, Alessandro}, \au{B{\'e}nichou,
  Olivier} \& \au{Illien, Pierre}} \yr{2022}  \at{Microscopic theory for the
  diffusion of an active particle in a crowded environment}.  \jt{Physical
  Review Letters}  \bvol{128}~(3),  \pg{038001}.

\bibitem[Rub{\'\i} {\em et~al.\/}(2017)Rub{\'\i}, Lervik, Bedeaux \&
  Kjelstrup]{rubi2017entropy}
{\sc \au{Rub{\'\i}, JM}, \au{Lervik, Anders}, \au{Bedeaux, Dick} \&
  \au{Kjelstrup, Signe}} \yr{2017}  \at{Entropy facilitated active transport}.
  \jt{The Journal of Chemical Physics}  \bvol{146}~(18),  \pg{185101}.

\bibitem[Rubi(2019)]{rubi2019entropic}
{\sc \au{Rubi, J~Miguel}} \yr{2019}  \at{Entropic diffusion in confined
  soft-matter and biological systems}.  \jt{EPL (Europhysics Letters)}
  \bvol{127}~(1),  \pg{10001}.

\bibitem[Sarfati {\em et~al.\/}(2021)Sarfati, Calderon \&
  Schwartz]{sarfati2021enhanced}
{\sc \au{Sarfati, Rapha{\"e}l}, \au{Calderon, Christopher~P} \& \au{Schwartz,
  Daniel~K}} \yr{2021}  \at{Enhanced diffusive transport in fluctuating porous
  media}.  \jt{ACS nano}  \bvol{15}~(4),  \pg{7392--7398}.

\bibitem[Scala {\em et~al.\/}(2007)Scala, Voigtmann \&
  De~Michele]{scala2007event}
{\sc \au{Scala, A}, \au{Voigtmann, Th} \& \au{De~Michele, C}} \yr{2007}
  \at{Event-driven brownian dynamics for hard spheres}.  \jt{The Journal of
  chemical physics}  \bvol{126}~(13),  \pg{134109}.

\bibitem[Simonnin {\em et~al.\/}(2017)Simonnin, Noetinger, Nieto-Draghi, Marry
  \& Rotenberg]{simonnin2017diffusion}
{\sc \au{Simonnin, Pauline}, \au{Noetinger, Beno{\^{i}}t}, \au{Nieto-Draghi,
  Carlos}, \au{Marry, Virginie} \& \au{Rotenberg, Benjamin}} \yr{2017}
  \at{Diffusion under confinement: Hydrodynamic finite-size effects in
  simulation}.  \jt{Journal of Chemical Theory and Computation}  \bvol{13}~(6),
   \pg{2881--2889}.

\bibitem[Solon {\em et~al.\/}(2015)Solon, Fily, Baskaran, Cates, Kafri, Kardar
  \& Tailleur]{solon_pressure_2015}
{\sc \au{Solon, A.~P.}, \au{Fily, Y.}, \au{Baskaran, A.}, \au{Cates, M.~E.},
  \au{Kafri, Y.}, \au{Kardar, M.} \& \au{Tailleur, J.}} \yr{2015}  \at{Pressure
  is not a state function for generic active fluids}.  \jt{Nature Physics}
  \bvol{11}~(8),  \pg{673--678}.

\bibitem[Su{\'a}rez {\em et~al.\/}(2016)Su{\'a}rez, Hoyuelos \&
  M{\'a}rtin]{suarez2016current}
{\sc \au{Su{\'a}rez, Gonzalo}, \au{Hoyuelos, Miguel} \& \au{M{\'a}rtin,
  H{\'e}ctor}} \yr{2016}  \at{Current of interacting particles inside a channel
  of exponential cavities: Application of a modified fick-jacobs equation}.
  \jt{Physical Review E}  \bvol{93}~(6),  \pg{062129}.

\bibitem[Su{\'a}rez {\em et~al.\/}(2015)Su{\'a}rez, Hoyuelos \&
  M{\'a}rtin]{suarez2015transport}
{\sc \au{Su{\'a}rez, Gonzalo~Pablo}, \au{Hoyuelos, Miguel} \& \au{M{\'a}rtin,
  Hector~Omar}} \yr{2015}  \at{Transport of interacting particles in a chain of
  cavities: Description through a modified fick-jacobs equation}.  \jt{Physical
  Review E}  \bvol{91}~(1),  \pg{012135}.

\bibitem[Taylor(1953)]{taylor1953}
{\sc \au{Taylor, Geoffrey~Ingram}} \yr{1953}  \at{Dispersion of soluble matter
  in solvent flowing slowly through a tube}.  \jt{Proceedings of the Royal
  Society of London. Series A. Mathematical and Physical Sciences}
  \bvol{219}~(1137),  \pg{186--203},  \arxiv{arXiv:
  https://royalsocietypublishing.org/doi/pdf/10.1098/rspa.1953.0139}.

\bibitem[Tomkins {\em et~al.\/}(2021)Tomkins, Hughes \&
  Morris]{tomkins2021update}
{\sc \au{Tomkins, Melissa}, \au{Hughes, Aoife} \& \au{Morris, Richard~J}}
  \yr{2021}  \at{An update on passive transport in and out of plant cells}.
  \jt{Plant Physiology}  \bvol{187}~(4),  \pg{1973--1984}.

\bibitem[Vilquin {\em et~al.\/}(2022)Vilquin, Bertin, Rapha{\"e}l, Dean, Salez
  \& Mcgraw]{vilquin2022nanoparticle}
{\sc \au{Vilquin, Alexandre}, \au{Bertin, Vincent}, \au{Rapha{\"e}l, Elie},
  \au{Dean, David~S}, \au{Salez, Thomas} \& \au{Mcgraw, Joshua~D}} \yr{2022}
  \at{Nanoparticle taylor dispersion near charged surfaces with an open
  boundary}.  \jt{arXiv preprint arXiv:2206.07413} .

\bibitem[Werber {\em et~al.\/}(2016)Werber, Osuji \&
  Elimelech]{werber2016materials}
{\sc \au{Werber, Jay~R}, \au{Osuji, Chinedum~O} \& \au{Elimelech, Menachem}}
  \yr{2016}  \at{Materials for next-generation desalination and water
  purification membranes}.  \jt{Nature Reviews Materials}  \bvol{1}~(5),
  \pg{1--15}.

\bibitem[Yang {\em et~al.\/}(2017)Yang, Liu, Li, Marchesoni, H{\"a}nggi \&
  Zhang]{yang2017hydrodynamic}
{\sc \au{Yang, Xiang}, \au{Liu, Chang}, \au{Li, Yunyun}, \au{Marchesoni,
  Fabio}, \au{H{\"a}nggi, Peter} \& \au{Zhang, HP}} \yr{2017}  \at{Hydrodynamic
  and entropic effects on colloidal diffusion in corrugated channels}.
  \jt{Proceedings of the National Academy of Sciences}  \bvol{114}~(36),
  \pg{9564--9569}.

\bibitem[Yoshida {\em et~al.\/}(2018)Yoshida, Kaiser, Rotenberg \&
  Bocquet]{yoshida2018dripplons}
{\sc \au{Yoshida, Hiroaki}, \au{Kaiser, Vojt{\v{e}}ch}, \au{Rotenberg,
  Benjamin} \& \au{Bocquet, Lyd{\'e}ric}} \yr{2018}  \at{Dripplons as localized
  and superfast ripples of water confined between graphene sheets}.  \jt{Nature
  communications}  \bvol{9}~(1),  \pg{1496}.

\bibitem[Zakine {\em et~al.\/}(2020)Zakine, Zhao, Kne\ifmmode \check{z}\else
  \v{z}\fi{}evi\ifmmode~\acute{c}\else \'{c}\fi{}, Daerr, Kafri, Tailleur \&
  van Wijland]{zakine_surfaceTension2020}
{\sc \au{Zakine, R.}, \au{Zhao, Y.}, \au{Kne\ifmmode \check{z}\else
  \v{z}\fi{}evi\ifmmode~\acute{c}\else \'{c}\fi{}, M.}, \au{Daerr, A.},
  \au{Kafri, Y.}, \au{Tailleur, J.} \& \au{van Wijland, F.}} \yr{2020}
  \at{Surface tensions between active fluids and solid interfaces: Bare vs
  dressed}.  \jt{Phys. Rev. Lett.}  \bvol{124},  \pg{248003}.

\bibitem[Zhang {\em et~al.\/}(2020)Zhang, Bertin, Arshad, Raphael, Salez \&
  Maali]{zhang2020direct}
{\sc \au{Zhang, Zaicheng}, \au{Bertin, Vincent}, \au{Arshad, Muhammad},
  \au{Raphael, Elie}, \au{Salez, Thomas} \& \au{Maali, Abdelhamid}} \yr{2020}
  \at{Direct measurement of the elastohydrodynamic lift force at the
  nanoscale}.  \jt{Physical review letters}  \bvol{124}~(5),  \pg{054502}.

\bibitem[Zwanzig(1992)]{zwanzig1992diffusion}
{\sc \au{Zwanzig, Robert}} \yr{1992}  \at{Diffusion past an entropy barrier}.
  \jt{The Journal of Physical Chemistry}  \bvol{96}~(10),  \pg{3926--3930}.

\end{thebibliography}



\section*{Supplementary material}

# Supplementary Material

Y. Wang, D. S. Dean, S. Marbach, and R. Zakine

## Supplementary Discussion

# 1 Boundary conditions in moving channels

We first establish the boundary conditions at the upper and lower moving walls. To do this we consider the condition of conservation of particle number in the total system volume, denoted here by $V$. The total particle number is given by

$$N = \int_V d^2\mathbf{r}\, \rho(\mathbf{r}, t) = \int dx \int_0^{h(x,t)} dy\, \rho(x, y, t) \tag{S1}$$

where $\rho(x, y, t)$ is any density field (tracer density or density of particles) that satisfies a generic conservation equation

$$\frac{\partial \rho(x, y, t)}{\partial t} + \nabla \cdot \mathbf{j}(x, y, t) = 0 \tag{S2}$$

The condition $dN/dt = 0$ then gives

$$\begin{aligned}
0 &= \int dx \frac{\partial h(x,t)}{\partial t} \rho(x, h(x,t)) + \int dx \int_0^{h(x,t)} dy \frac{\partial \rho(x, y, t)}{\partial t} \\
&= \int dx \frac{\partial h(x,t)}{\partial t} \rho(x, h(x,t)) - \int dx \int_0^{h(x,t)} dy\, \nabla \cdot \mathbf{j}(x, y, t) \\
&= \int dx \frac{\partial h(x,t)}{\partial t} \rho(x, h(x,t)) - \int_{\partial V} \mathbf{j}(\mathbf{r}, t) \cdot \mathbf{n}\, dS.
\end{aligned} \tag{S3}$$

Here $\partial V$ denotes the surface of the volume $V$, $\mathbf{n}$ is the outward surface normal and $dS$ is the element of surface. We assume that the system is in such a state that the flux of particles through the surfaces at the horizontal edges, in $x = -L_{\max}/2$ and that in $x = +L_{\max}/2$, vanish or are equal to one another, which is the case with periodic boundary conditions as is performed in numerics. Here $L_{\max}$ is the horizontal size of the periodic simulation box.

The upper surface normal is given by

$$\mathbf{n}|_{y=h(x,t)} = \frac{(-\partial h(x,t)/\partial x, 1)}{\sqrt{(\partial h(x,t)/\partial x)^2 + 1}}, \tag{S4}$$

and $dS = \sqrt{(\partial h(x,t)/\partial x)^2 + 1}\, dx$. The conservation of particle number can thus be written as

$$\int dx \left[ \frac{\partial h(x,t)}{\partial t} \rho(x, h(x,t)) - \left( \mathbf{j}(x, y, t) \cdot \mathbf{n}|_{y=h(x,t)} - \mathbf{j}(x, y, t) \cdot \mathbf{n}|_{y=0} \right) \sqrt{(\partial h(x,t)/\partial x)^2 + 1} \right] = 0. \tag{S5}$$

The boundary condition above must hold for any configuration of the density field and thus should hold at each point $\mathbf{x}$ on both the upper and lower surfaces. It should also hold for any functional form of the interface $h(x, t)$ which allows to split the terms into 2 boundary conditions

$$\mathbf{j}(x, y, t) \cdot \mathbf{n}|_{y=h(x,t)} \sqrt{(\partial h(x,t)/\partial x)^2 + 1} - \frac{\partial h(x,t)}{\partial t} \rho(x, h(x,t)) = 0 \tag{S6}$$

$$\mathbf{j}(x, y, t) \cdot \mathbf{n}|_{y=0} = 0. \tag{S7}$$

Using the expression of the upper surface normal (S4), then gives the upper boundary condition as

$$j_z(x, h(x,t), t) - \frac{\partial h(x,t)}{\partial x} j_x(x, h(x,t), t) = \frac{\partial h(x,t)}{\partial t} \rho(x, h(x,t)) = 0 \tag{S8}$$

that is reported in the main text.

## 2 Flow field calculations

Here we derive the expressions (1.17) and (1.18) from the main text used for the flow field within a channel with a fluctuating interface $h(x,t)$, when the embedded fluid is incompressible.

### 2.1 Notations

The flow field $\boldsymbol{U} = (u, v)$ satisfies the Stokes equations and local conservation of mass (incompressibility)

$$\begin{cases} \frac{\partial^2 u}{\partial y^2} + \frac{\partial^2 u}{\partial x^2} = -\frac{\partial P}{\partial x} \\ \frac{\partial^2 v}{\partial y^2} + \frac{\partial^2 v}{\partial x^2} = -\frac{\partial P}{\partial y} \\ \frac{\partial u}{\partial x} + \frac{\partial v}{\partial y} = 0 \end{cases} \tag{S9}$$

with boundary conditions valid at any boundaries:

$$\begin{cases} (\boldsymbol{U} - \boldsymbol{U}_{\text{wall}}) \cdot \boldsymbol{t} = -2b[\boldsymbol{t} \cdot \boldsymbol{\sigma}(\boldsymbol{U}) \cdot \boldsymbol{n}], \\ (\boldsymbol{U} - \boldsymbol{U}_{\text{wall}}) \cdot \boldsymbol{n} = 0, \end{cases} \tag{S10}$$

which correspond to partial slip on the tangential direction with slip length $b$, and no penetration on the normal direction, respectively. The stress tensor $\boldsymbol{\sigma}(\boldsymbol{U})$ of the fluid is defined as

$$\boldsymbol{\sigma}(\boldsymbol{U}) = \begin{pmatrix} \frac{\partial u}{\partial x} & \frac{1}{2}\left(\frac{\partial u}{\partial y} + \frac{\partial v}{\partial x}\right) \\ \frac{1}{2}\left(\frac{\partial u}{\partial y} + \frac{\partial v}{\partial x}\right) & \frac{\partial v}{\partial y} \end{pmatrix}. \tag{S11}$$

The boundary conditions have to be verified at each boundary (upper and lower walls). The velocity of the top wall/the boundary is $\boldsymbol{U}_{\text{wall}}(x, y = h(x,t)) = (u_{\text{wall,x}}, u_{\text{wall,y}})$. And we will specify the values of the components later on, because there is an important distinction to be made. Note that here $\boldsymbol{U}_{\text{wall}}$ corresponds to the velocity of a wall atom at position $(x, y = h(x,t))$.

The bottom wall is taken immobile and flat. We thus have $\boldsymbol{n} = (0, -1)$ (outwards normal) and $\boldsymbol{t} = (1, 0)$. Since the wall is impermeable, $v(x, y=0) = 0$, and hence $\frac{\partial v}{\partial x}(x, y=0) = 0$. As a consequence we find immediately the classical result at the bottom:

$$\begin{cases} u(x, y=0) = b_0 \frac{\partial u}{\partial y}(x, y=0) \\ v(x, y=0) = 0. \end{cases} \tag{S12}$$

### 2.2 Further simplifications: Lubrication approximation

Now we use the lubrication approximation to find the flow field in its approximate form. In general we will consider that variations along the $x$ axis are "slow" whilst those along $y$ are fast. Formally this can be expressed by taking non dimensional scales $x \to \tilde{x}L$ and $y \to \tilde{y}H$ and $H/L = \delta \ll 1$. As a consequence (in non dimensional scales) $\frac{\partial}{\partial x} = O(1)$ while $\frac{\partial}{\partial y} = O(1/\delta)$. Since local conservation of mass (incompressibility) in the bulk needs to be satisfied at first order (otherwise both flow fields are 0) we have $\frac{\partial u}{\partial x} = O(\frac{\partial v}{\partial y})$ and hence $u/v = O(1/\delta)$. This is physically intuitive, the flow field in the vertical direction is just a small correction compared to the dominant flow field.

Let's now "list" all terms that we will need to compare. First, $\frac{\partial^2 u}{\partial y^2} / \frac{\partial^2 u}{\partial x^2} = O(1/\delta^2)$. Next, we need the higher order terms to be related, so we have $\frac{\partial^2 u}{\partial y^2} = O(-\frac{\partial P}{\partial x})$. Then, since $v/u = O(\delta)$, we find $\frac{\partial^2 v}{\partial y^2} / \frac{\partial^2 u}{\partial y^2} = O(\delta)$. And finally, using $\frac{\partial P}{\partial y} / \frac{\partial P}{\partial x} = O(1/\delta)$, we find $\frac{\partial P}{\partial y} / \frac{\partial^2 u}{\partial y^2} = O(1/\delta)$, such that the NS equations simplify at highest order to

$$\begin{cases} \frac{\partial^2 u}{\partial y^2} = -\frac{\partial P}{\partial x} \\ 0 = -\frac{\partial P}{\partial y} \\ \frac{\partial u}{\partial x} + \frac{\partial v}{\partial y} = 0 \end{cases} \tag{S13}$$

These equations can be simply integrated, in the following order: $P = P(x,t)$ as there is no dependence on $y$ of pressure at lowest order. Then $u$ is integrated as $u = -(y^2/2)\frac{\partial P}{\partial x} + B(x)y + C(x)$ and finally $v = D + (y^3/6)\frac{\partial^2 P}{\partial x^2} - (y^2/2)\frac{\partial B}{\partial x} - \frac{\partial C}{\partial x}y$ where $B, C, D$ are to be determined, as well as $P$, from the boundary conditions.

For the top wall, $\boldsymbol{n} \simeq (-\frac{\partial h}{\partial x}, 1)$ (outwards normal) and $\boldsymbol{t} \simeq (1, \frac{\partial h}{\partial x})$ at lowest order in $\frac{\partial h}{\partial x} \sim h_0/L = O(\delta)$. Then we investigate the components of the stress tensor, when it is projected,

$$\boldsymbol{t} \cdot \boldsymbol{\sigma}(\boldsymbol{U}) \cdot \boldsymbol{n} = \boldsymbol{t} \cdot (-\frac{\partial u}{\partial x}\frac{\partial h}{\partial x} + \frac{1}{2}\left(\frac{\partial u}{\partial y} + \frac{\partial v}{\partial x}\right), -\frac{1}{2}\left(\frac{\partial u}{\partial y} + \frac{\partial v}{\partial x}\right)\frac{\partial h}{\partial x} + \frac{\partial v}{\partial y})$$
$$= -\frac{\partial u}{\partial x}\frac{\partial h}{\partial x} + \frac{1}{2}\left(\frac{\partial u}{\partial y} + \frac{\partial v}{\partial x}\right) - \frac{1}{2}\left(\frac{\partial u}{\partial y} + \frac{\partial v}{\partial x}\right)(\frac{\partial h}{\partial x})^2 + \frac{\partial v}{\partial y}(\frac{\partial h}{\partial x})$$
(S14)

It is clear that in all of these terms the highest order term is in fact only $\frac{\partial u}{\partial y}/2$ hence the boundary conditions at the top boundary sum up to

$$\begin{cases} u(y = h) - u_{\text{wall,x}} = -b_h \frac{\partial u}{\partial y}(y = h) \\ v(y = h) - u(y = h)\frac{\partial h}{\partial x} = u_{\text{wall,y}} - u_{\text{wall,x}}\frac{\partial h}{\partial x}. \end{cases}$$
(S15)

## 2.3 What is the velocity of the wall particles?

We consider in the simulations that the boundary's position obeys is $h(x,t) = H + h_0 \cos(kx - \omega t) = H + h_0 \cos\left(\frac{2\pi}{L}(x - v_{\text{wall}}t)\right)$. While this explicitly gives the *position* of the boundary it *does not* give the expression for the velocity of the wall atoms, which is what is needed to specify the boundary conditions via $\boldsymbol{U}_{\text{wall}}$. There are two possibilities. A first possibility would be that the wall atoms perform peristalsis or pumping (Chakrabarti & Saintillan, 2020, Marbach & Alim, 2019), *e.g.* atoms do not move in average but jiggle vertically. There is no average mass displacement in that case. This situation is quite common for phonons on carbon nanotubes (Ma *et al.*, 2015) or peristaltic waves of intestine walls (Codutti *et al.*, 2022) or in vasculature networks (Alim *et al.*, 2013). In that case $\boldsymbol{U}_{\text{wall}} = (0, \partial_t h)$. The other possibility is that there is average mass displacement of wall atoms, but the atoms do not move relative to one another. In that case, $\boldsymbol{U}_{\text{wall}} = (v_{\text{wall}}, 0)$. We will focus on peristalsis here.

## 2.4 Integration process to obtain the flow field

The boundary conditions thus obey

$$\text{partial slip: } \begin{cases} u(y=0) = b_0 \frac{\partial u}{\partial y}(y=0) \\ u(y=h) = -b_h \frac{\partial u}{\partial y}(y=h) \end{cases}$$
(S16)

$$\text{non penetration: } \begin{cases} v(y=h) - u(y=h)\frac{\partial h}{\partial x} = \partial_t h \\ v(y=0) = 0 \end{cases}$$
(S17)

which correspond respectively to partial slip on both walls, and non penetration boundary conditions. The quantities $b_0$ (respectively $b_h$) indicate the slip lengths on the bottom (respectively top) walls.

Integrating for $u$ yields

$$u(x,y,t) = 6U_0(x,t)\frac{\frac{b_0}{h}\left(\frac{2b_h}{h}+1\right) + \left(\frac{2b_h}{h}+1\right)\frac{y}{h} - \left(\frac{b_0}{h}+\frac{b_h}{h}+1\right)\frac{y^2}{h^2}}{12\frac{b_0}{h}\frac{b_h}{h} + 4\left(\frac{b_0}{h}+\frac{b_h}{h}\right)+1} = 6U_0(x,t)f(h(x,t),y)$$
(S18)

where $U_0(x,t) = \frac{1}{h}\int_0^h u\,dy$ is the average flow field in the $x$ direction. The expression for $u$ is slightly cumbersome but it has natural limits with limiting regimes of the slip lengths $b_0$ and $b_h$. In particular, in our case where $b_0 = b_h = \infty$ (perfect slip at both walls) we obtain (1.17). Because of mass conservation we also have

$$\frac{\partial(hU_0)}{\partial x} = -\frac{\partial h}{\partial t}$$
(S19)

4

such that $\frac{\partial U}{\partial x}\big|_0 = -\frac{1}{h}(\frac{\partial h}{\partial t} + U_0 \frac{\partial h}{\partial x})$. We can therefore express

$$\frac{\partial v}{\partial y} = -\frac{\partial u}{\partial x} = -\left(6\frac{\partial U_0}{\partial x}f(h(x,t),y) + 6U_0(x,t)\frac{\partial h}{\partial x}\frac{\partial f(h(x,t),y)}{\partial h}\right) \tag{S20}$$

and integration and use of boundary conditions finally yield a general expression for $v(x,y,t)$ that is quite lengthy and that we do not report here. Again, for $b_0 = b_h \to \infty$, we obtain (1.18).

Now what about the value of $U_0(x,t)$? With conservation of mass we obtain

$$U_0(x,t) = U_0(x=0,t)\frac{h(x=0,t)}{h(x,t)} + \frac{h_0}{h}v_{\text{wall}}\left[\cos\left(\frac{2\pi}{L}(x-v_{\text{wall}}t)\right) - \cos\left(\frac{2\pi}{L}v_{\text{wall}}t\right)\right] \tag{S21}$$

but we are missing the boundary term $U_0(x=0,t)$. To be coherent with our assumption that the wall atoms do not move in average, namely of peristalsis, then it is natural to expect that there is no pressure driven flow and hence that the average pressure drop has to vanish $\int_0^L \frac{\partial P}{\partial x}dx = 0$ (the average force on the fluid is 0, see Chakrabarti & Saintillan (2020)). Expanding $\int_0^L \frac{\partial P}{\partial x}dx$ in powers of $\epsilon$ we simply obtain at lowest order that $\int_0^L U_0(x,t)dx = 0$. After an easy integration we obtain that $U_0(x=0,t) = v_{\text{wall}}$ and hence $U_0(x,t) \simeq v_{\text{wall}}\frac{h_0}{H}\cos\left(\frac{2\pi}{L}(x-v_{\text{wall}}t)\right)$.

5

# 3 Long time transport coefficients for space and time varying diffusivity

In this section we derive the long time transport coefficients for tracers in a bath of interacting particle when we do not neglect the contribution of the mean density to the local diffusion coefficient.

Our starting point, as in the main paper, is to obtain the stationary density of particles $\bar{\rho}(x)$ in the frame of reference where the wall is immobile, but now without neglecting the dependence of the diffusion coefficient $\bar{D}_0(\bar{\rho}(x))$ on density. The steady state flux of particles is obtained by assuming in (3.7) of the main text that $\bar{\rho}(x,y) \simeq \bar{\rho}(x)$, such that

$$J = h(x)\left(-D(x)\frac{\partial \bar{\rho}(x)}{\partial x} - v_0 D(x)\bar{\rho}(x)\frac{\partial \bar{\rho}(x)}{\partial x} + \bar{\rho}(x)v_{\text{wall}}\right), \tag{S22}$$

where we reversed the direction of $v_{\text{wall}}$ compared to the main text to obtain $J > 0$, we introduced $v_0 \equiv E_0/(k_B T \rho_0)$ a characteristic interaction volume, and we introduced the shorthand notation $D(x) \equiv \bar{D}_0(\bar{\rho}(x))$. For simplicity we also introduce $f(x)$ such that $h(x) = H(1 + \varepsilon f(x))$ (in the frame of reference where the channel is fixed), and where $\int_0^L f(x)\mathrm{d}x = 0$ by definition. To pursue the derivation we will need to obtain the expression of $J$ at order $\varepsilon^2$ as well as that of $\bar{\rho}(x) = \rho_0 + \varepsilon\rho_1(x) + \varepsilon^2\rho_2(x)$. Notice, compared to the main text, that now we assume $D(x)$ depends on space as

$$D(x) = \bar{D}_0(\rho_0) + \bar{D}_0'(\rho_0)(\varepsilon\rho_1(x) + \varepsilon^2\rho_2(x))) + \frac{1}{2}\bar{D}_0''(\rho_0)(\varepsilon\rho_1(x) + \varepsilon^2\rho_2(x)))^2 + ..., \tag{S23}$$

and where the $'$ denotes the derivative with respect to $\rho_0$. In all of these cases note that the solutions $\rho_i(x)$ should still be periodic. We will also make use of the conservation of mass

$$\int_0^L \rho(x)h(x)\mathrm{d}x = \int_0^L (\rho_0 + \varepsilon\rho_1(x) + \varepsilon^2\rho_2(x))H(1 + \varepsilon f(x))\mathrm{d}x = \rho_0 L H \tag{S24}$$

that we can expand at all orders in $\varepsilon$ to yield (using the fact that $\int f(x)\mathrm{d}x = 0$),

$$\int_0^L \rho_1(x)\mathrm{d}x = 0 \;,\; \int_0^L (\rho_2(x) + \rho_1(x)f(x))\,\mathrm{d}x = 0. \tag{S25}$$

In the following we will only need integrals $\int_0^L \rho_2(x)\mathrm{d}x$, and not the expression of $\rho_2(x)$, which means we do not have to explicitly calculate $\rho_2(x)$ and rather we can simply express its integral with respect to $\rho_1(x)$.

Now we also expand $J = J_0 + \varepsilon J_1 + \varepsilon^2 J_2$. Carefully expanding at all orders in $\varepsilon$, we obtain

$$\begin{cases} J_0 &= H\rho_0 v_{\text{wall}}, \\ J_1 &= Hf(x)\rho_0 v_{\text{wall}} - H\bar{D}_0\rho_1'(x) - Hv_0\bar{D}_0\rho_0\rho_1'(x) + H\rho_1(x)v_{\text{wall}}, \\ J_2 &= Hf(x)\rho_1(x)v_{\text{wall}} - Hf(x)\bar{D}_0\rho_1'(x) - Hf(x)v_0\bar{D}_0\rho_0\rho_1'(x) \\ &\quad - H\bar{D}_0'(\rho_0)\rho_1\rho_1'(x) - Hv_0\bar{D}_0'(\rho_0)\rho_1\rho_0\rho_1'(x) - Hv_0\bar{D}_0\rho_1(x)\rho_1'(x) \\ &\quad - H\bar{D}_0\rho_2'(x) - Hv_0\bar{D}_0\rho_0\rho_2'(x) + H\rho_2(x)v_{\text{wall}}. \end{cases} \tag{S26}$$

We can simplify these expressions by taking their averages over a period, and using periodicity and conservation of mass, we find

$$\begin{cases} J_0 &= H\rho_0 v_{\text{wall}}, \\ J_1 &= 0, \\ J_2 &= -\frac{H}{L}D^\star \int_0^L \mathrm{d}x f(x)\rho_1'(x) \end{cases} \tag{S27}$$

where we recall that $D^\star = \bar{D}_0(\rho_0)(1 + v_0\rho_0)$.

6

We can obtain the analytical expression for $\rho_1(x)$ by solving the equation for $J_1$ before doing the average,

$$J_1 = 0 = Hf(x)\rho_0 v_{\text{wall}} - HD^\star \rho_1'(x) + H\rho_1(x)v_{\text{wall}}, \qquad (S28)$$

which does not change compared to the main text and is simply

$$\rho_1(x) = \rho_0 \frac{\text{Pe}^\star}{1 + (\text{Pe}^\star)^2} \left(\sin(k_0 x) - \text{Pe}^\star \cos(k_0 x)\right). \qquad (S29)$$

Notice that compared to the solution of the main text some signs are reversed that correspond to the reversal of $v_{\text{wall}}$.

Finally, we can insert the expression for $\rho_1(x)$ to obtain the expression of the flux at second order

$$J = H\rho_0 v_{\text{wall}} - \frac{H}{L}D^\star \int_0^L dx f(x)\rho_1'(x) = H\rho_0 v_{\text{wall}}\left(1 - \frac{\varepsilon^2}{2}\frac{1}{1 + (\text{Pe}^\star)^2}\right). \qquad (S30)$$

The subsequent step is to analyze equation (3.14) of the main manuscript, on the marginal density $p_{\text{tr}}(x, t)$ to find the tracer, in the frame of reference where the wall is fixed

$$\frac{\partial p_{\text{tr}}(x,t)}{\partial t} = \frac{\partial}{\partial x}\left(\overline{D}_0(\overline{\rho}(x))\left[\frac{\partial p_{\text{tr}}}{\partial x} - p_{\text{tr}}\frac{\partial \ln h(x,t)}{\partial x} + \frac{E_0}{k_B T}\frac{p_{\text{tr}}(x,t)}{\rho_0}\frac{\partial \overline{\rho}}{\partial x}\right] - v_{\text{wall}} p_{\text{tr}}\right). \qquad (S31)$$

Notice that compared to (3.14) we have reversed the sign of $v_{\text{wall}}$.

Now we will first recast the above equation with easier notations

$$\frac{\partial p_{\text{tr}}(x,t)}{\partial t} = \frac{\partial}{\partial x}\left(D(x)\frac{\partial p_{\text{tr}}(x,t)}{\partial x} - D(x)p_{\text{tr}}(x,t)\frac{\partial \ln h(x)}{\partial x} + D(x)v_0 p_{\text{tr}}(x,t)\frac{\partial \rho(x)}{\partial x} - v p_{\text{tr}}(x,t)\right) \qquad (S32)$$

where $D(x)$ contains all the spatial dependence of the diffusion coefficient induced by the locally varying density and we abbreviate $v \equiv v_{\text{wall}}$.

Now recall that

$$J \equiv h(x)\left(D(x)\frac{\partial \rho(x)}{\partial x} + v_0 D(x)\rho(x)\frac{\partial \rho(x)}{\partial x} - \rho(x)v\right) \qquad (S33)$$

is the integrated longitudinal flux of the steady state local density. We can then recast (S32) again into

$$\frac{\partial p_{\text{tr}}(x,t)}{\partial t} = \frac{\partial}{\partial x}\left(D(x)\frac{\partial p_{\text{tr}}(x,t)}{\partial x} - D(x)p_{\text{tr}}(x,t)\frac{\partial \ln h(x)\rho(x)}{\partial x} - \frac{J}{\rho(x)h(x)}p_{\text{tr}}(x,t)\right). \qquad (S34)$$

To obtain the long time diffusion and drift of this equation, we finally need to put it under the form used in Ref. Guérin & Dean (2015)

$$\frac{\partial p_{\text{tr}}(x,t)}{\partial t} = \frac{\partial}{\partial x}\left(\frac{\partial}{\partial x}(\kappa(x)p_{\text{tr}}(x,t)) - u(x)p_{\text{tr}}(x,t)\right). \qquad (S35)$$

We see immediately that $\kappa(x) \equiv D(x)$ and

$$u(x) \equiv \frac{\partial D(x)}{\partial x} + \frac{J}{\rho(x)h(x)} + D(x)\frac{\partial \ln(\rho(x)h(x))}{\partial x} = \frac{1}{\rho(x)h(x)}\left(J + \frac{\partial D(x)\rho(x)h(x)}{\partial x}\right). \qquad (S36)$$

We now can use exact results for one dimensional systems in Ref. Guérin & Dean (2015), which give

$$V_{\text{eff}} = \frac{L}{\int_0^L dx I_+(x)} \qquad (S37)$$

and

$$D_{\text{eff}} = \frac{L^2 \int_0^L dx D(x) I_+(x)^2 I_-(x)}{[\int_0^L dx I_+(x)]^3}, \qquad (S38)$$

7

where
$$I_+(x) = \frac{1}{\kappa(x)} e^{\Gamma(x)} \int_x^{+\infty} e^{-\Gamma(x')} dx' \tag{S39}$$

and symmetrically
$$I_-(x) = \frac{1}{\kappa(x)} e^{-\Gamma(x)} \int_{-\infty}^x e^{\Gamma(x')} dx' \tag{S40}$$

with
$$\Gamma(x) = \int_0^x \frac{u(x')}{\kappa(x')} dx'. \tag{S41}$$

Notice that the above integrals indeed converge because the average direction of the drift is towards the right ($u$ is positive). In the case where the drift is reversed, one would simply have to swap integral bounds in the definitions of $I_+(x)$ and $I_-(x)$.

Our aim is noe the drift is w to simplify the above expressions for $V_{\text{eff}}$ and $D_{\text{eff}}$. We start by simplifying the expressions for $I_+(x)$ and $I_-(x)$. Notice that

$$\begin{aligned}\Gamma(x) &= \int_0^x \frac{1}{D(x')\rho(x')h(x')} \left(J + \frac{\partial D(x')\rho(x')h(x')}{\partial x'}\right) dx' \\ &= \ln\left(\frac{D(x)\rho(x)h(x)}{D(0)\rho(0)h(0)}\right) + \int_0^x \frac{J}{D(x')\rho(x')h(x')} dx'\end{aligned} \tag{S42}$$

and therefore we can simplify

$$\begin{aligned}I_+(x) =& \frac{1}{D(x)} \frac{D(x)\rho(x)h(x)}{D(0)\rho(0)h(0)} \exp\left(J \int_0^x \frac{dx'}{h(x')\rho(x')D(x')}\right) \times \\ & \int_x^\infty dx' \frac{D(0)\rho(0)h(0)}{h(x')\rho(x')D(x')} \exp\left(-J \int_0^{x'} \frac{dx''}{h(x'')\rho(x'')D(x'')}\right) \\ =& \rho(x)h(x) \exp\left(J \int_0^x \frac{dx'}{h(x')\rho(x')D(x')}\right) \times \\ & \int_x^\infty dx' \frac{1}{h(x')\rho(x')D(x')} \exp\left(-J \int_0^{x'} \frac{dx''}{h(x'')\rho(x'')D(x'')}\right) \\ =& -\frac{1}{J} \rho(x)h(x) \exp\left(J \int_0^x \frac{dx'}{h(x')\rho(x')D(x')}\right) \times \\ & \left[\exp\left(-J \int_0^\infty \frac{dx''}{h(x'')\rho(x'')D(x'')}\right) - \exp\left(-J \int_0^x \frac{dx''}{h(x'')\rho(x'')D(x'')}\right)\right].\end{aligned} \tag{S43}$$

Finally, notice that $\int_0^\infty \frac{dx''}{h(x'')\rho(x'')D(x'')} \to +\infty$ since all of the functions $h(x)$, $\rho(x)$ and $D(x)$ are all positive and periodic. We therefore obtain, since $J > 0$ (since we reversed the sign of $v_{\text{wall}}$ in this entire section),

$$I_+(x) = +\frac{\rho(x)h(x)}{J}. \tag{S44}$$

The term $I_+(x)$ is thus proportional to the steady state density.

On the other hand we have

$$I_-(x) = \frac{1}{D(x)h(x)\rho(x)} \exp\left(-J \int_0^x \frac{dx'}{h(x')\rho(x')D(x')}\right) \int_{-\infty}^x dx' h(x')\rho(x') \exp\left(J \int_0^{x'} \frac{dx''}{h(x'')\rho(x'')D(x'')}\right), \tag{S45}$$

which does not simplify as easily as $I_+(x)$.

We can now turn to simplifying $V_{\text{eff}}$. It is obtained via (S37) and (S44)

$$V_{\text{eff}} = \frac{LJ}{\int_0^L dx h(x)\rho(x)} = \frac{J}{\rho_0 H}, \tag{S46}$$

8

which is clearly a result that can be argued from physical grounds. Naturally, the flux of particles $J = \rho_0 H V_{\text{eff}}$, is the mean velocity multiplied by the average density. This result can be written using (S30) as

$$V_{\text{eff}} = v_{\text{wall}}\left(1 - \frac{\varepsilon^2}{2}\frac{1}{1 + (\text{Pe}^\star)^2}\right), \qquad (S47)$$

which means that going back to the rest frame of the fluid we find

$$V_{\text{eff}} = v_{\text{wall}}\frac{\varepsilon^2}{2}\frac{1}{1 + (\text{Pe}^\star)^2}. \qquad (S48)$$

We now make progress on the calculation of $D_{\text{eff}}$. From (S38) and our expanded expression for $I_+(x)$ we obtain

$$D_{\text{eff}} = \frac{JL^2 \int_0^L dx\, D(x)\rho^2(x)h^2(x)I_-(x)}{[\int_0^L dx\, \rho(x)h(x)]^3} = \frac{J \int_0^L dx\, D(x)\rho^2(x)h^2(x)I_-(x)}{L[\rho_0 H]^3}. \qquad (S49)$$

We must make more progress on $I_-(x)$ to obtain a more explicit result. Let

$$R(x) \equiv \frac{J}{h(x)\rho(x)D(x)}. \qquad (S50)$$

$R(x)$ is not periodic however we can split it up between a periodic and a non periodic part by defining

$$\overline{R} = \frac{1}{L}\int_0^L R(x)dx \text{ and } r(x) = \int_0^x (R(x') - \overline{R})dx' \qquad (S51)$$

such that

$$\int_0^x dx'\, R(x') = r(x) + \overline{R}x. \qquad (S52)$$

The function $r(x)$ is now clearly periodic. Now we use

$$\begin{aligned}
I_-(x)\frac{J}{R(x)} &= \exp(-r(x) - \overline{R}x)\int_{-\infty}^x dx'\, h(x')\rho(x')\exp(r(x') + \overline{R}x') \\
&= \exp(-r(x) - \overline{R}x)\sum_{n=0}^\infty \int_{x-(n+1)L}^{x-nL} dx'\, h(x')\rho(x')\exp(r(x') + \overline{R}x') \\
&= \exp(-r(x) - \overline{R}x)\sum_{n=0}^\infty \int_x^{x+L} dx'\, h(x')\rho(x')\exp(r(x') + \overline{R}(x' - (n+1)L)) \\
&= \exp(-r(x) - \overline{R}x)\int_x^{x+L} dx'\, h(x')\rho(x')\exp(r(x') + \overline{R}x')\sum_{n=0}^\infty e^{-\overline{R}(n+1)L} \\
&= \exp(-r(x) - \overline{R}x)\int_x^{x+L} dx'\, h(x')\rho(x')\exp(r(x') + \overline{R}x')\frac{1}{\exp(\overline{R}L) - 1} \\
&= \frac{1}{\exp(\overline{R}L) - 1}\exp(-r(x) - \overline{R}x)\int_0^L dx'\, h(x + x')\rho(x + x')\exp(r(x + x') + \overline{R}(x + x')) \\
&= \frac{1}{\exp(\overline{R}L) - 1}\int_0^L dx'\, h(x + x')\rho(x + x')\exp(r(x + x') - r(x) + \overline{R}x'). \qquad (S53)
\end{aligned}$$

We can now go back to $D_{\text{eff}}$,

$$D_{\text{eff}} = J\frac{\int_0^L dx \int_0^L dx'\, h(x)\rho(x)h(x + x')\rho(x + x')\exp(r(x + x') - r(x) - \overline{R}x')}{L\rho_0^3 H^3(\exp(\overline{R}L) - 1)}. \qquad (S54)$$

We must now expand $D_{\text{eff}}$ with respect to $\varepsilon$. Before moving on, we observe that

$$\begin{aligned}
\overline{R} &= \frac{1}{L}\int_0^L \frac{V_{\text{eff}}}{D(x)}\frac{\rho_0 H}{\rho(x)h(x)}dx \\
&= \ldots \\
&= \frac{v_{\text{wall}}}{\overline{D}_0}\left[1 - \frac{\varepsilon^2}{2}\frac{(\text{Pe}^\star)^2}{1+(\text{Pe}^\star)^2}\left(\varepsilon_D + \varepsilon_{D,2} - \varepsilon_D^2\right)\right] \\
&\equiv \overline{\text{Pe}}\frac{2\pi}{L}\left(1 + \varepsilon^2 R_2\right)
\end{aligned} \quad (S55)$$

where we used any formal analysis software to obtain the systematic expansion of $1/(D(x)\rho(x)h(x))$ with respect to $\varepsilon$ and then its integral and introduced

$$\varepsilon_D = \frac{\overline{D}_0'(\rho_0)\rho_0}{\overline{D}_0(\rho_0)} \quad \text{and} \quad \varepsilon_{D,2} = \frac{\overline{D}_0''(\rho_0)\rho_0^2}{2\overline{D}_0(\rho_0)}. \quad (S56)$$

It is thus clear that $r(x)$ is of order $\varepsilon$ at least, and we can write

$$\begin{aligned}
r(x) &= -\varepsilon\frac{v_{\text{wall}}}{\overline{D}_0}\int_0^x dx'\left(f(x') + \frac{\rho_1(x')}{\rho_0} + \frac{\overline{D}_0'(\rho_0)}{\overline{D}_0}\rho_1(x') + O(\varepsilon)\right) \\
&= -\varepsilon\frac{\overline{\text{Pe}}}{1+(\text{Pe}^\star)^2}\left(-2(1+\varepsilon_D)\text{Pe}^\star \sin(\pi x/L)^2 + (\varepsilon_D(\text{Pe}^\star)^2 - 1)\sin(2\pi x/L)\right) + O(\varepsilon) \\
&\equiv \varepsilon r_1(x) + O(\varepsilon^2).
\end{aligned} \quad (S57)$$

In what follows, we show that it is sufficient to work at order $O(\varepsilon)$ on $r(x)$ to obtain the relevant results. Using this $O(\varepsilon)$ approximation on $r_1$ we obtain

$$D_{\text{eff}} = \frac{J}{L\rho_0^3 H^3(\exp(\overline{R}L) - 1)}\int_0^L dx \int_0^L dx' h(x)\rho(x)h(x+x')\rho(x+x')\exp(\varepsilon r_1(x+x') - \varepsilon r_1(x) + \overline{R}x'). \quad (S58)$$

Let $I(x,x') = h(x)\rho(x)h(x+x')\rho(x+x')\exp(\varepsilon r_1(x+x') - \varepsilon r_1(x) + \overline{R}x')$ be the integrand, and we now expand it to order $\varepsilon^2$,

$$\begin{aligned}
I(x,x') &= H^2(1+\varepsilon f(x))(1+\varepsilon f(x+x'))(\rho_0 + \varepsilon\rho_1(x) + \varepsilon^2\rho_2(x))(\rho_0 + \varepsilon\rho_1(x+x') + \varepsilon^2\rho_2(x+x')) \\
&\quad \times \exp(\overline{R}x')\left(1 + \varepsilon(r_1(x+x') - r_1(x)) + \frac{\varepsilon^2}{2}(r_1(x+x') - r_1(x))^2\right).
\end{aligned} \quad (S59)$$

Before expanding more, to make progress we will already remove all the terms that integrate to 0, keeping in mind that

$$0 = \int_0^L dx f(x+a) = \int_0^L dx \rho_1(x+a) = \int_0^L dx\left(\rho_1(x+a)f(x+a) + \rho_2(x+a)\right) \quad (S60)$$

for any $a$ as well as the identity

$$\int_0^L dx(g(x) - g(x+a)) = 0, \quad (S61)$$

for any periodic function $g$, and in particular for $r_1(x)$,

$$\begin{aligned}
I(x,x') &= H^2 \exp(\overline{R}x')\Big(\rho_0^2 + \varepsilon^2\Big[\rho_0^2 f(x)f(x+x') + \rho_1(x)\rho_1(x+x') + \rho_1(x)\rho_0 f(x+x') + \rho_1(x+x')\rho_0 f(x) \\
&\quad + (f(x)\rho_0 + f(x+x')\rho_0 + \rho_1(x) + \rho_1(x+x'))(r_1(x+x') - r_1(x)) \\
&\quad + \frac{1}{2}(r_1(x+x') - r_1(x))^2\Big]\Big) \\
&= H^2 \exp(\overline{R}x')\Big(\rho_0^2 + \varepsilon^2\Big[\rho_0^2 f(x)f(x+x') + \rho_1(x)\rho_1(x+x') + \rho_1(x)\rho_0 f(x+x') + \rho_1(x+x')\rho_0 f(x) \\
&\quad + (f(x)\rho_0 + \rho_1(x)) r_1(x+x') - (f(x+x')\rho_0 + \rho_1(x+x'))r_1(x) \\
&\quad + \frac{1}{2}(r_1(x+x') - r_1(x))^2\Big]\Big) \\
&\equiv H^2 \rho_0^2 \exp(\overline{R}x')\Big(1 + \varepsilon^2 \mathcal{I}(x,x')\Big). \quad (\text{S}62)
\end{aligned}$$

This approach eliminates the appearance of the function $\rho_2(x)$. In addition, we find that all terms in the expansion of $r_1(x)$ appear at order $\varepsilon^2$ in $I(x,x')$ and therefore we indeed only need the order 1 terms in $r(x)$.

$$D_{\text{eff}} = \frac{J}{\rho_0 H} \frac{1}{\overline{R}} + \frac{J}{\rho_0 H} \varepsilon^2 \frac{\int_0^L dx' \int_0^L dx' \mathcal{I}(x,x')}{L\left(\exp(\overline{R}L) - 1\right)}. \quad (\text{S}63)$$

We recall that

$$\frac{J}{\rho_0 H} = V_{\text{eff}} = v_{\text{wall}}\left(1 - \frac{\varepsilon^2}{2} \frac{1}{1 + (\text{Pe}^\star)^2}\right) \quad (\text{S}64)$$

and $\overline{R} = \frac{v_{\text{wall}}}{\overline{D}_0}\left(1 + \varepsilon^2 R_2\right)$ such that

$$D_{\text{eff}} = \overline{D}_0\left(1 - \varepsilon^2 R_2 - \frac{\varepsilon^2}{2} \frac{1}{1 + (\text{Pe}^\star)^2}\right) + v_{\text{wall}} \varepsilon^2 \frac{1}{L} \int_0^L dx' \int_0^L dx' \mathcal{I}(x,x'). \quad (\text{S}65)$$

Carrying on the integrals with the use of an automated software for example, it is quite straightforward to show

$$D_{\text{eff}} = \overline{D}_0\left(1 + \frac{\varepsilon^2}{2} \frac{3\overline{\text{Pe}}^2 - 1 - (\text{Pe}^\star)^2 \varepsilon_D^2 + (1 + \overline{\text{Pe}}^2)(\text{Pe}^\star)^2 [\varepsilon_D + \varepsilon_{D,2}]}{(1 + (\text{Pe}^\star)^2)(1 + \overline{\text{Pe}}^2)}\right) \quad (\text{S}66)$$

where we recall that

$$\varepsilon_D = \frac{\overline{D}_0'(\rho_0)\rho_0}{\overline{D}_0(\rho_0)} \quad \text{and} \quad \varepsilon_{D,2} = \frac{\overline{D}_0''(\rho_0)\rho_0^2}{2\overline{D}_0(\rho_0)}. \quad (\text{S}67)$$

In this compact form notice that we recover the limit where $D_{\text{eff}}$ is given by (3.19) of the main text when the diffusion coefficient is assumed to be uniform in space, namely when $\varepsilon_D = 0$ and $\varepsilon_{D,2} = 0$.

In summary we found that the effective drift is not changed by a varying diffusion coefficient in this context. Likely the largest contribution to the effective diffusion coefficient is that proportional to $\varepsilon_D$, such that (S66) can be approximated by

$$D_{\text{eff}} \simeq \overline{D}_0\left(1 + \frac{\varepsilon^2}{2} \frac{3\overline{\text{Pe}}^2 - 1}{(1 + (\text{Pe}^\star)^2)(1 + \overline{\text{Pe}}^2)} + \varepsilon_D \frac{\varepsilon^2}{2} \frac{(\text{Pe}^\star)^2}{1 + (\text{Pe}^\star)^2}\right). \quad (\text{S}68)$$

In our simulations, at most $\varepsilon_D \simeq 0.1$ which yields negligible corrections numerically. Notice that, based on a general perturbation theory that derives the long-time diffusion coefficient with spatiotemporal varying local drift and diffusivity, we can recover the same results found here. The details of this perturbation theory will be published elsewhere.

# Supplementary References

Alim, Karen, Amselem, Gabriel, Peaudecerf, François, Brenner, Michael P & Pringle, Anne 2013 Random network peristalsis in physarum polycephalum organizes fluid flows across an individual. *Proceedings of the National Academy of Sciences* **110** (33), 13306–13311.

Chakrabarti, Brato & Saintillan, David 2020 Shear-induced dispersion in peristaltic flow. *Physics of Fluids* **32** (11), 113102.

Codutti, Agnese, Cremer, Jonas & Alim, Karen 2022 Changing flows balance nutrient absorption and bacterial growth along the gut. *bioRxiv* .

Guérin, T. & Dean, D. S. 2015 Kubo formulas for dispersion in heterogeneous periodic nonequilibrium systems. *Physical Review E* **92** (6), 062103.

Ma, Ming, Grey, François, Shen, Luming, Urbakh, Michael, Wu, Shuai, Liu, Jefferson Zhe, Liu, Yilun & Zheng, Quanshui 2015 Water transport inside carbon nanotubes mediated by phonon-induced oscillating friction. *Nature nanotechnology* **10** (8), 692–695.

Marbach, Sophie & Alim, Karen 2019 Active control of dispersion within a channel with flow and pulsating walls. *Physical Review Fluids* **4** (11), 114202.



\end{document}